\documentclass{JHEP3}

  \usepackage{latexsym,bm,amsmath,amssymb,amsfonts}
  \usepackage{epsfig,graphics,graphicx}
  \usepackage{slashed}

  \long\def\comment#1{ }

  \newcommand{\eqnum}[1]{Eq.~\eqref{#1}}
  \newcommand{\dif}{{\rm d}}

  \newcommand{\del}{\partial}

  \newcommand{\mcal}{\mathcal}
  \newcommand{\rme}{{\rm e}}
  \newcommand{\rmd}{{\rm d}}   

  \newcommand{\Lam}{\Lambda_{{\rm QCD}}}
  
  \newcommand{\nn}{\nonumber\\}
   
  \newcommand{\order}[1]{\mcal{O}{(#1)}}

  \newcommand{\beq}{\begin{eqnarray}}
  \newcommand{\eeq}{\end{eqnarray}}
  
 \def\simge{\mathrel{%
   \rlap{\raise 0.511ex \hbox{$>$}}{\lower 0.511ex \hbox{$\sim$}}}}
\def\simle{\mathrel{
   \rlap{\raise 0.511ex \hbox{$<$}}{\lower 0.511ex \hbox{$\sim$}}}}

\providecommand{\href}[2]{#2}

\title{\rm \LARGE Partons and jets in a strongly--coupled plasma from AdS/CFT\footnote{
Based on lectures presented at the XLVIII Cracow School of Theoretical Physics,
{\em Aspects of Duality},
Zakopane, Poland, June 13–-22, 2008.}}

\author{Edmond Iancu\\Institut de Physique Th\'eorique de Saclay,
 F-91191 Gif-sur-Yvette, France\\
        E-mail: \email{Edmond.Iancu@cea.fr}}

\abstract{We give a pedagogical review of recent progress towards understanding
the response of a strongly coupled plasma at finite temperature to a hard probe.
The plasma is that of the ${\mathcal N}=4$
supersymmetric Yang--Mills theory and the hard probe is a virtual photon, or,
more precisely, an ${\mathcal R}$--current. Via the gauge/gravity duality, the
problem of the current interacting with the plasma is mapped onto the
gravitational interaction between a Maxwell field and a black hole embedded in
the $AdS_5\times S^5$ geometry. The physical interpretation of the AdS/CFT
results can be then reconstructed with the help of the ultraviolet/infrared
correspondence. We thus deduce that, for sufficiently high energy, the photon
(or any other hard probe: a quark, a gluon, or a meson) disappears into the
plasma via a universal mechanism, which is quasi--democratic
parton branching: the current develops a parton cascade such that, at any step
in the branching process, the energy is almost equally divided among the
daughter partons. The branching rate is controlled by the plasma which acts
on the colored partons with a constant force $\sim T^2$. When reinterpreted in
the plasma infinite momentum frame, the same AdS/CFT results suggest a
parton picture for the plasma structure functions, in which all the
partons have fallen at very small values of Bjorken's $x$. For a
time--like current in the vacuum, quasi--democratic branching implies that
there should be no jets in electron--positron annihilation at strong coupling,
but only a spatially isotropic distribution of hadronic matter.

}

\begin{document}

\section{Introduction: From RHIC physics and lattice QCD to AdS/CFT}
\label{Intro}

One of the most interesting suggestions emerging from the heavy ion
program at RHIC is the fact that the deconfined, `quark--gluon' matter
produced in the early stages of an ultrarelativistic nucleus--nucleus
collision might be {\em strongly interacting} (see the summary of the
experimental results in the ``white papers'' of the four experiments at
RHIC \cite{Arsene:2004fa,Back:2004je,Adams:2005dq,Adcox:2004mh} and the
review articles
\cite{Shuryak:2003xe,Gyulassy:2004zy,Muller:2007rs,CasalderreySolana:2007zz,Heinz:2008tv}
for discussions of their theoretical interpretations). This represents an
important paradigm shift, since the prevalent opinion for quite some time
was that this form of hadronic matter should be weakly coupled, because
of its high density and of the asymptotic freedom of QCD. This shift of
paradigm intervened only a few years after the recognition of the {\em
AdS/CFT correspondence}
\cite{Maldacena:1997re,Gubser:1998bc,Witten:1998zw,MaldPR}
--- a theoretical revolution which offered a whole new framework, based
on string theory, to address problems in strongly coupled gauge theories.
The advent of the RHIC data has motivated an intense theoretical activity
over the last few years, aiming at using the AdS/CFT correspondence to
understand properties of QCD--like matter at finite temperature and/or
high energy (see, e.g., the recent review paper \cite{SONREV} and Refs.
therein).

One should emphasize here that the experimental evidence in favour of
strong--coupling dynamics at RHIC is rather indirect
--- its physical interpretation also involves theoretical
assumptions which are generally model--dependent ---, but some of the
data seem quite robust and compelling. This is especially the case for
those which reflect the long--range, collective properties of the
hadronic matter. For instance, the RHIC data exhibit a form of collective
motion called `elliptic flow' \cite{Ollitrault:1992bk}, which
demonstrates that the partonic matter produced in the early stages of a
Au+Au collision behaves like a fluid. Remarkably, these data can be well
accommodated within theoretical analyses using hydrodynamics, which
assume early thermalization and nearly zero viscosity --- or, more
precisely, a very small viscosity to entropy--density ratio $\eta/s$.
These features are hallmarks of a system with very strong interactions:
indeed, at weak coupling $g\ll 1$, the equilibration time and the ratio
$\eta/s$ are both parametrically large, since proportional to the mean
free path $\sim 1/g^4$. On the other hand, AdS/CFT calculations for gauge
theories with a gravity dual \cite{Policastro:2001yc} suggest that, in
the limit of an infinitely strong coupling, the ratio $\eta/s$ should
approach a universal lower bound which is $\hbar/4\pi$
\cite{Kovtun:2004de}. (The existence of such a bound is also required by
the uncertainty principle.) Interestingly, it appears that, within the
error bars, the ratio $\eta/s$ extracted from the RHIC data
\cite{Teaney:2003kp,Luzum:2008cw} is roughly consistent with this lower
bound, thus supporting the new paradigm of a {\em strongly coupled
Quark--Gluon Plasma} (sQGP).

\FIGURE[t] {\centerline{\includegraphics[width=.75\textwidth]{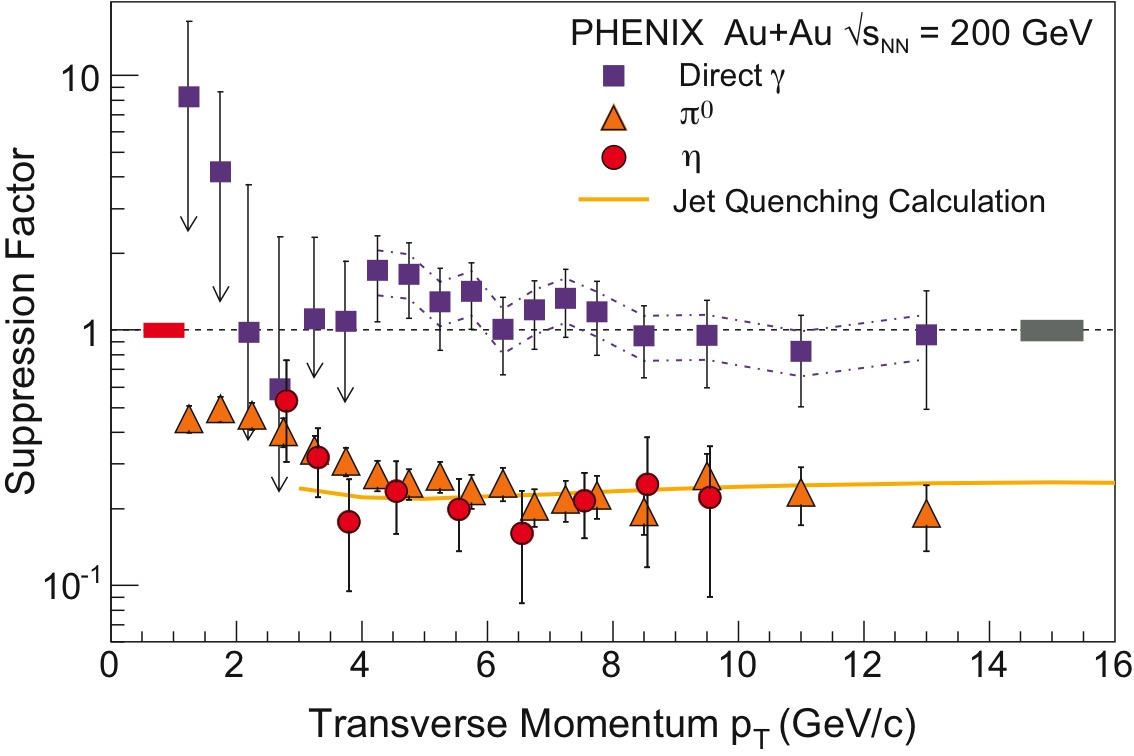}}
\caption{\sl The ratio $R_{AA}$ of measured versus expected yield of
various particles ($\pi^0, \eta, \gamma$) in Au+Au collisions at
$\sqrt{s_{\rm NN}} = 200$ GeV as function of the transverse momentum
$p_T$ (RHIC, PHENIX collaboration). Unlike the direct photons, the mesons
shows a strong amount of suppression at high $p_T$, which is moreover the
same for pions and $\eta$--mesons. This suggests that the suppression is
an effect related to the absorption (energy loss) of energetic partons in
the medium.} \label{fig:RAA} }

But experimental indications in favour of strong interactions have also
emerged from different type of data --- those associated with {\em hard
probes}. A `hard process' in QCD is a scattering involving a large
momentum exchange, $Q\gg \Lam\sim 200$ MeV. In the context of heavy ion
collisions, the `hard probes' are highly energetic `jets' (partons,
virtual photons, dileptons, heavy--quark mesons), which are produced by
the hard scattering of the incoming quarks and gluons, and which on their
way towards the detectors measure the properties of the surrounding
matter with a high resolution, meaning on very short space--time scales.
One would expect such hard interactions to lie within the realm of
perturbative QCD, yet some of the experimental results at RHIC seem
difficult to explain by perturbative calculations at weak coupling. One
of these results is the ratio $R_{AA}$ between the particle yield in
Au+Au collisions and the respective yield in proton--proton collisions
rescaled by the number of participating nucleons. This ratio would be one
if a nucleus--nucleus collision was the incoherent superposition of
collisions between the constituents nucleons (protons and neutrons) of
the two incoming nuclei. But the RHIC measurements show that $R_{AA}$ is
close to one only for direct photon production, whereas for hadron
production it is strongly suppressed (roughly, by a factor of 5; see
Fig.~\ref{fig:RAA}). This suggests that, after being produced through a
hard scattering, the partonic jets are somehow absorbed by the
surrounding medium.
 \FIGURE[t]{ \centerline{
\includegraphics[width=.4\textwidth]{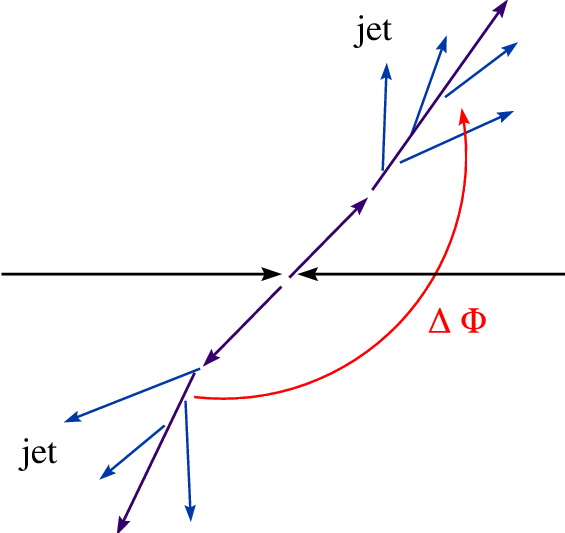}\qquad
\includegraphics[width=.5\textwidth]{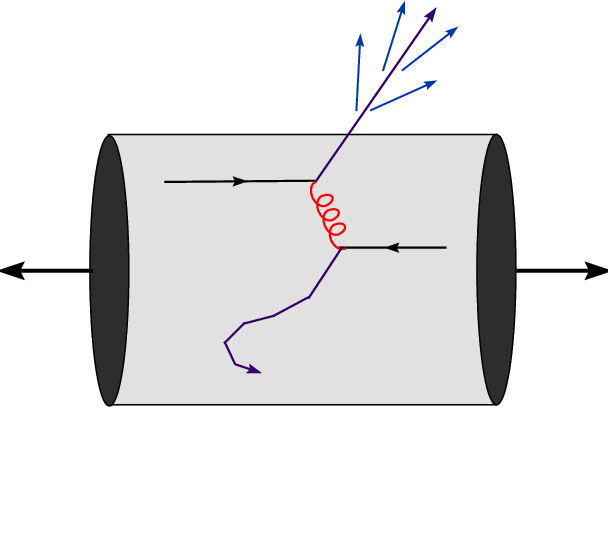}}
\caption{\sl Jet production in high--energy scattering. Left: the typical
situation in a proton--proton collision: the leading partons fragment
into two back--to--back hadronic jets, which are both observed by the
detector. Right: a nucleus--nucleus collision: one of the leading partons
escapes the interaction region and yields a jet in the detector, but the
other one is absorbed by the surrounding matter.} \label{Fig:JETS}}

Additional evidence in that sense comes from studies of jets and, more
precisely, of the angular correlation of the radiation associated with a
trigger particle with high transverse momentum (the `near side jet'). A
high--energy proton--proton (or electron--positron) collision generally
produces a pair of partons whose subsequent evolution (through
fragmentation and hadronisation) leaves two jets of hadrons which
propagate back--to--back in the center of mass frame (see
Fig.~\ref{Fig:JETS} left). Hence, if one uses a hard particle in one of
these jets to trigger the detector, then the distribution of radiation in
the azimuthal angle $\Delta\Phi$ shows two well pronounced peaks, at
$\Delta\Phi=0$ and $\Delta\Phi=\pi$, as shown in Fig.~\ref{Fig:JetCorr}
(the curve denoted there as `p+p min. bias'). A similar distribution is
seen in deuteron--gold collisions (the points d+Au in
Fig.~\ref{Fig:JetCorr}), but not in central Au+Au collisions, where the
peak at $\Delta\Phi=\pi$ (the `away side jet') has disappeared, as shown
by the respective RHIC data in Fig.~\ref{Fig:JetCorr}. It is then natural
to imagine that the hard scattering producing the jets has occurred near
the edge of the interaction region, so that the near side jet has escaped
to the detector, while the away side jet has been absorbed while crossing
through the medium (see Fig.~\ref{Fig:JETS} right).

The Au+Au results in Figs.~\ref{fig:RAA} and \ref{Fig:JetCorr} show that
the matter produced right after a heavy ion collision is {\em opaque},
which may well mean that this matter is dense, or strongly--coupled, or
both. The theoretical way to describe the disappearance of a parton in
this matter is by computing the rate for energy loss $\rmd E/\rmd t$,
which is proportional to a specific transport coefficient
--- the `jet quenching parameter' $\hat q$ --- which characterizes the parton
interactions in the medium (see, e.g., \cite{CasalderreySolana:2007zz}
and Refs. therein). This extraction of this parameter from the RHIC data
is accompanied by large uncertainties, so the obtained values lies within
a wide window: $\hat q\,\simeq\,0.5 \div 15$ ${\rm GeV}^2/{\rm fm}$ (see,
e.g., \cite{Eskola:2004cr,Dainese:2004te}). It is often stated that this
value is too large to be accommodated by weak coupling calculations, but
this is still under debate \cite{Baier:2006fr}. What is clear, however,
is that a complimentary analysis of these phenomena in the
non--perturbative regime at strong coupling would be highly valuable, and
this is where the AdS/CFT correspondence comes into the play.

\FIGURE[t]{ \centerline{
\includegraphics[width=.75\textwidth]{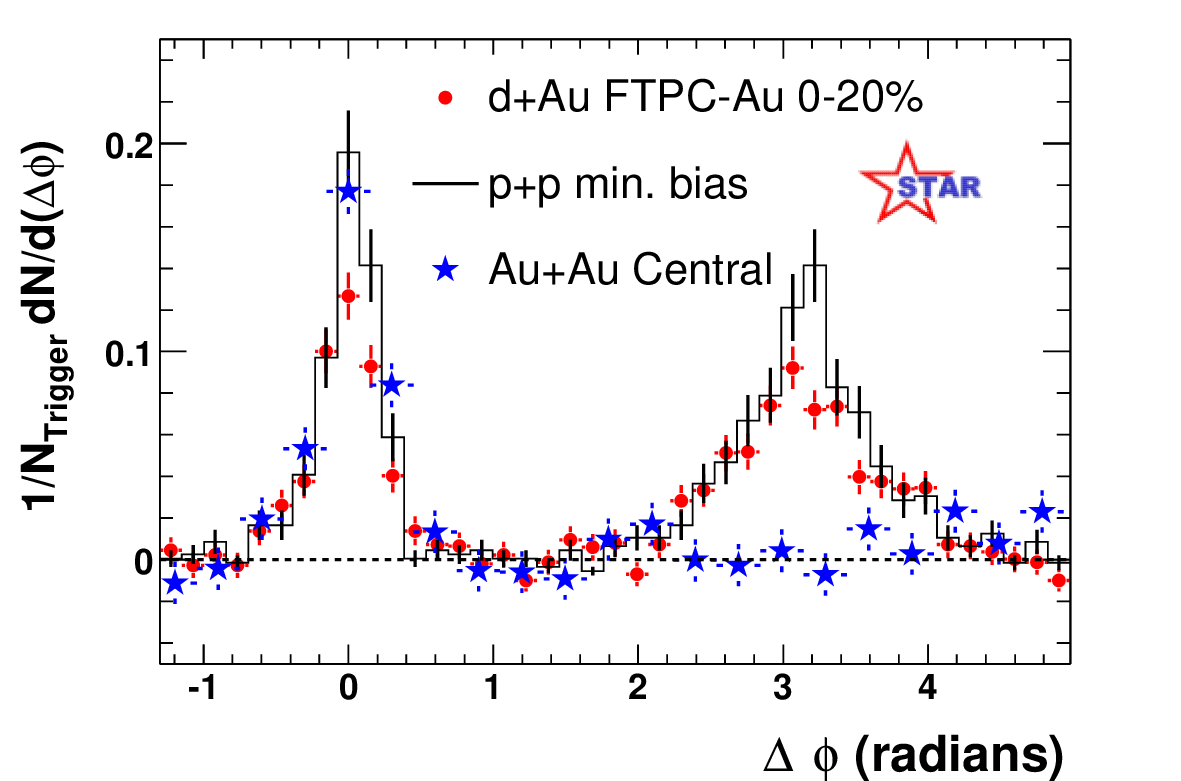}}
\caption{\sl Azimuthal correlations for jet measurements at RHIC (STAR
collaboration) in p+p, d+Au, and Au+Au collisions. The selected events
are such that the trigger particle has transverse momentum 4 GeV $ < p_T
<$ 6 GeV and the associated radiation has $p_T > $ 2 GeV.}
\label{Fig:JetCorr}}

The standard non--perturbative technique in QCD, which is lattice gauge
theory, is not applicable (at least, in its current formulation) for {\em
dynamical} observables, so like real--time evolution, transport
coefficients, or interaction rates. On the other hand, such problems can
be addressed via the AdS/CFT correspondence, but the applicability of the
latter is restricted to the limit where the `t Hooft coupling is strong
$\lambda\equiv g^2 N_c\gg 1$ (which in practice means a large number of
colors: $N_c\gg 1$), and to special gauge theories which are more
symmetric than QCD and for which a `gravity dual' (i.e., an alternative
representation as a string theory living in a curved space--time geometry
in $D=1+9$ dimensions) has been identified. The original, and so far best
established, such duality is that between the ${\mathcal N}=4$
supersymmetric Yang--Mills (SYM) theory and the type IIB superstring
theory living in a background geometry which is asymptotically
$AdS_5\times S^5$ \cite{Maldacena:1997re,Gubser:1998bc,Witten:1998zw}
(see Sect.~3 below). The ${\mathcal N}=4$ theory is {\em a priori} quite
different from her QCD `cousin' : it is maximally supersymmetric, it has
conformal symmetry at quantum level (meaning that the coupling is fixed),
it has no confinement (and hence no hadronic asymptotic states), and the
fields in the Lagrangian are all in the adjoint representation of the
`colour' gauge group SU$(N_c)$ (unlike QCD, where the fermions lie in the
fundamental representation). So the relevance of the AdS/CFT results for
our real world is generally far from being clear. Yet, the particular
context of ultrarelativistic heavy ion collisions, as explored at RHIC
and in the near future at LHC, is quite exceptional in that respect,
because many of the limitations of the AdS/CFT correspondence become less
important in this context.

\FIGURE[t] {\centerline{
\includegraphics[width=.5\textwidth]{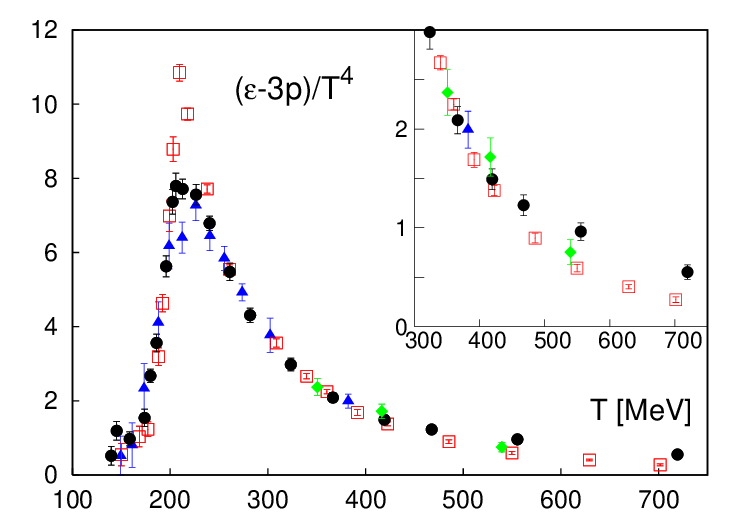}\qquad
\includegraphics[width=.5\textwidth]{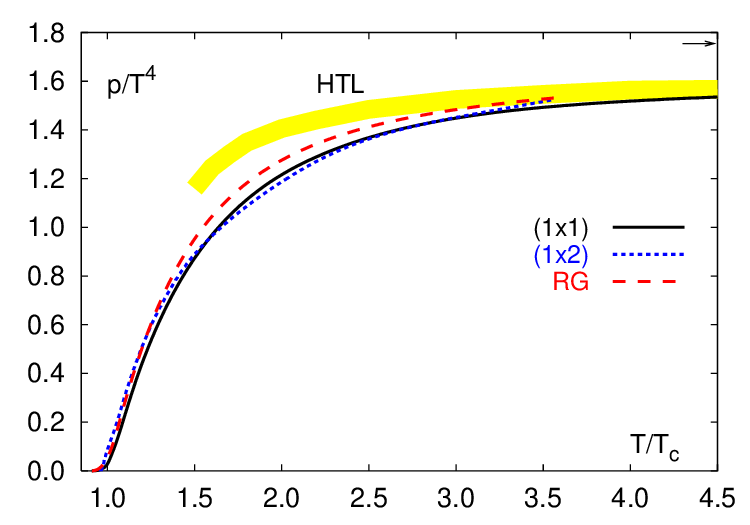}}
\caption{\sl Lattice results for the trace anomaly, $T^{\mu}_{\mu}=
\epsilon -3p$, in units of $T^4$ (left) \cite{latttrace} and for the
pressure of the SU$(3)$ gauge theory (right) \cite{latteos}. In the right
figure, different lines correspond to different gauge actions, whereas
the upper band denoted as `HTL' (for `Hard Thermal Loop') represents the
results of a parameter--free resummation of perturbation theory
\cite{BIR1}. The small arrow in the upper right corner indicated the
pressure of an ideal gas, i.e., the zero--coupling limit $g\to 0$.}
\label{fig:eos}}

Indeed, the QCD matter of interest is anyway in a deconfined,
quark--gluon plasma, phase, for which the `conformal anomaly' --- the
breaking of the conformal symmetry of the QCD Lagrangian by the running
of the coupling
--- appears to be relatively small. This is confirmed by lattice
simulations for the QCD thermodynamics within the temperature range
corresponding to the energy density produced at RHIC and (in perspective)
LHC: the relevant range for $T$ is $2 T_c\le T \le 5T_c$, where
$T_c\simeq\Lam\simeq 200$ MeV is the critical temperature for the
deconfinement phase transition. The running of the QCD
coupling\footnote{It is meaningful to choose the renormalization scale
$\mu$ as the `first Matsubara frequency' $\mu=2\pi T$, since this is the
value which minimizes the logarithms of $\mu$ in perturbation theory.}
$g(\mu)$ is negligible within such a restricted range and, besides, the
relevant value turns out to be quite large: $g\gtrsim 1.5$, meaning
$\lambda\gtrsim 6$, which leaves the hope for a strong--coupling
behaviour. Moreover, the lattice calculation of the `trace anomaly'
$\langle T^{\,\mu}_\mu\rangle$, which is proportional to the QCD
$\beta$--function,
  \beq \langle T^{\,\mu}_\mu\rangle\,=\,\epsilon -3p
  \,=\,\beta(g)\,\frac{\rmd p}{\rmd g}
 \,,\eeq
($\epsilon$ is the energy density and $p$ is the pressure) yields a
relatively small result --- less than 10\% of the total energy density
--- for all temperatures above $2 T_c\simeq 400$ MeV (see
Fig.~\ref{fig:eos} left).

But lattice QCD at finite temperature also illustrates the difficulty to
decide whether the quark--gluon plasma is strongly--coupled, or not,
within the relevant range of temperatures. To explain this, consider the
lattice results for the pressure, as shown in Fig.~\ref{fig:eos} (right):
after a sharp increases around $T_c$, the QCD pressure is slowly
approaching, for temperatures $T\gtrsim 1.5T_c$, towards the
corresponding value $p_0$ for an ideal gas, which in Fig.~\ref{fig:eos}
(right) is indicated by the small arrow in the upper right corner. As
visible in this figure, the deviation $(p-p_0)/p_0$ is quite small, less
than $20\% $, for all temperatures $T\gtrsim 2 T_c$. One may thus
conclude that the QGP is {\em weakly coupled} at these temperatures. And,
indeed, a weak--coupling calculation \cite{BIR1}, based on a {\em
resummation} of the perturbation theory and whose results are indicated
by the upper, `HTL', band in Fig.~\ref{fig:eos} right, provides a rather
good description of the lattice results for $T\gtrsim 2.5T_c$. However,
this conclusion is challenged by the AdS/CFT calculation of the pressure
in the ${\mathcal N}=4$ SYM plasma in the strong coupling limit
$\lambda\to\infty$ \cite{Gubser:1998nz}, which yields a remarkable result
: the pressure at infinite coupling is exactly 3/4 of the corresponding
ideal--gas value $p_0$ :
 \beq\label{pstrong}
 p(\lambda\to\infty)\,=\,\frac{\pi^2}{8}\,N_c^2 T^4\,=\,\frac{3}{4}\,
 p_0\,.\eeq
This ratio $p/p_0=0.75$ is close to the value $p/p_0\approx 0.85$ found
in lattice QCD at $T=2.5T_c$ (see Fig.~\ref{fig:eos} right), so the
latter might be consistent with strong coupling as well !

Since the lattice QCD results cannot be unambiguously interpreted, it is
interesting to have a closer look at the ${\mathcal N}=4$ SYM theory at
large $N_c$, for which both weak--coupling and strong--coupling
calculations are possible. The corresponding expansions are known to
next--to--leading order, i.e., to $\order{\lambda^{3/2}}$ at weak
coupling \cite{Vazquez-Mozo:1999ic} and, respectively,
$\order{\lambda^{-3/2}}$ at strong coupling \cite{Gubser:1998nz}, and can
be summarized as follows (for the entropy density $s$, for convenience):
writing $s=f(\lambda)s_0$ with $s_0 = ({2\pi^2}/{3})N_c^2 T^3$ (the ideal
gas value), one finds
  \beq\label{SSYMweak} f(\lambda)&
  \,=\,&1-\,\frac{3}{2\pi^2}\lambda\,+\,
 \frac{\sqrt2+3}{\pi^3}\lambda^{3/2}\, +\ldots\quad\mbox{for small $\lambda$\,,} \\
\label{SSYMstrong}
 f(\lambda)&
  \,=\,&\frac{3}{4} \left( 1 + \frac{15\zeta(3)}{8}
\lambda^{-3/2} +\ldots\right)\quad\mbox{for large $\lambda$}\,.
 \eeq
These expansions are illustrated in Fig.~\ref{figsym} \cite{BIR2},
together with an interpolation between them which is nicely monotonic and
can be viewed as the `true' non--perturbative result (by lack of better
approximations for intermediate values of the coupling). Also shown in
Fig.~\ref{figsym} (the band denoted as `2PI' there) is the result of a
resummation of perturbation theory obtained via the same method as the
`HTL' band in Fig.~\ref{fig:eos} right. The resummation is necessary
since, as also manifest in Fig.~\ref{figsym}, the usual expansion in
powers of $g$ (or $\lambda$) is poorly convergent and has no predictive
power except at extremely small values of the coupling. This problem is
generic to field theories at finite temperature, and is associated with
collective phenomena which provide screening effects and thermal masses
proportional to powers of $g$\,; the `resummations' consist in keeping
such medium effects within dressed propagators and vertices, instead of
expanding them out in perturbation theory. (See the review papers
\cite{QGPrev} for more details and references.) As also visible in
Fig.~\ref{figsym}, the resummed perturbation theory yields a monotonic
curve which matches with the `true' result up to $\lambda\simeq 4$, where
$s/s_0\simeq 0.85$. By comparison, the strong coupling expansion in
\eqnum{SSYMstrong} approaches the `true' result only for $\lambda \gtrsim
8$. This suggests that a value $p/p_0\approx 0.85$ as found by lattice
QCD around $2.5T_c$ truly corresponds to an {\em intermediate} value of
the coupling (neither weak, nor strong), which is at least marginally
within the reach of (properly organized) perturbation theory.

 \FIGURE[t]{
\includegraphics[width=0.62\textwidth]{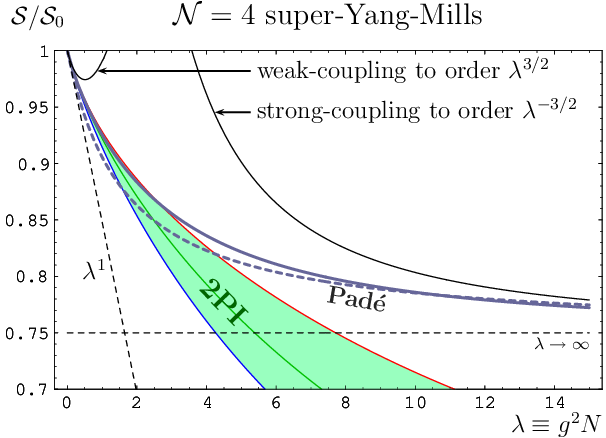}
\caption{\sl Weak and strong coupling results for the entropy density of
$\mathcal N=4$ SYM theory together with the result of the resummed
perturbation theory (the band denoted as `2PI'). The dashed and full
heavy gray lines represent the Pad\'e approximants $R_{[1,1]}$ and
$R_{[4,4]}$ which interpolate between weak and strong coupling results to
leading and next-to-leading orders, respectively. \label{figsym}} }

To summarize, the lattice results for QCD thermodynamics at $T\gtrsim
2T_c$ do not provide strong evidence in favour of a strong--coupling
dynamics, but they do not exclude it either. Moreover, the $\mathcal N=4$
SYM theory together with the AdS/CFT correspondance offers an unique
opportunity to perform explicit calculations at both weak and strong
coupling, with conclusions which may guide our interpretation of the
corresponding results from lattice QCD. The purpose of these lectures is
to present a similar guidance, but for a different physical problem: that
of a `hard probe' (a high--energy parton) propagating through a
strongly--coupled $\mathcal N=4$ SYM plasma at finite temperature. There
are clearly many differences between this idealized problem and the
corresponding one in the phenomenology of heavy--ion collisions (like the
replacement of QCD by the $\mathcal N=4$ SYM theory, or the assumption
that the deconfined matter is at thermal equilibrium), but the crucial
assumption in our opinion is that the coupling is strong. Thus, by
comparing the conclusions of this AdS/CFT analysis with the respective
data at RHIC (and in perspective LHC), and may hope to answer the
following, fundamental question: {\em is this particular regime of QCD
mostly on the strong--coupling side, or on the weak--coupling one ?}

In the recent literature, the problem of a hard probe propagating through
a strongly coupled plasma has been addressed from different perspectives
and within different approaches, depending upon the nature of the hard
probe and of its string theory `dual'. The results of these various
approaches appear to be consistent with each other at a fundamental
level, and they point towards a {\em universal mechanism for parton
energy loss at strong coupling.} Our main purpose in what follows will be
to explain how this mechanism emerges from the results of the AdS/CFT
calculations. To that aim we shall focus on the case where the `hard
probe' is a {\em virtual photon} (more precisely, an $\mathcal
R$--current; see below) \cite{HIM2,HIM3}. This choice is motivated by
simplicity: from the experience with QCD one knows that an
electromagnetic current is the simplest device to produce and study
hadronic jets. In deep inelastic scattering (DIS), the exchange of a
highly virtual space--like photon between a lepton and a hadron acts as a
probe of the hadron parton structure on the resolution scales set by the
process kinematics. Also, the partonic fluctuation of a space--like
current can mimic a quark--antiquark `meson', which is nearly on--shell
in a frame in which the current has a high energy. Furthermore, the decay
of the time--like photon produced in electron--positron annihilation is
the simplest device to produce and study hadronic jets in QCD. Thus, the
propagation of an energetic current through the plasma gives access to
quantities like the plasma parton distributions, the meson screening
length, or the jet energy loss. The relation between our results for the
virtual photon and the corresponding ones for other `hard probes'
--- a heavy quark
\cite{Herzog:2006gh,Gubser:2006bz,Broad,Gubser:2006nz,Friess:2006fk,Gubser:2007xz,Chesler:2007sv,QSAT},
a quark--antiquark meson (built with heavy quarks)
\cite{Kruczenski:2003be,Peeters:2006iu,Liu:2006nn,Chernicoff:2006hi,Caceres:2006ta,Ejaz:2007hg,Myers:2008cj,Faulkner:2008qk},
or a massless gluon \cite{GubserGluon,Karch,Chesler3} --- will be
described at appropriate places.

Within the ${\mathcal N}=4$ SYM theory, the role of the electromagnetic
current is played by the `${\mathcal R}$--current' --- a conserved
Abelian current whose charge is carried by fermion and scalar fields in
the adjoint representation of the color group (see Sect. 3 for more
details). Thus, DIS at strong coupling can be formulated as the
scattering between this ${\mathcal R}$--current and some appropriate
`hadronic' target. The first such studies \cite{Polchinski,HIM1} have
addressed the zero--temperature problem, where the target was a `dilaton'
--- a massless string state `dual' to a gauge--theory `hadron', whose
existence requires the introduction of an infrared cutoff $\Lambda$ to
break down conformal symmetry. These studies led to an interesting
picture for the partonic structure at strong coupling: through successive
branchings, all partons end up `falling' below the `saturation line',
i.e., they occupy --- with occupation numbers of order one --- the
phase--space at transverse momenta below the saturation
scale\footnote{Here, $x$ is the Bjorken variable for DIS, which is
roughly proportional to the inverse energy squared: $x\simeq Q^2/s$; see
Sect.~2.2 for details.} $Q_s(x)$. This scale rises with $1/x$ as
$Q_s^2(x)\sim 1/x$ which is much faster than for the corresponding scale
in perturbative QCD \cite{CGC}. This comes about because the high--energy
scattering at strong coupling is governed by a spin $j\simeq 2$
singularity (corresponding to graviton exchange in the dual string
theory), rather than the usual $j\simeq 1$ singularity associated with
gluon exchange at weak coupling.

In Refs. \cite{HIM2,HIM3} these studies and the corresponding partonic
picture have been extended to a finite--temperature ${\mathcal N}=4$ SYM
plasma and also to the case of a time--like current (the strong--coupling
analog of $e^+e^-$ annihilation). Note that this finite--$T$ case is
conceptually clearer than the zero--temperature one, in that it does not
require any `deformation' of the gauge theory, like an IR cutoff. It is
also technically simpler, in that the calculations can be performed in
the strong 't Hooft coupling limit $\lambda\equiv g^2N_c\to \infty$ at
fixed $g^2\ll 1$ (meaning $N_c\to\infty$). This is so since the large
number of degrees of freedom in the plasma, of order $N_c^2$ per unit
volume, compensates for the $1/N_c^2$ suppression of the individual
scattering amplitudes; hence, a strong--scattering situation can be
achieved even in the strict large--$N_c$ limit. The AdS/CFT calculation
shows that the saturation momentum of the plasma rises with the energy
even faster than for a hadronic target, namely like $Q_s^2(x)\sim 1/x^2$.
This difference is easily understood: the additional factor of $1/x$ is
associated with the longitudinal extent of the interaction region, which
for an infinite target (so like the plasma) grows with the energy, by
Lorentz time dilation.

The results of Refs. \cite{HIM2,HIM3} will be described in Sects. 4 and 5
below, together with their physical interpretations. But before that, in
Sect. 2, we shall briefly remind the perturbative QCD viewpoint on the
simplest processes mediated by a virtual photon --- $e^+e^-$ annihilation
and DIS ---, which will serve as a level of comparison for the
corresponding discussion at strong coupling. Then, in Sect. 3, we shall
give a succinct introduction to the AdS/CFT correspondence, whose purpose
is not to be exhaustive --- more details can be found in the review
papers and textbooks listed in the references
\cite{MaldPR,SONREV,Becker,Kiritsis} --- but merely to present in a
minimal but self--contained way that part of the formalism which is
needed for our present purposes.

But more than describing the formalism, our main objective in these
lectures is to present a physical picture for the dynamics at strong
coupling, as originally proposed in Refs. \cite{HIM2,HIM3}. Building such
a picture is generally difficult and in any case ambiguous, because of
the lack of a direct connection between the AdS/CFT approach and the
standard tools of quantum field theory, like Feynman diagrams. For the
problem at hand, we shall rely on the intuition coming from perturbative
QCD in order to propose a physical interpretation for the AdS/CFT
results. But the most important tool in that sense will be the
ultraviolet--infrared correspondence
\cite{UVIR,Polchinski,Brodsky:2008pg,HIM3}, which relates the radial
distance in $AdS_5$ to the virtuality of the partonic fluctuation created
by the ${\mathcal R}$--current in the gauge theory. We feel that a more
systematic use of this duality could provide more physical insight into
other related calculations in the literature. For the same purpose, it
turns out to be very useful to have a {\em space--time representation}
for the dual processes in AdS/CFT, in addition to the more standard
momentum--space picture, which is used to compute correlations. As we
shall explain, via the UV/IR correspondence the space--time picture on
the string theory side can be mapped onto an intuitive physical picture
for the strong--coupling dynamics on the gauge theory side.

Our main physical conclusion is that a {\em partonic interpretation} for
the high--energy processes makes sense even at strong coupling, and that
the main mechanism for parton evolution in this regime is {\em
quasi--democratic parton branching}, i.e., a successive branching process
through which the energy of the incoming current, or parton, is rapidly
and quasi--democratically divided among the daughter partons. This
process takes place both in the vacuum (where, for instance, it leads to
an isotropic distribution of particles in the final state of $e^+e^-$
annihilation, instead of the jet structure familiar in QCD), and in the
finite--temperature plasma, where the rate for branching is influenced by
the medium properties (we shall then speak of {\em medium--induced parton
branching}). This branching process which, in the case of a plasma,
continues until the partons have energies and virtualities of order $T$,
represents the dominant mechanism for energy loss in the large--$N_c$
limit, where other mechanisms, like thermal rescattering, are suppressed.

\section{Partons and jets in QCD at weak coupling}

Before we turn to our main goal, which is a study of hard probes
propagating through a strongly--coupled plasma, let us briefly discuss
the situation in QCD, where hard scattering is rather associated with
weak coupling. (More details on these pQCD topics can be found in
textbooks like \cite{Peskin,Dokshitzer:1991wu}.) By ``hard scattering''
we mean that the momentum transfer $Q$ in the collision (the scale which
determines the relevant value of the QCD running coupling) is much larger
than $\Lam\sim 200$ MeV, so that $\alpha_s(Q^2)$ is reasonably small. (In
practice, $Q^2\sim 4$ GeV$^2$ is already a `hard scale', in which case
$\alpha_s\simeq 0.25$.) We shall focus on processes which are mediated by
a hard, virtual, electromagnetic current, since these are the processes
that we shall later be interested in at strong coupling. At weak coupling
at least, these are the processes in which the partonic picture of QCD is
most directly revealed. In our subsequent discussion, we shall briefly
review this picture and in particular emphasize those aspects which
transcend a purely perturbative point of view, and hence may be expected
to survive at strong coupling.

\subsection{Electron--positron annihilation}

\FIGURE[t]{\centerline{
\includegraphics[width=0.5\textwidth]{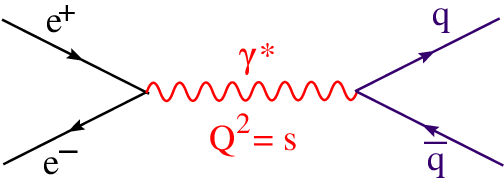}}
\caption{\sl Electron--positron annihilation to lowest order in
perturbative QCD \label{fig:epem}} }

The simplest process in perturbative QCD is electron--positron ($e^+e^-$)
annihilation into hadrons. To lowest order in the electromagnetic
($\alpha_{\rm em}$) and strong ($\alpha_s$) coupling constants, this
process proceeds as depicted in Fig.~\ref{fig:epem}: the electron and
positron annihilate with each other into a {\em time--like} virtual
photon, with positive virtuality\footnote{Throughout these lectures, we
shall use the 4--dimensional Minkowski metric with signature
$\eta_{\mu\nu}=(-1,1,1,1)$ (since this is the usual convention in the
context of gravity and string theory). Accordingly, the scalar product of
two vectors $a^\mu$ and $b^\mu$, with $a^\mu=(a^0, {\bm a})$ etc., reads
$a\cdot b\equiv \eta_{\mu\nu} a^\mu b^\nu=  a^\mu b_\mu =-a^0b^0+{\bm
a}\cdot{\bm b}$, and hence $q^2\equiv q^\mu q_\mu=-q_0^2+{\bm q}^2$.}
 $Q^2\equiv -q^\mu q_\mu = s$ (with
$s=(E_{e^+}+E_{e^-})^2$ the total energy squared in the center--of--mass
(COM) frame and $q^\mu$ the 4--momentum of the photon), which then decays
into a quark--antiquark ($q\bar q$) pair. This process is `hard' provided
the energy is high enough : $\sqrt{s}\gg \Lam$. In a confining theory
like QCD, quarks cannot appear in the final state, which must involve
only hadrons. Hence, the structure of the final state, as seen by a
detector, will be determined by the subsequent evolution of the quark and
the antiquark via {\em parton branching} (see Fig.~\ref{fig:epembranch}),
with the emerging partons eventually combining into hadrons. Since
hadronisation is a non--perturbative process, one may wonder whether it
makes any sense at all to use a partonic picture (which is rooted in
perturbation theory), even for the early and the intermediate stages of
the collision. This is however justified by the separation of time scales
in the problem: quantum processes are not instantaneous, rather it takes
some time to emit a parton --- the more so the softer the parton. Hard
processes occur very fast and determine the probability for a scattering
to happen, i.e., the total cross--section for $e^+e^-$ annihilation,
which is therefore computable in perturbation theory. The processes
responsible for hadronisation involve `soft' quanta with momenta $k\sim
\Lam$, hence they occur relatively late and affect only the precise
structure of the final state in terms of hadrons, but not the total
cross--section.

\FIGURE[h]{
\includegraphics[width=0.7\textwidth]{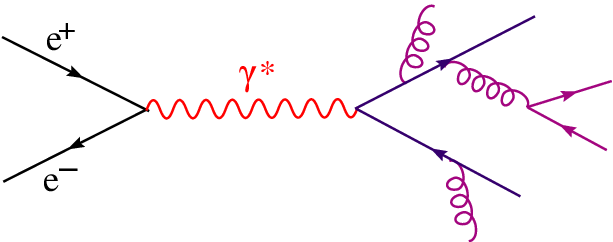}
\caption{\sl Parton evolution in the final state of $e^+e^-$
annihilation. \label{fig:epembranch}} }

Let us be more specific about these lifetime arguments, as they will play
an important role in what follows. One can estimate the duration of a
process from the uncertainty principle. The fastest process is the one
depicted in Fig.~\ref{fig:epem} --- the $e^+e^-$ annihilation into a
$q\bar q$ pair --- which in the COM frame lasts for a time $\Delta
t_0\sim 1/Q=1/\sqrt{s}$. The emitted quark and antiquark are themselves
off--shell --- each of them carries roughly half of the energy of the
virtual photon and half of its virtuality --- so they will decay by
radiating softer gluons (cf. Fig.~\ref{fig:epembranch}). The lifetime of
a time--like quark (more generally, parton) with 4--momentum
$p^\mu=(\omega,{\bf p})$ is estimated as
 \beq\label{Deltat}
 \Delta t\,\sim\,\frac{1}{P}\, \gamma\,=\,
 \frac{\omega}{P^2}\,,\eeq
where the first factor $1/P$ (with $P^2=\omega^2-p^2$ and $p=|{\bf p}|$)
is the parton lifetime in its own rest frame and the second factor
$\gamma=1/\sqrt{1-v^2}={\omega}/{P}$, with $v=p/\omega$, is the Lorentz
factor for the relativistic time dilation. This $\Delta t$ can be also
interpreted as the {\em formation time} of the radiated gluon, and can be
alternatively expressed in terms of the kinematics of the latter (see
Fig.~\ref{fig:epembranch}). A simple calculation yields (we assume here
that $k_\parallel\ll p$)
 \beq\label{Deltatk}
 \Delta t\,\sim\,\frac{k_\parallel}{k_\perp^2}\,,\eeq
where $k_\parallel$ and $k_\perp$ are the components of the gluon spatial
momentum which are parallel and, respectively, perpendicular to the
3--momentum ${\bf p}$ of the parent quark. As anticipated, it takes
longer time to emit softer gluons, i.e., gluons with lower transverse
momenta $k_\perp$. In particular, the hadronisation time is estimated as
$t_{\rm hadr} \sim{k_\parallel}/\Lam^2$ with $k_\parallel\lesssim
\sqrt{s}$. This means that, at high energy, there exists a parametrically
wide interval, namely,
 \beq \frac{1}{\sqrt{s}}\   <  \ t \ <  \frac{\sqrt{s}}{\Lam^2}\,,
 \eeq
during which the effects of confinement can be safely neglected and a
parton description applies. Note that the value of the coupling constant
did not play any role in this argument, which is rather controlled by the
kinematics via the uncertainty principle. On the other hand, the details
of the partonic pictures are very different at weak and, respectively,
strong coupling, as we shall later discover.

\FIGURE[t]{ \centerline{
\includegraphics[width=0.6\textwidth]{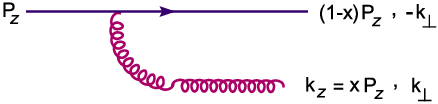}}
\caption{Gluon bremsstrahlung out of a parent quark to lowest order in
pQCD. \label{fig:onegluon}} }

Sticking to weak coupling for the time being, parton branching is
controlled by {\em bremsstrahlung}, which to lowest order in pQCD yields
the following rate for emitting a gluon out of a parent quark or gluon
(see also Fig.~\ref{fig:onegluon} and, e.g., \cite{Peskin} for details):
  \beq\label{brem} \rmd \mathcal{P}_{\rm Brem}\,\simeq\,\frac{\alpha_s
 C_R}{\pi^2}\,\frac{\rmd^2k_\perp}{k_\perp^2}\,\frac{\rmd x}{x}\,,\eeq
where $k_\perp$ is the gluon transverse momentum and $x=k_\parallel/p$ is
the fraction of the parent parton longitudinal momentum which is taken
away by the gluon. $C_R$ is the Casimir for the SU$(N_c)$ representation
pertinent to the parent parton: $C_F=(N_c^2-1)/N_c$ for a quark, or
$C_A=N_c$ for a gluon. In writing Eq.~(\ref{brem}) we have specialized to
$x\ll 1$ since this is the most interesting regime at high energy and
weak coupling: as manifest on this equation, the bremsstrahlung favors
the emission of relatively soft gluons, with small longitudinal fractions
$x\ll 1$ and transverse momenta logarithmically distributed within the
range $\Lam < k_\perp < k_\parallel$, since the corresponding
phase--space is large and compensates for the smallness of the
coupling\footnote{The fact that the running coupling is to be evaluated
at the hard scale $Q^2=s$ follows via an analysis of virtual, loop,
corrections to the tree diagram in Fig.~\ref{fig:epem}.}:
  \beq \Lam\,\ll\,k_\perp\,\ll\,k_\parallel=xp \,\ll \,\sqrt{s}
\ \ \Longrightarrow\ \ \mathcal{P}_{\rm
 soft}\,\sim\,\alpha_s(Q^2)\,\ln^2\frac{ \sqrt{s}}{\Lam}\,.\eeq
(The softest among these gluons are responsible for hadronisation.)
However, such soft gluons are quasi--collinear with their parents
partons, so their emission does not significantly alter the topology of
the final state: instead of a pair of bare quarks, the detector will see
a pair of well collimated hadronic jets (see Fig.~\ref{fig:jets} left).
Harder emissions leading to multi--jets events (see Fig.~\ref{fig:jets}
right) are possible as well, and actually seen in the experiments, but
they are comparatively rare since they occur with a small probability
$\mathcal{P}_{\rm hard}\sim\alpha_s(Q^2)\ll 1$ with $Q^2=s$. The total
cross--section for $e^+e^-$ annihilation can be computed in pQCD as a
series in powers of $\alpha_s(s)$, with the different terms in this
series roughly corresponding to different numbers of jets in the final
state:
 \beq\label{sigmaepem}\sigma(s)\,=\,\sigma_{\rm QED}\,\times\, \big(3\sum_f
e_f^2\big)\left(1\,+\,\frac{\alpha_s(s)}{\pi}
 \,+\,\order{\alpha_s^2(s)}\right)\,,\eeq
where $\sigma_{\rm QED}= 4\pi\alpha_{\rm em}^2/3s$ is the QED
cross--section for $e^+ e^-\to\mu^+ \mu^-$, the factor of $N_c=3$ is the
number of color degrees of freedom for quarks in SU(3), and $e_f$ is the
electric charge of the quarks with flavor $f$ (in units of the electron
charge $e$). The experimental verification of \eqnum{sigmaepem}
represents one of the most solidly established tests of pQCD so far.

\FIGURE[t]{ \centerline{
\includegraphics[width=0.5\textwidth]{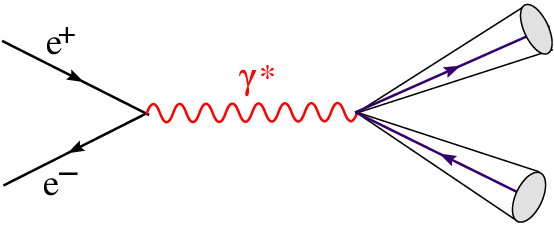}\qquad
\includegraphics[width=0.5\textwidth]{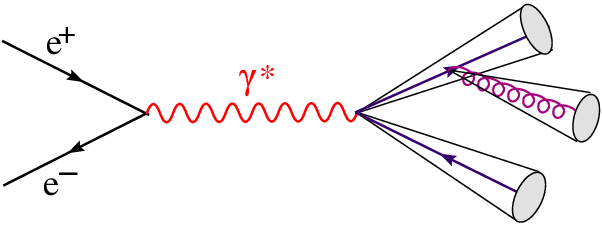}}
\caption{\sl Jet structure in the final state for $e^+e^-$ annihilation.
\label{fig:jets}} }

\FIGURE[h]{ \centerline{
\includegraphics[width=0.5\textwidth]{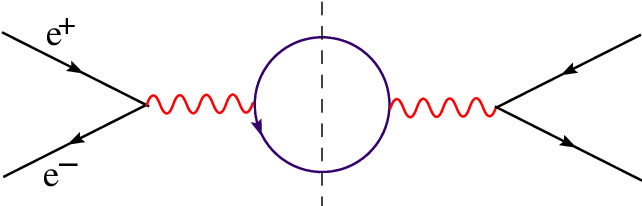}\qquad
\includegraphics[width=0.5\textwidth]{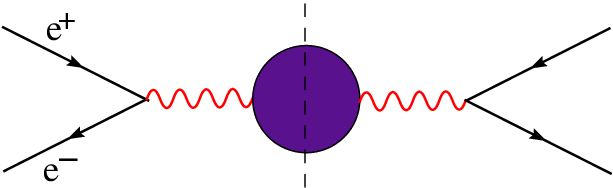}}
\caption{\sl Total cross--section for $e^+e^-$ annihilation as given by
the optical theorem. \label{fig:optical}} }

To conclude this discussion of $e^+e^-$ annihilation, let us describe a
recipe for computing the corresponding cross--section which goes beyond
perturbation theory, and thus also applies in the strong--coupling regime
to be considered later on. By the optical theorem, this cross--section
can be related to the imaginary part of the forward scattering amplitude
$e^+e^-\to e^+e^-$. For instance, to lowest order in $\alpha_s$, the
cross--section for the process $e^+e^-\to q\bar q$ illustrated in
Fig.~\ref{fig:epem} can be expressed as a cut through the
one--quark--loop contribution to the forward amplitude, cf.
Fig.~\ref{fig:optical} (left). More generally, the following formula
holds to leading order in $\alpha_{\rm em}$ but to all orders in
$\alpha_s$ :
 \beq\label{sigmaPI}
 \sigma(e^+e^-)\,=\,\frac{1}{2s}\ \ell^{\mu\nu}
 \ {\rm Im}\,\Pi_{\mu\nu}(q)\,,\eeq
where $\ell^{\mu\nu}$ is a leptonic tensor associated with the external
electron and positron lines and $\Pi_{\mu\nu}(q)$ is the (retarded)
vacuum polarization tensor for the virtual photon, and can in turn be
computed as the following current--current correlator in the vacuum (or
`vacuum polarization tensor')
\beq\label{JJTL}
 \Pi_{\mu\nu}(q)\,\equiv \int \rmd^4x\,\rme^{-iq\cdot x}\,i\theta(x_0)\,
 \langle 0\,|\, [J_\mu(x), J_\nu(0)]\,|\,0\rangle\,,\eeq
where $J^\mu$ is the electromagnetic current density of the quarks :
 \beq
 J^\mu\,=\,\sum_f e_f\,\bar q_f\,\gamma^\mu\,q_f\,,
 \eeq
that is, the operator which couples to the photon: $\mathcal{L}_{\rm int}
= e A_\mu J^\mu$. Current conservation $q^\mu \Pi_{\mu\nu}=0$ together
with Lorentz symmetry imply that $\Pi_{\mu\nu}$ has only one independent
scalar component (recall that $Q^2\equiv -q^\mu q_\mu > 0$ and
$\eta_{\mu\nu}=(-1,1,1,1)$)
 \beq\label{PiTL}
 \Pi_{\mu\nu}(q)\,=\,
  \left(\eta_{\mu\nu}+\frac{q_\mu q_\nu}{Q^2} \right)\Pi(Q^2)\,.\eeq

\FIGURE[t]{ \centerline{
\includegraphics[width=0.6\textwidth]{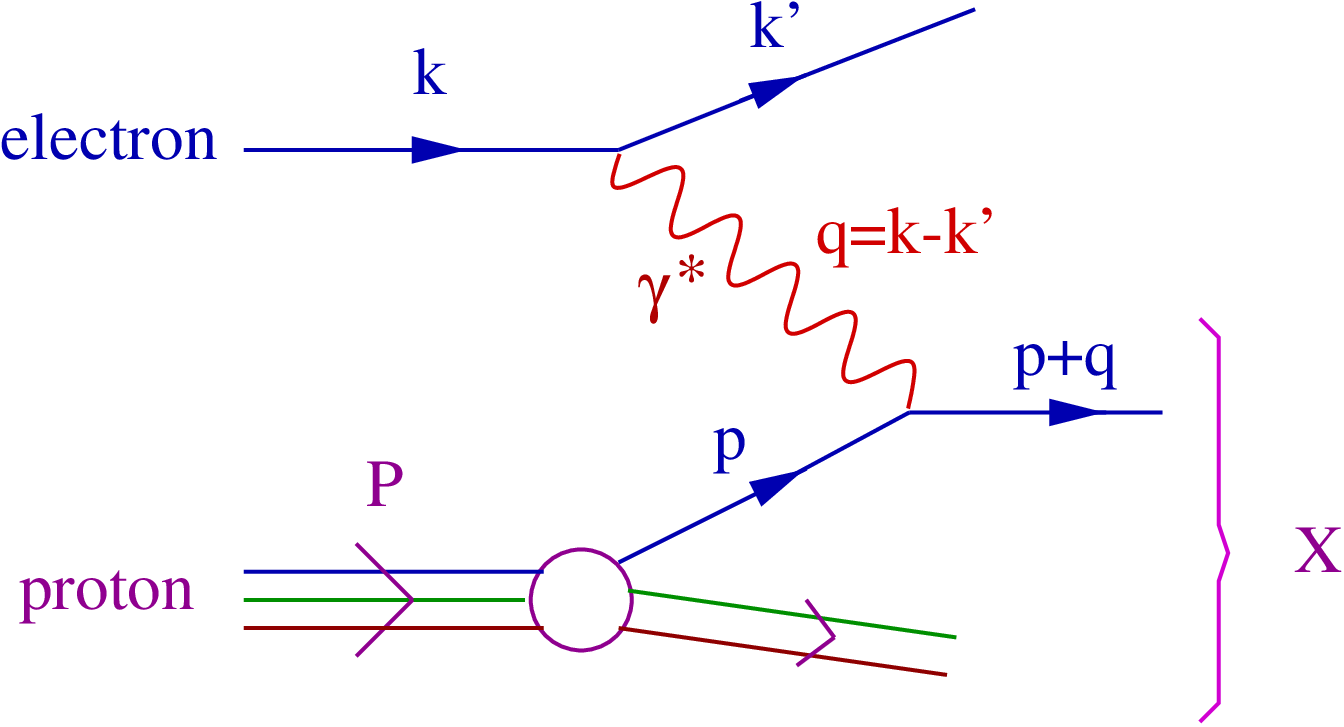}}
\caption{\sl Deep inelastic electron--proton scattering: general
kinematics.\label{fig:DIS}} }

\subsection{Deep inelastic scattering}

Another important hadronic process which is mediated by a virtual photon
is the deep inelastic scattering (DIS) between a lepton (say, electron)
and a hadron (say, the proton), as illustrated in Fig.~\ref{fig:DIS}. In
DIS, the exchanged photon is space--like: $-q^\mu q_\mu < 0$, and then it
is convenient to use the notation $Q^2$ for the positive quantity
$Q^2\equiv q^\mu q_\mu > 0$ (i.e., {\em minus} the photon virtuality).
The photon couples to the electromagnetic current of the quarks inside
the proton. By the optical theorem, the total cross--section
$\sigma(ep\to eX)$ can be written similarly to \eqnum{sigmaepem}, but
with the current--current correlator now computed as an expectation value
over the proton wavefunction:
 \beq\label{JJDIS}
 \Pi_{\mu\nu}(q,P)\,\equiv \int \rmd^4x\,\rme^{-iq\cdot x}\,i\theta(x_0)\,
 \langle P\,|\, [J_\mu(x), J_\nu(0)]\,|\,P\rangle\,,\eeq
where the proton state $|\,P\rangle$ is denoted by its 4--momentum
$P^\mu$. The latter introduces a privileged direction in space, so the
tensorial structure of $\Pi_{\mu\nu}$ is more complicated than in the
vacuum: it now involves two scalar functions, which both depend upon two
kinematical invariants. It is customary to write
 \beq \label{Pitensor} \Pi_{\mu\nu}(q,P)=
  \left(\eta_{\mu\nu}-\frac{q_\mu q_\nu}{Q^2} \right)\Pi_1(x,Q^2)+
  \left(P_\mu-q_\mu \frac{P\cdot q}{Q^2}\right)
  \left(P_\nu-q_\nu \frac{P\cdot q}{Q^2}\right)\Pi_2(x,Q^2)\,,\eeq
and express the cross--section in terms of the following {\em structure
functions}
 \beq  \label{F12} F_1(x,Q^2)\,=\,\frac{1}{2\pi}\, {\rm Im} \,\Pi_1,\qquad
  F_2(x,Q^2)\,=\,\frac{-(P\cdot q)}{2\pi}\, {\rm Im}\, \Pi_2\,, \eeq
which are dimensionless\footnote{Note that the polarization tensor
carries a different dimension in the case of the vacuum, where
$\Pi_{\mu\nu}(q)$ has mass dimension 2 (as clear from its definition
(\ref{JJTL})), and in the case of DIS off a hadron, where
$\Pi_{\mu\nu}(q)$ is dimensionless. This difference arises from the
normalization of the proton wavefunction in \eqnum{JJDIS}.}. We have here
used the following kinematic invariants
 \beq\label{DISkinem}
 Q^2\,\equiv\, q^\mu q_\mu \,=\,-q_0^2 + {\bm q}^2\, \ge\, 0,\qquad
 x\,\equiv\,\frac{Q^2}{-2(P\cdot q)}\,=\,\frac{Q^2}{s+Q^2-M^2}\,,\eeq
where $M$ is the mass of the proton (hence, $P^2=-M^2$), and
$s\equiv -(P+q)^2$ is the invariant energy squared of the photon+proton
system, and is the same as the invariant mass squared $M_X^2$ of the
hadronic system $X$ produced by the collision, cf. Fig.~\ref{fig:DIS}.
Note that $M_X^2\ge M^2$ and hence $x\le 1$. The `deep inelastic' regime
corresponds to large virtuality $Q^2\gg M^2$ (`hard photon'), and the
`high energy' one to small $x$ :  $s\gg Q^2\,\Longrightarrow \,x\simeq
Q^2/s\,\ll\,1$.

The kinematical variables in \eqnum{DISkinem} are particularly convenient
as they have a direct physical interpretation: they mesure the resolution
of the virtual photon as a probe of the internal structure of the proton.
More precisely, in a frame in which the proton has a large longitudinal
momentum $P\gg M$ (`infinite momentum frame', or IMF), the scattering
consists in the absorbtion of the virtual photon by a quark excitation
which has a longitudinal momentum fraction $k_z/P$ equal to $x$ and
occupies an area $\sim 1/Q^2$ in the transverse plane $(x,y)$ (the plane
normal to the collision axis, chosen here to be $z$). This can be
understood with reference to Figs.~\ref{fig:onegluon}, \ref{fig:abs}, and
\eqnum{Deltatk} : a partonic excitation with longitudinal momentum $k_z$
and transverse momentum $k_\perp$ has a lifetime
 \beq\label{partlife}
 \Delta t_{\rm part} \,\sim\,\frac{k_z}{k_\perp^2}\,
   =\,\frac{xP}{k_\perp^2}\,.\eeq
For this parton to be `seen' in DIS, it must live longer than the
interaction time with the virtual photon, in turn estimated as ($q_0$ is
the energy of $\gamma^*$ in the IMF)
 \beq\label{colltime}
  \Delta t_{\rm col} \, \sim \, \frac{1}{q_0}
 \, \sim \, \frac{xP}{Q^2}\,.\eeq
 \FIGURE[h]{ 
\includegraphics[width=0.35\textwidth]{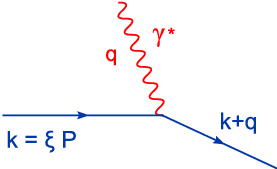}
\caption{\sl The virtual photon absorbtion by a nearly on--shell quark
with longitudinal momentum fraction $\xi$. \label{fig:abs}} }
 \noindent This condition requires
$k_\perp^2\lesssim Q^2$, which via the uncertainty principle implies that
the parton is localized within an area $\gtrsim 1/Q^2$. Furthermore, in
the IMF, partons are quasi--free and hence nearly on--shell, and their
longitudinal momenta are much larger than the transverse ones (they are
nearly collinear with the proton). With reference to Fig.~\ref{fig:abs},
these conditions imply
 \beq k^2\,\approx\,0\ \& \
 (k+q)^2\,\approx\,0 &\ \Longrightarrow \ & Q^2 +
2\xi P\cdot q \,\approx\,0  \nn &\ \Longrightarrow\ &
 \xi\,=\,\frac{Q^2}{-2(P\cdot q)}\,=\,x\nonumber\eeq
Note that the choice of the IMF is crucial for the validity of this
interpretation: it is only in this frame that the virtual excitations of
the proton (quarks and gluons) live long enough --- by Lorentz time
dilation --- to be unambiguously distinguished from vacuum fluctuations
with the same quantum numbers and momenta, and to be treated as
quasi--free during the comparatively short duration of the scattering
with the external probe (here, the virtual photon).

Then the DIS cross--section can be factorized as the elementary
cross--section for the photon absorbtion by a quark times a `parton
distribution function' which describes the probability to find a quark
with longitudinal momentum fraction equal to $x$ and transverse area
$1/Q^2$. This correspondence is such that the structure function
$F_2(x,Q^2)$ introduced in \eqnum{F12} is a direct measure of the quark
and antiquark distribution functions:
 \beq
 F_2(x,Q^2)\,=\,\sum_{f} e_f^2\,\big[xq_f(x,Q^2) + x\bar
 q_f(x,Q^2)\big]\,,\eeq
where $q_f(x,Q^2)$ is the number of quarks of flavor $f$ with
longitudinal momentum fraction $x$ and transverse size $1/Q$. Thus, the
experimental measurement of $F_2(x,Q^2)$ gives us a direct access to the
phase--space distribution of quarks within the proton wavefunction and in
the infinite momentum frame. This gives us furthermore access to the
gluon distribution, albeit indirectly, modulo our theoretical
understanding of {\em parton evolution}.

 \FIGURE[t]{ \centerline{
 \includegraphics[width=0.37\textwidth]{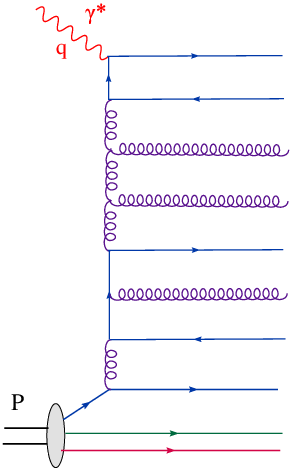}\qquad\qquad
 \includegraphics[width=0.4\textwidth]{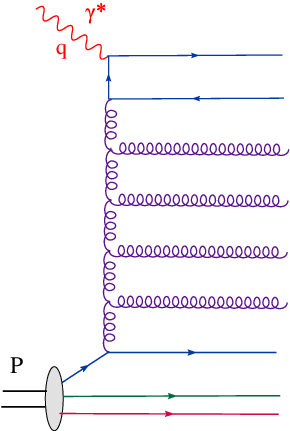}}
 \caption{\sl Parton evolution in perturbative QCD. The parton cascade
 on the right involves only gluons and is a part of the BFKL resummation
 at small $x$. \label{fig:cascade}} }

Namely, quark and gluons can transform into each other via parton
branching, so in general the quark struck by the virtual photon in DIS is
a `sea' quark, i.e., a quark from a partonic cascade initiated by one of
the valence quarks, as illustrated in Fig.~\ref{fig:cascade}. At weak
coupling, this branching proceeds through bremsstrahlung and favors an
evolution in which the virtuality is strongly increasing when moving up
(from the target proton towards the projectile photon) along the cascade.
That is, after each individual splitting, the daughter parton emitted in
the $t$--channel has either a much larger transverse momentum than its
parent parton, or a much smaller longitudinal--momentum fraction (and
then it is generally a gluon), or both. This is so since, according to
Eq.~(\ref{brem}), such emissions are favored by the large available phase
space, which equals $\ln(Q^2/\Lam^2)$ for the emission of a parton (quark
or gluon) with transverse momentum $k_\perp\ll Q$ and, respectively,
$\ln(1/x)$ for that of a gluon with longitudinal momentum fraction $\xi$
within the range $x\ll \xi\ll 1$. Depending upon the relevant values of
$Q^2$ and $x$, one can write down evolution equations which resum either
powers of $\alpha_s\ln Q^2$, or of $\alpha_s\ln(1/x)$, to all orders; the
coefficients in these equations, which represent the elementary splitting
probability can be computed as power series in $\alpha_s$ starting with
the leading--order result in Eq.~(\ref{brem}). As obvious from the
previous considerations, the $Q^2$--evolution (as encoded in the DGLAP
equation \cite{DGLAP}) mixes the quark and gluon distribution functions
(see Fig.~\ref{fig:cascade}.a), and this allows us to reconstruct the
gluon distribution from the $Q^2$--dependence of the experimental results
for $F_2$. The small--$x$ evolution, on the other hand, which is
described by the BFKL equation \cite{BFKL} and its non--linear
generalizations \cite{CGC} (see below), involves only gluons and
corresponds to resumming ladder diagrams like those in
Fig.~\ref{fig:cascade}.b in which successive gluons are strongly ordered
in $x$.

  \FIGURE[t]{ \centerline{
 \includegraphics[width=0.55\textwidth]{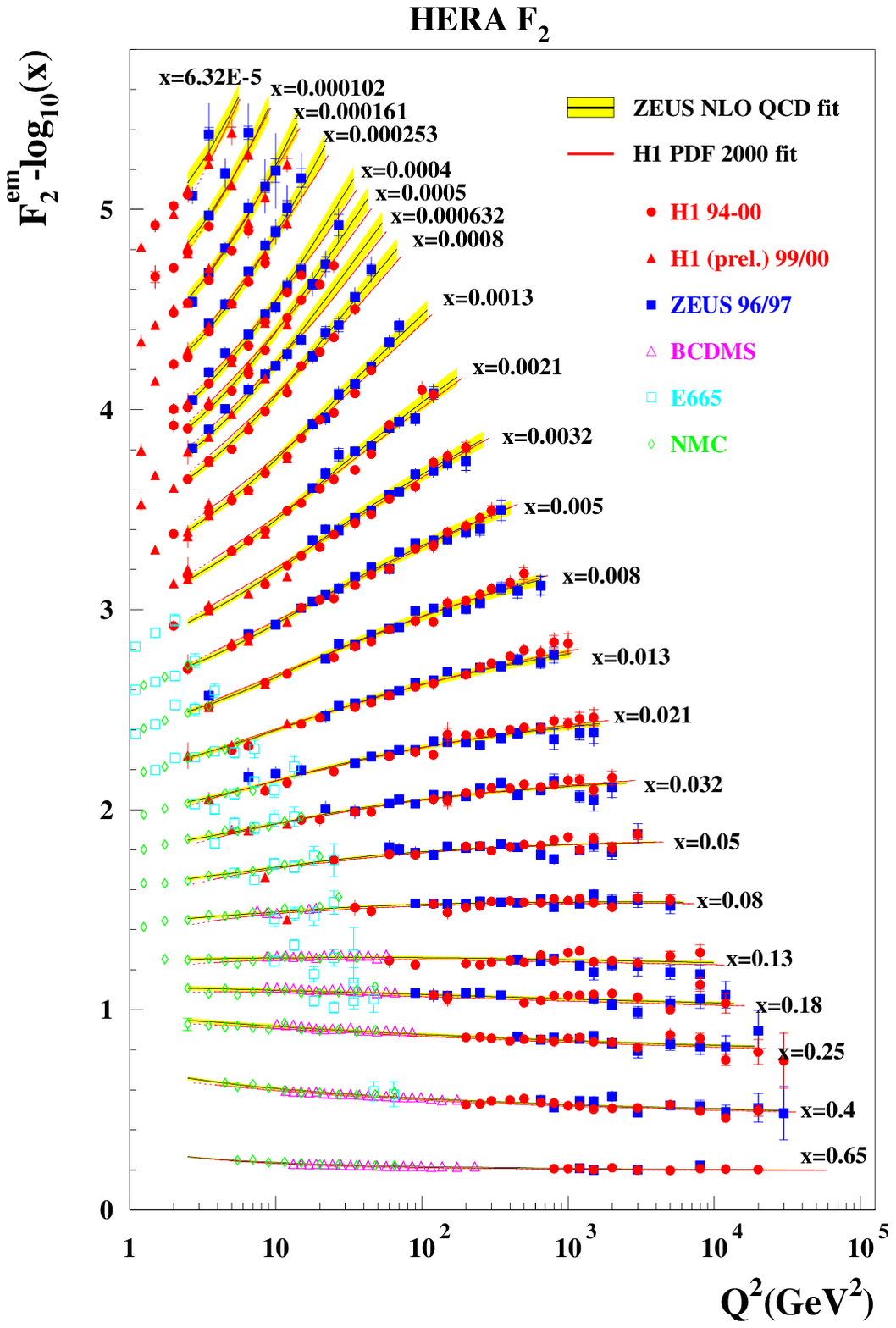}\qquad
 \includegraphics[width=0.5\textwidth]{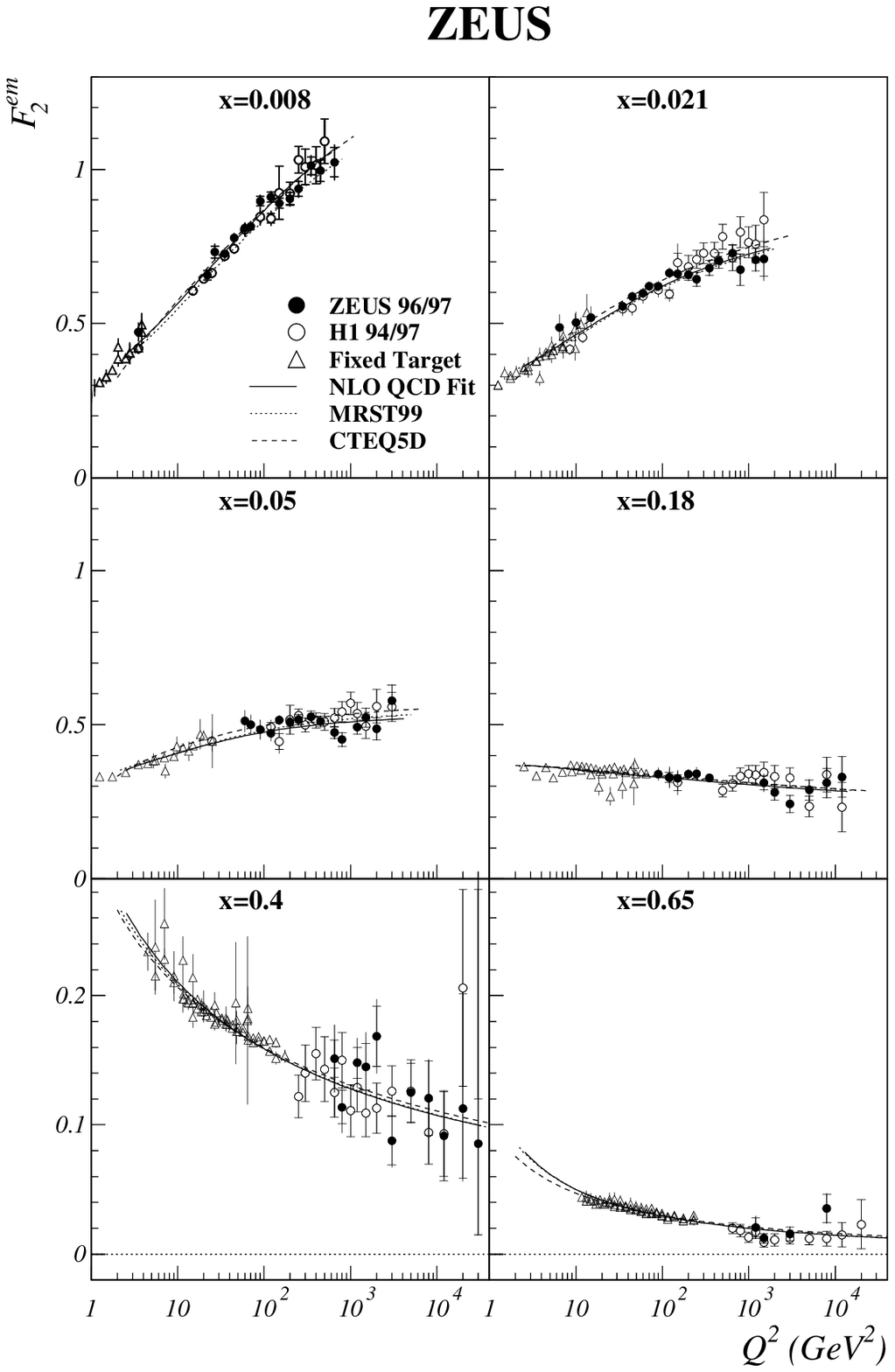}}
 \caption{\sl HERA results for $F_2$ (combined results from ZEUS
 and H1) which illustrate the effects of the evolution with
 increasing $Q^2$ for different values of $x$ : $F_2$ is increasing
 with $Q^2$ at all the values of $x$ except the very large ones
 $x\gtrsim 0.2$, where $F_2$ is decreasing. \label{fig:F2}} }

Here, we shall not discuss the perturbative evolution in more detail, but
merely emphasize some features which are interesting for comparison with
the situation at strong coupling, to be described later on. First note
that the parton lifetime, cf. \eqnum{partlife}, is strongly decreasing
when moving up along the cascade (for both the $Q^2$ and the small--$x$
evolutions), so that the cascade is frozen ---  the parton distribution
is fixed within it --- during the relatively short duration of the
collision with $\gamma^*$, cf. \eqnum{colltime}, which is the same as the
lifetime of the struck quark. Second, after each splitting, the energy of
the parent parton gets divided among the two daughter ones, so we expect
the evolution to increase the number of partons at small values of $x$
and decrease that at larger values. Moreover, the gluon distribution
should rise faster with decreasing $x$, so the small--$x$ partons should
be predominantly gluons. These expectations are indeed confirmed by the
experimental results at HERA displayed in Figs.~\ref{fig:F2} and
\ref{fig:HERA} \cite{Nagano} (and Refs. therein).

 \FIGURE[t]{ \centerline{
 \includegraphics[width=0.54\textwidth]{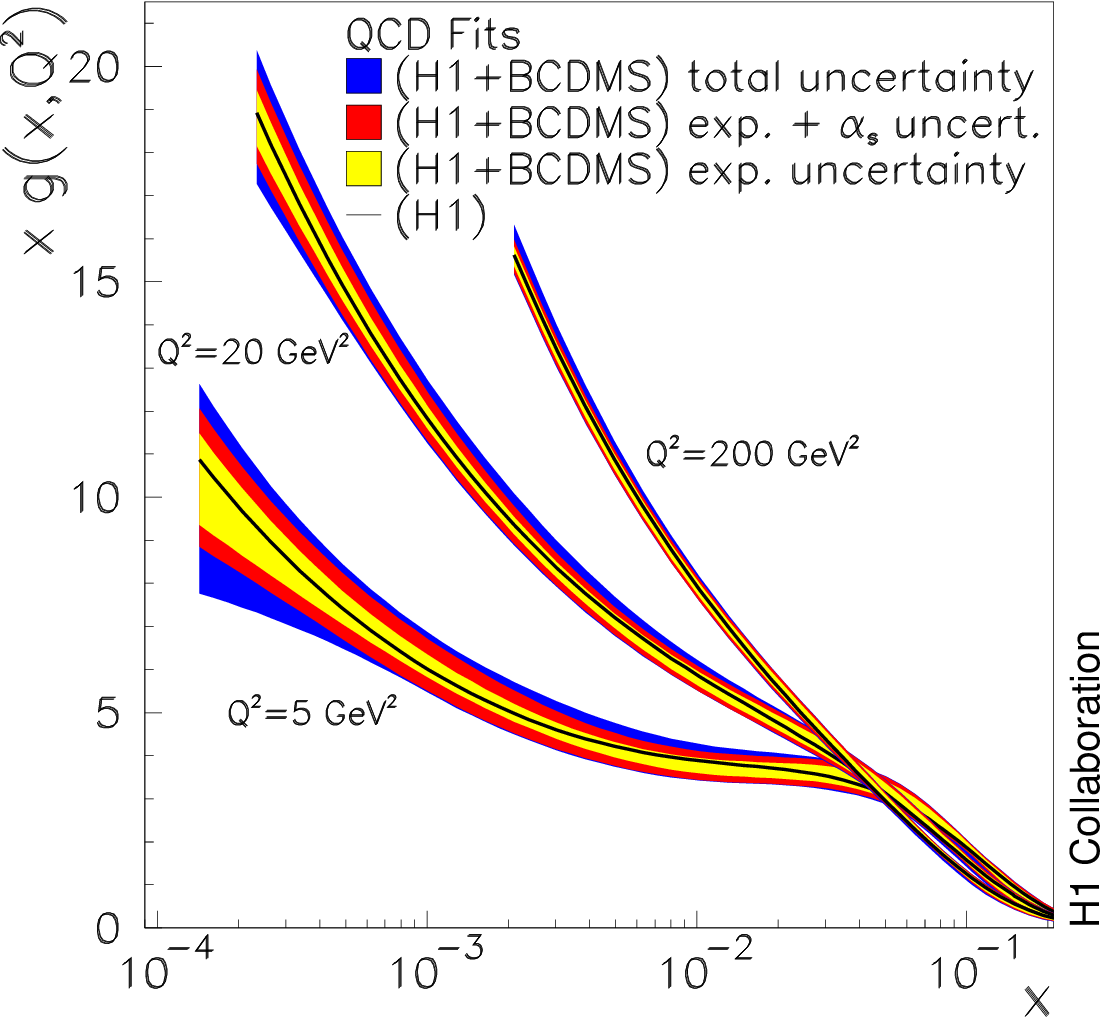}\qquad
 \includegraphics[width=0.55\textwidth]{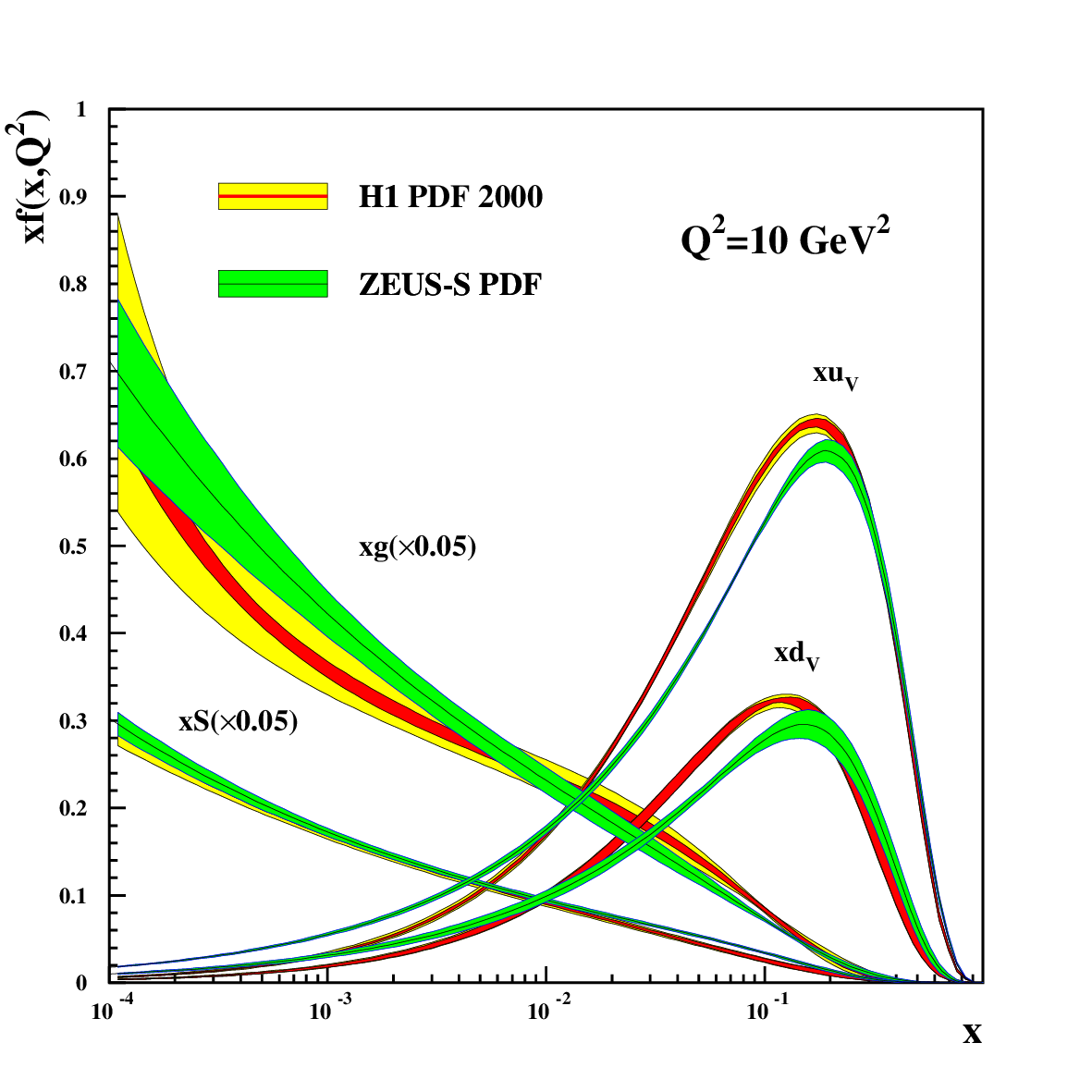}}
 \caption{\sl Parton distributions extracted from
 combined fits to the H1 and ZEUS data at HERA, which illustrate
 the evolution with decreasing $x$ at fixed $Q^2$. Left: the rise in the
 gluon distribution. Right: the $1/x$--evolution of the gluon, sea quark,
 and valence quark distributions for $Q^2=10$ GeV$^2$ (note that the
 gluon and sea quark distributions have been reduced by a factor of 20 to
 fit inside the figure).
  \label{fig:HERA}} }

But although they are less numerous, the few partons remaining at larger
values of $x$ do still carry most of the total energy of the proton, and
that even for very large $Q^2$. This is so since the dominant evolution
is such that the daughter gluon takes away only a small fraction of the
longitudinal momentum of its parent parton, so the latter `survives' (as
one of the $s$--channel partons in the cascades in
Fig.~\ref{fig:cascade}) with a relatively large momentum. To see this
more quantitatively, consider the following `energy sum--rule', which is
the condition that the ensemble of partons (quarks, antiquarks, and
gluons) which exist on a given resolution scale $Q^2$ carry the totality
of the proton longitudinal momentum:
 \beq\label{energysum}
 \int_0^1 \rmd x \,x
 \big[q(x,Q^2)+{\bar q}(x,Q^2)
 +g(x,Q^2)\big]\,=\,1\,.
 \eeq
The HERA data show that the `gluon distribution' $xg(x,Q^2)$ rises with
$1/x$ roughly like $xg(x,Q^2)\sim 1/x^\omega$ for $x\le 0.01$, but the
exponent $\omega$ is small enough, namely $\omega=0.2\div 0.3$ (it slowly
varies with $Q^2$), for the integral in \eqnum{energysum} to be dominated
by large values $x\sim 1$. This value $\omega=0.2\div 0.3$ is indeed
consistent with predictions of the QCD evolution equations at
next--to--leading--order (NLO) accuracy.

However, such a power increase with $1/x$ cannot continue forever, i.e.,
not up to arbitrarily high energies, since this would enter in conflict
with the unitarity constraint for DIS and other hadronic processes. For
instance, the cross--section for the virtual photon absorbtion by the
proton in DIS is related to $F_2$ :
 \beq
 \sigma_{\gamma^* p}(x,Q^2)\,=\,
 \frac{4\pi^2 \alpha_{\rm em}}{Q^2}\,F_2(x,Q^2)\,.\eeq
In the high--energy limit $x\to 0$ we expect this cross--section to grow,
at most, like a power of $\ln(1/x)$; this is Froissart bound and is a
consequence of the unitarity of the $S$--matrix. (A similar bound holds
for the $pp$ collisions to be studied at LHC.) There are also physical
arguments which are supported by explicit calculations within pQCD and
which are telling us what should be the physical mechanism responsible
for taming this growth: this is {\em gluon saturation}. With increasing
energy, the gluon density increases as well and eventually it becomes so
high that the gluon start interacting with each other --- meaning that
the evolution starts to be {\em non--linear}
--- and these interactions limit the further growth of the {\em gluon
occupation number}.

To understand the relevance of the occupation number --- a concept that
will be important at strong coupling as well --- notice that, in order to
interact with each other, the gluons must overlap, meaning that not only
their {\em number}, but also their (longitudinal and transverse) {\em
sizes}, should be large enough. At high--energy, the proton is Lorentz
contracted --- it looks to the virtual photon like a pancake --- so all
the partons within a longitudinal tube at a given impact parameter can
interact with the photon and also with each other. This argument must be
corrected for the uncertainty principle, but it is essentially correct:
the small--$x$ partons, with longitudinal momenta $k_z\simeq xP$, are
delocalized in $z$ over a distance $\Delta z\sim 1/xP$, which is of the
same order as the longitudinal wavelength of the virtual
photon\footnote{The last statement is strictly true in the Breit frame to
be introduced in Sect. 5.4.}. Incidentally, this argument also shows that
the longitudinal phase--space for DIS at high energy is measured by the
{\em rapidity} $Y\equiv \ln(1/x)\simeq \ln(s/Q^2)$ :
 \beq\label{rapid} \Delta k_z\,\Delta z\,\sim\,\frac{\Delta(xP)}{xP}
 \,\sim\,\frac{\Delta x}{x}\,\sim\,\Delta Y\,.\eeq
Indeed, the parton distributions are defined as the number of partons
{\em per unit rapidity}\,; e.g.,
 \beq\label{xgx}
 x g(x,Q^2)\,\equiv\,x\,\frac{\rmd N_{g}}{\rmd x}(Q^2)\,=\,
  \int\rmd^2 b_\perp\
 \int^{Q}\!\! \rmd^2k_\perp\
\frac{\rmd N_g}{\rmd Y \rmd^2b_\perp \rmd^2k_\perp}\,,
 \eeq
where the first integral runs over all impact parameters within the
proton transverse area and the second one over all the transverse momenta
up to $Q$ (cf. the discussion after \eqnum{colltime}).

 \FIGURE[t]{
\centerline{
\includegraphics[width=0.7\textwidth]{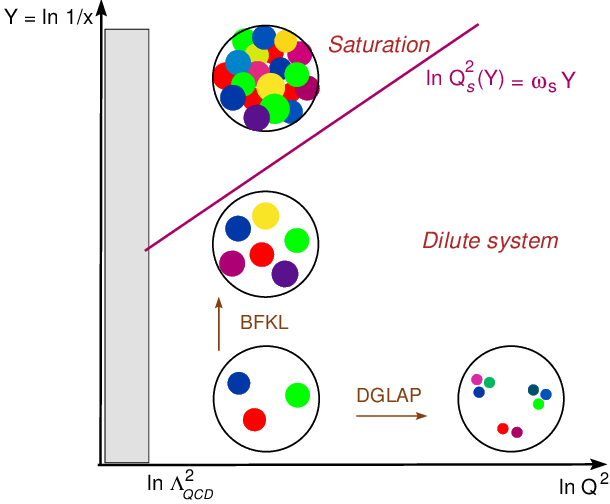}}
\caption{\sl The phase--space for parton evolution in the kinematical
variables appropriate for DIS ($\ln Q^2$ and $Y=\ln 1/x$), which
illustrates the distribution of partons (shown as colored blobs with area
$\sim 1/Q^2$) within the proton disk, and the saturation line $\ln
Q^2_s(Y)=\omega_s Y$. \label{fig:QSAT}} }

Consider now the gluon overlap in the two--dimensional transverse space.
As illustrated in Fig.~\ref{fig:QSAT}, when $Q^2$ is high, the gluons
form a dilute system (although they are relatively numerous) because each
of them occupies only a small area $\sim 1/Q^2$. But when decreasing $x$
at fixed $Q^2$, one emits more and more gluons having (almost) the same
area, so these gluons will eventually start overlapping. We see that,
what controls the gluon interactions with each other, is not their number
density ${xg(x,Q^2)}/{\pi R^2}$ ($R$ is the proton radius), but rather
their occupation number
 \beq\label{ngdef}
n_g(Y,b_\perp,k_\perp)\,\equiv\,\frac{(2\pi)^3}{2(N_c^2-1)}\,
\,\frac{{\rm
 d} N_g}{\rmd Y\, \rmd^2b_\perp \rmd^2k_\perp}\,\sim \,
 \frac{1}{Q^2}\times
 \frac{xg(x,Q^2)}{\pi R^2 (N_c^2-1)}
 \,.\eeq
As shown by the last estimate, $n_g$ measures the `fraction' of the
proton area which is covered with gluons of a given color. This
`fraction' can be bigger than one since the gluons can overlap with each
other. In fact, at weak coupling, the gluon interactions become an effect
of $\order{1}$ when $n_g\sim 1/(\alpha_s N_c)\sim 1/\lambda$, since in
that case the overlap is strong enough to compensate for the smallness of
the coupling. This condition defines a critical line in the kinematical
plane $(x,Q^2)$ --- the {\em saturation line} --- which separates between
a dilute region where $n_g\ll 1/\lambda$ and a dense region where the
occupation number saturates at a value $n_g\sim \order{1/\lambda}$ (see
Fig.~\ref{fig:QSAT}). One can solve this condition for $Q^2$ and thus
deduce the {\em saturation momentum}
 \beq\label{Qsat}
 Q^2_s(x)\, \sim \,
 {\lambda }\, \frac{x g(x,Q^2_s)}{R^2 (N_c^2-1)}\,\,\sim\,\,
 \frac{1}{x^{\omega}}\,, 
 \eeq
which is the value of the transverse momentum around which non--linear
effects become important for a given value of $x$. Alternatively, this is
the photon virtuality at which unitarity corrections become important in
DIS. As shown in \eqnum{Qsat}, $Q^2_s(x)$ rises with $1/x$ roughly like
the gluon distribution, i.e., as a power $1/x^{\omega}$ with
${\omega}\simeq 0.2\div 0.3$ from fits to the HERA data. (In logarithmic
coordinates $(Y,\ln Q^2)$, this yields a saturation line which is a
straight line, as shown in Fig.~\ref{fig:QSAT}.) Thus, with increasing
energy, the saturation region extends to higher and higher values of
$Q^2$, i.e., to smaller and smaller gluons.

These conclusions are supported by more refined analyses within pQCD,
which succeeded in resumming the non--linear effects associated with
gluon saturation within the evolution equations at high energy. This led
to non--linear generalizations of the BFKL equation --- the functional
JIMWLK equation and its mean--field (or large--$N_c$) approximation known
as BK --- which describe the transition towards saturation with
increasing energy and thus permit the calculation of the saturation line
(see the review papers \cite{CGC} and references therein). So far, the
full non--linear equations are known only to leading--order accuracy at
weak coupling, but the asymptotic form of the saturation line at high
energy is also known to NLO accuracy \cite{DT02}. Interestingly, such
analyses confirm the power--law behaviour $Q^2_s(x)\sim
{1}/{x^{\omega_s}}$ (at least, as an approximation valid in a limited
range in $Y$), but the value of the saturation exponent $\omega_s$ is
strongly reduced by NLO corrections: one finds $\omega_s\simeq
4.88(\alpha_s N_c/\pi)\simeq 0.12\lambda$ at LO (which would yield
$\omega_s\sim 1$ for $\alpha_s=0.2$ and $N_c=3$),  but $\omega_s\simeq
0.3$ at NLO. Note that this NLO value is roughly consistent with the
experimental results at HERA, thus suggesting that the (unknown)
corrections of higher order should be rather small. In fact, a
substantial fraction of the NLO corrections comes from the running of the
coupling \cite{DT02}.

\section{Current--current correlator from AdS/CFT: General formalism}

With this section, we begin the study of the main problem of interest
tous here, which is the propagation of a high--energy abelian current
through a strongly coupled plasma at temperature $T$. As mentioned in the
Introduction, our plasma will not be that of QCD, but rather the one
described by the maximally supersymmetric ${\mathcal N}=4$ Yang--Mills
theory, which is conformally invariant (so, in particular, the coupling
is fixed), and for which the AdS/CFT correspondence is most firmly
established. Since we shall not perform calculations directly in the
gauge theory (but only in the `dual' superstring theory), there is no
need to exhibit the Lagrangian of ${\mathcal N}=4$ SYM. (This can be
found in the textbooks listed in the References \cite{Becker,Kiritsis}.)
For our purposes, it suffices to recall that this Lagrangian involves 3
types of massless fields --- gluons, 4 Majorana fermions, and 6 real
scalars
--- which all transform under the adjoint representation of the colour
group SU$(N_c)$. Besides the Lagrangian has a global SU(4) ${\mathcal
R}$--symmetry (that is, a symmetry which does not commute with the
supersymmetry generators), under which the gluons are neutral, the four
fermions transform as a ${\bf 4}$ or $\bar{\bf 4}$ (depending upon their
chirality), and the six scalars transform as a ${\bf 6}$. This global
symmetry is interesting for our purposes as it allows one to introduce an
analog of the electromagnetism: to that aim, we shall pick one of the
U(1) subgroups of  SU(4) and gauge it, that is, replace the ordinary
derivatives by covariant derivatives: $\partial_\mu\to
\partial_\mu - iet^{a_0}_R A_\mu^{a_0}$ where $a_0$ is the SU(4)--index
of the chosen U(1) subgroup, $t^{a_0}_R$ is the respective generator in
the appropriate representation, and $A_\mu^{a_0}$ is an Abelian gauge
field endowed with the standard, Maxwell--like, kinetic term in the
action. Furthermore, $e$ is the analog of the electric charge, that we
shall take to be arbitrarily small. In the subsequent formulae, the
charge $e$ and the index $a_0$ will be always omitted. Associated to
$A_\mu^{a_0}\equiv A_\mu$ there is a conserved `electric current'
$J^\mu$, obtained by rewriting the interaction terms in the action
as\footnote{Strictly speaking, there is also a interaction piece in the
action which is quadratic in $A_\mu$, as coming from the scalar sector;
this will be neglected in what follows since $A_\mu$ can be taken to be
arbitrarily small.} $A_\mu J^\mu$. This current is built with selected
fermionic and scalar fields (see e.g. \cite{CaronHuot:2006te} for an
explicit construction). We shall refer to it as the `${\mathcal
R}$--current'.

The problem that we shall consider will be the scattering between this
${\mathcal R}$--current and the ${\mathcal N}=4$ SYM plasma in the
high--energy regime (the kinematics will be shortly specified) and in the
strong t' Hooft coupling limit taken as
 \beq\label{LNC}
 N_c\,\to\, \infty\quad\mbox{and}\quad\lambda\equiv g^2 N_c
 \,\to\,\infty\quad\mbox{with}\quad g^2\,\ll\,1.\eeq
That is, $N_c$ is taken to be arbitrarily large whereas the gauge
coupling $g$ is fixed and small. This limit is convenient for
applications of the AdS/CFT correspondence, as we now explain.

The AdS/CFT conjecture establishes a correspondence, or `duality',
between the ${\mathcal N}=4$ SYM theory (the `Conformal Field Theory')
with arbitrary values for the parameters $g$ and $N_c$ and the type IIB
superstring theory living in a $D=10$ curved space--time which is
$AdS_5\times S^5$ (hence, the `AdS'). This duality means that the
background geometry for the string theory corresponds to the vacuum of
the gauge theory, and that all the observables (like gauge--covariant
correlation functions) in one description can be equivalently calculated
--- after appropriate identifications --- in the other description. The
duality extends to finite temperature by adding a `black hole' to
$AdS_5$. One thus obtains the $AdS_5\times S^5$--Schwarzschild metric,
for which a common parametrization reads
 \beq
 \rmd s^2
 =\frac{r^2}{R^2}(-f(r)\rmd t^2+\rmd \bm{x}^2)+\frac{R^2}{r^2f(r)}
 \rmd r^2+R^2\rmd \Omega_5^2\,,
 \label{met2} \eeq
where $t$ and $\bm{x}=(x,y,z)$ are the time and spatial coordinates of
the physical Minkowski world, $r$ (with $0\le r < \infty$) is the radial
coordinate on $AdS_5$ (or `5th dimension'),  and $\rmd\Omega^2_5$ is the
angular measure on $S^5$. Furthermore, $R$ is the common radius of
$AdS_5$ and $S^5$, and
 \beq\label{def} f(r)\,=\,1-\frac{r_0^4}{r^4}\,=\,1-u^2\,=\,
 1-\frac{\chi^4}{\chi_0^4}\,, \eeq
where $r_0$ is the Black Hole (BH) horizon and is related to its
temperature $T$ (the same as for the ${\mathcal N}=4$ SYM plasma) via
$r_0=\pi R^2 T$. (Note that this BH is homogeneous in the four physical
dimensions but has an horizon in the fifth dimension which encloses the
real singularity at $r=0$.) When $r\to\infty$, $f(r)\to 1$ and $\rmd
s^2\propto (-\rmd t^2+\rmd \bm{x}^2)$ is conformal to the flat Minkowski
metric. Hence, the boundary of $AdS_5$ at $r\to\infty$ will be referred
to as the `Minkowski boundary'. In fact, we have $f(r)\approx 1$ whenever
$r\gg r_0$, so far away from the horizon the geometry is $AdS_5\times
S^5$. As shown in \eqnum{def}, some other radial coordinates will be also
used in what follows: these are defined as $u\equiv (r_0/r)^2$ and
$\chi\equiv R^2/r = \sqrt{u}/(\pi T)$, and in terms of them the Minkowski
boundary lies at $u=\chi=0$ and the BH horizon at $u_0=1$ and,
respectively, $\chi_0=1/\pi T$.

Besides $R$, the superstring theory involves two more parameters, the
(dimensionless) string coupling constant $g_s$ and the string length
$\ell_s$, which is the characteristic scale on which the string structure
(as opposed to a point--like particle) can be resolved, and is related to
the Planck length in ten dimensions by $\ell_{\rm P}=g_s^{1/4}\ell_s$.
The AdS/CFT correspondence makes the following identification between the
free parameters of the two dual descriptions:
 \beq\label{AdS}
   4\pi g_s\,=\,g^2\,,\qquad (R/\ell_s)^4\,=\,g^2N_c
 \,\equiv\,\lambda\,.\eeq
The first relation tells us that when the Yang--Mills coupling $g^2$ is
small, so is also the string coupling, hence one can neglect quantum
corrections (string loops) on the string theory side. The second relation
shows that when $\lambda$ is large, the geometry of the string theory is
weakly curved, so that the massive string excitations (with mass $m\sim
R/\ell_s^2$) can be reliably decoupled from the low--energies ones, and
then the superstring theory reduces to type IIB supergravity. Hence, when
we have both $g^2\ll 1$ and $\lambda\to\infty$ --- this corresponds to
the strong coupling limit of the ${\mathcal N}=4$ SYM theory in the sense
of \eqnum{LNC} ---, the dual superstring theory reduces to {\em classical
supergravity} in ten dimensions. After also performing a Kaluza--Klein
reduction around $S^5$ and keeping only the lowest harmonics, one finally
obtains a classical theory in five dimensions which involves massless
fields, among which the (5-dimensional) graviton, the dilaton, and a
SO(6) $\simeq$ SU(4) non--Abelian gauge field. The quantum correlation
functions in the strongly coupled CFT can now be computed from solutions
to the classical equations of motion for these massless fields with
appropriate boundary conditions.

In what follows, we shall describe this calculation for the problem of
interest here, namely the correlation functions of the ${\mathcal
R}$--current $J_\mu$. Let $Z_{4D}[A_\mu]$ denote the respective
generating functional in the 4--dimensional gauge theory ($A^\mu(x)$ is a
`dummy' source field for $J_\mu$). Within AdS/CFT, the current $J_\mu$ is
viewed as a perturbation of the supergravity fields acting at the
Minkowski boundary ($r\to\infty$, or $u=0$). Recall that $J_\mu$ carries
a hidden SU(4)--group index $a_0$, in addition to the manifest 4D vector
index $\mu$. Thus, by covariance, it is natural that this current induces
a non--zero expectation value for the respective component
$A_m^{a_0}\equiv A_m$ of the SO(6) vector field in 5D supergravity. (We
use $m,\,n,\,p,...$ to denote vector indices on $AdS_5$: $m=0,1,2,3,u$.)
For more clarity, let us temporarily denote by $A^m_{\rm cl}$ the
solution to the supergravity equations of motion obeying the appropriate
boundary conditions, that will be shortly specified. In the strong
coupling limit of \eqnum{LNC}, $Z_{4D}[A_\mu]$ can be computed as
 \beq\label{ZADS}
 Z_{4D}[A_\mu]\,\equiv\,\big\langle \rme^{i\int \rmd^4x J_\mu A^\mu}
 \big\rangle\ = \ \rme^{iS_{\rm SUGRA}[A_{\rm cl}]}\,,\eeq
where $S_{\rm SUGRA}[A_{\rm cl}]$ is the supergravity action evaluated
with the classical solution $A^m_{\rm cl}$ which in turn obeys the
boundary condition (BC)
 \beq\label{BCM}
 A_\mu^{\rm cl}(x,u=0) = A_\mu(x),\qquad A_u^{\rm cl}(x,u=0) =0\,,
 \eeq
and hence it is a functional of the 4D `source' field $A_\mu(x)$. The
classical EOM being second order differential equations, a second
boundary condition is needed to uniquely specify their solutions. As a
general rule, we shall require the solution to be regular everywhere in
the `bulk' (i.e., away from the Minkowsky boundary) of $AdS_5$. As we
shall see, however, this condition is not always sufficient, especially
at finite temperature. Whenever the solution involves modes which are
propagating in the radial direction, and which in general can either
approach towards the boundary (`incoming'), or move away from it
(`outgoing'), we shall require the physical solution to involve {\em
outgoing} modes alone. In the finite $T$ case, this can be physically
understood as the condition that the modes be fully absorbed by the BH,
without reflecting wave. More generally, at both zero and non--zero $T$,
this `outgoing wave' prescription generates the retarded current--current
correlator \cite{Son:2002sd}, which at finite $T$ is defined as
 \beq
 \Pi_{\mu\nu}(q)\,\equiv\,i\int \rmd^4x\,\rme^{-iq\cdot x}\,\theta(x_0)\,
 \langle [J_\mu(x), J_\nu(0)]\rangle_T\,, \label{Rdef}  \eeq
with the brackets denoting the thermal expectation value. Note that, in
order to compute $\Pi_{\mu\nu}$, it is sufficient to know the classical
action to quadratic order in the source field $A_\mu$, meaning that we
can take the latter (and hence the field $A^m_{\rm cl}$ induced in the
bulk) to be arbitrarily weak. Accordingly, we need the supergravity
action only to quadratic order in $A^m$; not surprisingly, this is the
same as the Maxwell action in the $AdS_5\times S^5$--Schwarzschild
background geometry:
    \beq\label{action0} S\,=\,-\frac{N^2_c}{64\pi^2R}\int \rmd^4x
   \rmd u \sqrt{-g}\,g^{mp}g^{nq}\,F_{mn}F_{pq}\,, \eeq
where $F_{mn}=\partial_m A_n-\partial_n A_m$, $\partial_m=
\partial/\partial x^m$ with $x^m=(t,\bm{x},u)$, $g=\det(g_{mn})$ is the
determinant of the matrix made with the covariant components of the
metric on $AdS_5$, cf. Eq.~(\ref{met2}), and $g^{mn}$ are the respective
contravariant components, as obtained by inverting the matrix $(g_{mn})$.
The classical EOM generated by (\ref{action0}) are Maxwell equations in a
curved space--time:
 \beq\label{maxwell} \partial_m\big(\sqrt{-g}g^{mp}g^{nq} F_{pq})\,=\,0\,.
 \eeq
We shall work in the gauge $A_u=0$ (which is consistent with the BC in
\eqnum{BCM}) and choose the incoming perturbation as a plane wave
propagating in the $z$ direction, with longitudinal momentum $k$ and
energy $\omega$ in the plasma rest frame: that is, our source field reads
$A_\mu(x)= A_\mu^{(0)}\rme^{-i\omega t+ik z}$. \eqnum{maxwell} being
linear, the solution $A^{\rm cl}_\mu$ (that we shall simply denote as
$A_\mu$ from now on, and refer to as the ``Maxwell wave'') preserves this
plane--wave structure in the Minkowski directions
   \beq \label{pw}
   A_\mu(t,\bm{x},u)\,=\,\rme^{-i\omega t+ik z}\,A_\mu(u)\,, \eeq
so the only non--trivial dependence is that upon $u$. This is determined
by the following equations, as obtained from \eqnum{maxwell} (below,
$i=1,2$) :
   \beq \varpi A^\prime_0+\kappa fA_3^\prime \,=\,0 \label{11} \\[0.2cm]
   A_i^{\prime\prime}+\frac{f^\prime}{f}A_i^\prime +
   \frac{\varpi^2-\kappa^2f}{uf^2}A_i\,=\,0 \label{ai} \\
   A_0^{\prime\prime}-\frac{1}{uf}(\kappa^2A_0+\varpi k A_3)\,=\,0  \label{13} \eeq
where a prime on a field indicates a $u$--derivative and we have
introduced dimensionless, energy and longitudinal momentum, variables,
defined as
 \beq\label{dimko}
\varpi\equiv \frac{\omega}{2\pi T}\,, \qquad  \kappa\equiv \frac{k}{2\pi
T}\,.
 \eeq
Denoting $a(u)\equiv A_0^\prime (u)$, Eqs.~(\ref{11}) and (\ref{13}) can
be combined to give
 \beq
a^{\prime\prime}+\,\frac{(uf)^\prime}{uf}\,a^\prime +
\,\frac{\varpi^2-\kappa^2f}{uf^2}\, a\,=\,0\,. \label{key}
   \eeq
The boundary conditions (\ref{BCM}) together with Eq.~(\ref{13}) imply
 \beq A_\mu(u=0) = A_\mu^{(0)} \ \Longrightarrow \
 \lim_{u\to 0}\big[u a'(u)\big]\,=\,\kappa^2A_0^{(0)}+\varpi \kappa A_3^{(0)}\,.
 \label{bc} \eeq
The field $a$ describes a longitudinal wave, while $A_1$ and $A_2$ are
transverse wave.

Because of the assumed plane wave structure, the action density in
\eqnum{action0} is homogeneous in the physical Minkowski directions, so
the corresponding integrations simply yield the volume of the 4D
space--time: $S=\int\rmd^4 x\, \mathcal{S} = \Delta V\,\Delta
t\,\mathcal{S}$. When evaluated on the classical solution, the action
density $\mathcal{S}$ is quadratic in the boundary values $A_\mu^{(0)}$
and yields the retarded polarization tensor via differentiation (with
$q^\mu=(\omega,0,0,k)$):
 \beq\label{SPI}
 \Pi_{\mu\nu}(q)\,=\,\frac{\partial^2 \mathcal{S}}
{\partial A_\mu^{(0)}
 \partial A_\nu^{(0)}}\,.\eeq
To that aim, it is useful to notice that the classical action density can
be fully expressed in terms of the values of the field $\tilde A_\mu(u)$
and of its first derivative at $u=0$ :
 \beq \mathcal{S} =\,\frac{N^2_cT^2}{16}\,
 \Big[- A_0\partial_u A_0^* + f A_3 \partial_u A_3^*
  + f A_i \partial_u A_i^*
\Big]_{u=0}
    \,. \label{actioncl} \eeq
(The appearance of the factor $T^2$ in front of $\mathcal{S}$ is merely a
consequence of our definition of the variable $u$, which scales like
$T^2$, so $\partial_u\sim 1/T^2$.) Eq.~(\ref{actioncl}) follows from
(\ref{action0}) after using the EOM (\ref{maxwell}) to perform an
integration by parts over $u$ and dropping the contribution from the
upper limit $u=1$ (i.e., from the BH horizon), in accordance with the
prescription in Ref. \cite{Son:2002sd,Herzog:2002pc}. A star on a field
denotes complex conjugation: the classical solutions develop an imaginary
part (in spite of obeying equations of motion with real coefficients)
because of the outgoing--wave condition at large $u$. Via
Eq.~(\ref{SPI}), this introduces an imaginary part in $\Pi_{\mu\nu}(q)$
which physically describes the dissipation of the current in the original
gauge theory. In fact, the imaginary part of the expression within the
square brackets in Eq.~(\ref{actioncl}) is independent of $u$ and hence
it can be evaluated at any $u$ \cite{Son:2002sd}.

Eqs.~(\ref{11})--(\ref{actioncl}) encode various physical phenomena
depending upon the kinematics : When  $\omega$ and $k$ are relatively
small, $\omega,\,k\ll T$, with moreover $\omega\ll k$, these equations
describe the diffusion of the $\mathcal{R}$--charge in the
strongly--coupled plasma and can be used to compute the respective
transport coefficient; this has been studied at length in Refs.
\cite{Klebanov:1997kc,Son:2002sd,Herzog:2002pc,Teaney:2006nc,SONREV}.
When $\omega=k$, they describe the photon emission from the plasma (for
$\mathcal{R}$--photons, of course); this has been studied in Ref.
\cite{CaronHuot:2006te} for the case $\omega,\,k\sim T$. When $\omega$
and $k$ are large compared to $T$, the equations describe the
high--energy scattering between the $\mathcal{R}$--current (or the
virtual $\mathcal{R}$--photon) and the plasma. This is the problem
addressed in Refs. \cite{HIM2,HIM3} and to which we shall devote our
attention in what follows. More precisely, we are interested in `hard
probes', so we shall choose a current with relatively high virtuality :
$Q^2\equiv {|\omega^2-k^2|}\gg T^2$, which probes the structure of the
plasma on distances much shorter than the thermal wavelength $1/T$. For a
space--like current ($\omega < k$), this set--up describes DIS, whereas
for a time--like current ($\omega > k$), it describes the current decay
into partons and their subsequent evolution in the plasma. In what
follows, we shall mostly assume the high--energy kinematics $\omega\sim
k\gg Q$, since this is the most interesting one for our purposes.

To conclude this section, let us present a different form of the
equations of motion, obtained after some change of variables, which will
be useful later on. For definiteness, we concentrate on the longitudinal
mode, and denote
 \beq\label{change}
 a(u)\,\equiv\,\frac{1}{(2\pi T)^2}\,
 \frac{\psi(\chi)}{\sqrt{\chi}}\,.\eeq
(Recall that $\chi\equiv {R^2}/{r}={\sqrt{u}}/{\pi T}$.) Then \eqnum{key}
becomes
 \beq\label{SchL}
 \psi''\,+\,\frac{1}{4\chi^2}\,\psi
\,+\,\frac{\omega^2-k^2f}{f^2}\,\psi\,+\,\frac{f'}{f}\bigg(
 \psi'-\frac{1}{\chi}\,\psi\bigg)\,=\,0\,,\eeq
where the prime now denotes differentiation w.r.t. $\chi$. This form of
the equation is interesting since the last term, proportional to $f'$,
can be neglected in all cases of interest, as we shall later argue. If
so, then the above equation becomes formally identical to the
Schr\"odinger equation for a non--relativistic particle with mass $k$
which is in a stationary state with zero energy:
 \beq\label{SchStat}
 -\frac{1}{2k}\,
 \frac{\del^2 \psi}{\del \chi^2}\,+\,V(\chi)\psi
 \,=\,E\psi\,,\quad\mbox{with}\quad V(\chi)=
 -\frac{1}{8k\chi^2}
 -\frac{\omega^2-k^2f}{2kf^2} \quad\mbox{and}\quad E=0\,.\eeq
This representation will allow us to use the intuition developed with the
Schr\"odinger equation for studies of the Maxwell wave propagating in the
$AdS_5$ geometry. A time--dependent generalization of this equation will
be also useful. Namely, assume that, instead of being a pure plane--wave,
the incoming perturbation, and thus the induced field $A_\mu$, are
wave--packets in energy peaked around $\omega$. The corresponding
equations of motion are obtained by replacing
\beq
 -\omega^2\,\longrightarrow\,
  \frac{\del^2}{\del t^2}\,\approx\,
-\omega^2-2i\omega
 \frac{\del}{\del t}
\eeq
in equations like \eqnum{key}. (The approximate equality above holds
since the additional time dependence on top of the phase $\rme^{-i\omega
t}$ is weak.) Then \eqnum{SchStat} is replaced by the time--dependent
version of the Schr\"odinger equation, which reads (in the high--energy
kinematics $\omega\sim k\gg Q$)
 \beq\label{Schtime}
 i\frac{\del  \psi}{\del t}\,=\,
  -\frac{1}{2k}\,
 \frac{\del^2 \psi}{\del \chi^2}\,+\,V(\chi)\psi\,.\eeq
In what follows, it will be useful to consider solutions to this equation
with the initial condition that, at $t=0$, the field $\psi(t=0,\chi)$ is
localized near the Minkowski boundary at $\chi=0$. Physically, this
corresponds to a point--like current, as we shall see.

\section{The vacuum case as a warm up}

Let us first consider the zero--temperature case, i.e. the propagation of
the ${\mathcal R}$--current through the vacuum of the ${\mathcal N}=4$
SYM theory at infinite 't Hooft coupling (cf. \eqnum{LNC}). Although the
corresponding result for $\Pi_{\mu\nu}$ is {\em a priori} known, for
reasons to be later explained, it is nevertheless interesting to go
through the calculations and explicitly deduce this result, in order to
get acquainted with the AdS/CFT formalism in a relatively simple set--up.
Moreover, as explained in Sect. 2, this result covers an interesting
physical problem: via \eqnum{sigmaepem}, it provides the total
cross--section for the analog of electron--positron annihilation at
strong coupling. The most interesting conclusion which will emerge from
the present discussion is that the AdS/CFT calculation is {\em not}
merely a `black box': by using its results together with physical
intuition and general arguments (like the uncertainty principle), one can
develop some physical understanding of the underlying process and of the
structure of the final state. That is, one get some physical insight into
the `blob' on the photon line in the right--hand figure in
Fig.~\ref{fig:optical}.

In the dual, supergravity, calculation the Maxwell wave propagates
through pure $AdS_5$ (no black hole), according to equations which are
obtained by letting $f\to 1$ in the equations in the previous
section\footnote{In this zero--temperature context, it is understood that
the reference scale $T$ which enters the definition of dimensionless
variables like $u$, $\varpi$, and $\kappa$, is some arbitrary mass scale,
which drops out from the final results.}. With $f=1$,
Eqs.~(\ref{11})--(\ref{key}), or (\ref{SchL}), depend upon $\omega$ and
$k$ only via the Lorentz--invariant combination $\omega^2-k^2$, which
defines the virtuality of the ${\mathcal R}$--current: $q^\mu
q_\mu=k^2-\omega^2$. This is as it should, since there is no privileged
frame at $T=0$. Then current conservation implies that $\Pi_{\mu\nu}(q)$
has the transverse structure displayed in \eqnum{PiTL}, i.e.
 \beq\label{R0} \Pi_{\mu\nu}(q)=
  \left(\eta_{\mu\nu}-\frac{q_\mu q_\nu}{q^2} \right)\Pi(q^2)\,\qquad
 \mbox{(vacuum)}\,.\eeq
The scalar function $\Pi(q^2)$ can be computed from a study of the
longitudinal sector alone, that is, by solving the vacuum version of
\eqnum{SchL}, which reads
 \beq\label{SchLvac}
 -\frac{1}{2k}\,\psi''\,+\,\left(-\frac{1}{8k\chi^2}
 \,\pm\,\frac{Q^2}{2k}\right)\psi\,=\,0\,.\eeq
where we recall that $Q^2 \equiv |k^2-\omega^2|$ and the plus (minus)
sign in front of $Q^2/2k$ corresponds to a space--like (time--like)
current. As anticipated, this is of the Schr\"odinger type, with the
potential exhibited in Fig.~\ref{fig:Vvac}. Already this figure is
telling us a lot about the dynamics: \texttt{(i)} in the {\em
space--like} case, there is a potential barrier with height $\sim Q^2/k$,
so the wave can penetrate only in the `classically allowed region' on the
left of the barrier, at $\chi\lesssim 1/Q$; \texttt{(ii)} in the {\em
time--like} case, there is no such barrier, so the wave can penetrate up
to arbitrarily large values of $\chi$, where it moves freely (since the
potential becomes flat for $\chi \gg 1/Q$). These general features will
be substantiated by the explicit solutions that we now construct. To that
aim, it is useful to notice that \eqnum{SchLvac} is tantamount to a
Bessel equation for the function $\psi/\sqrt{\chi}$.

 \FIGURE[t]{
\centerline{\includegraphics[width=0.45\textwidth]{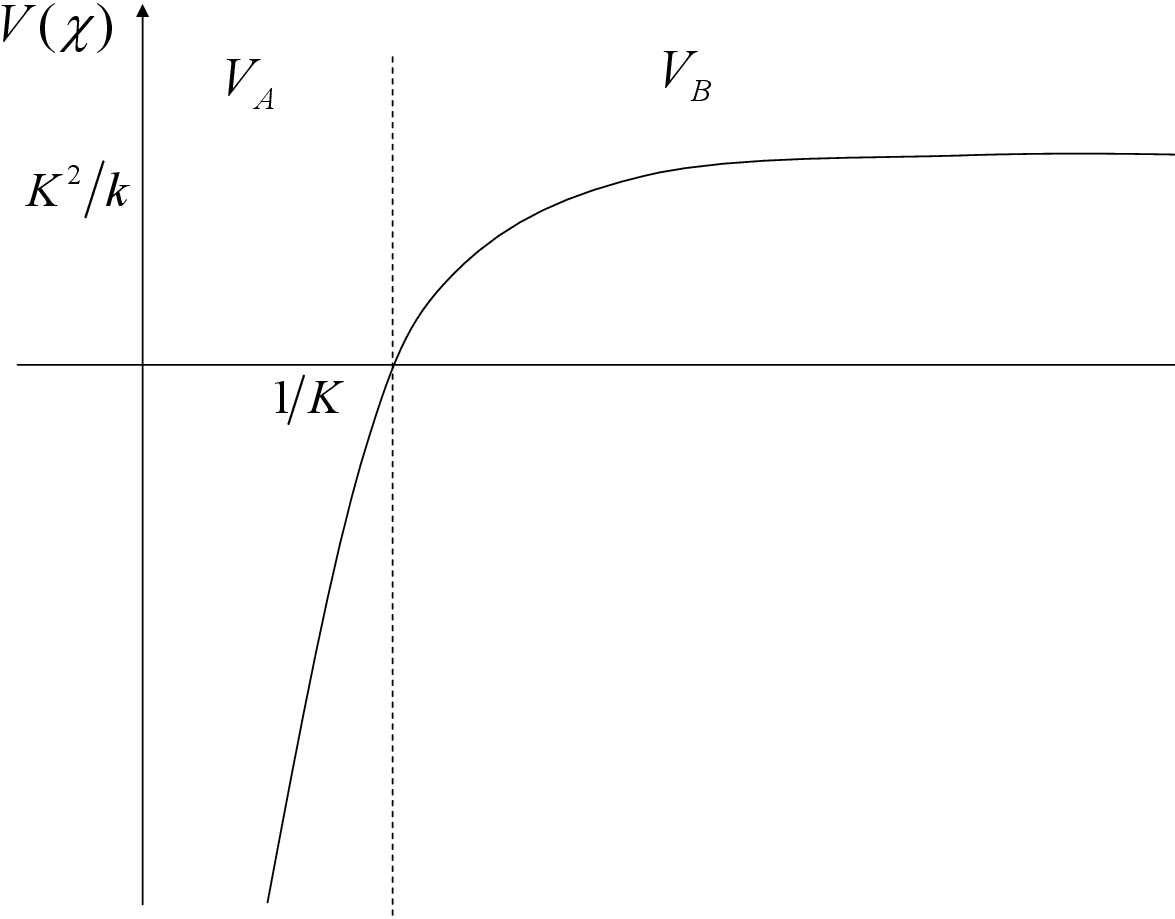}
\includegraphics[width=0.45\textwidth]{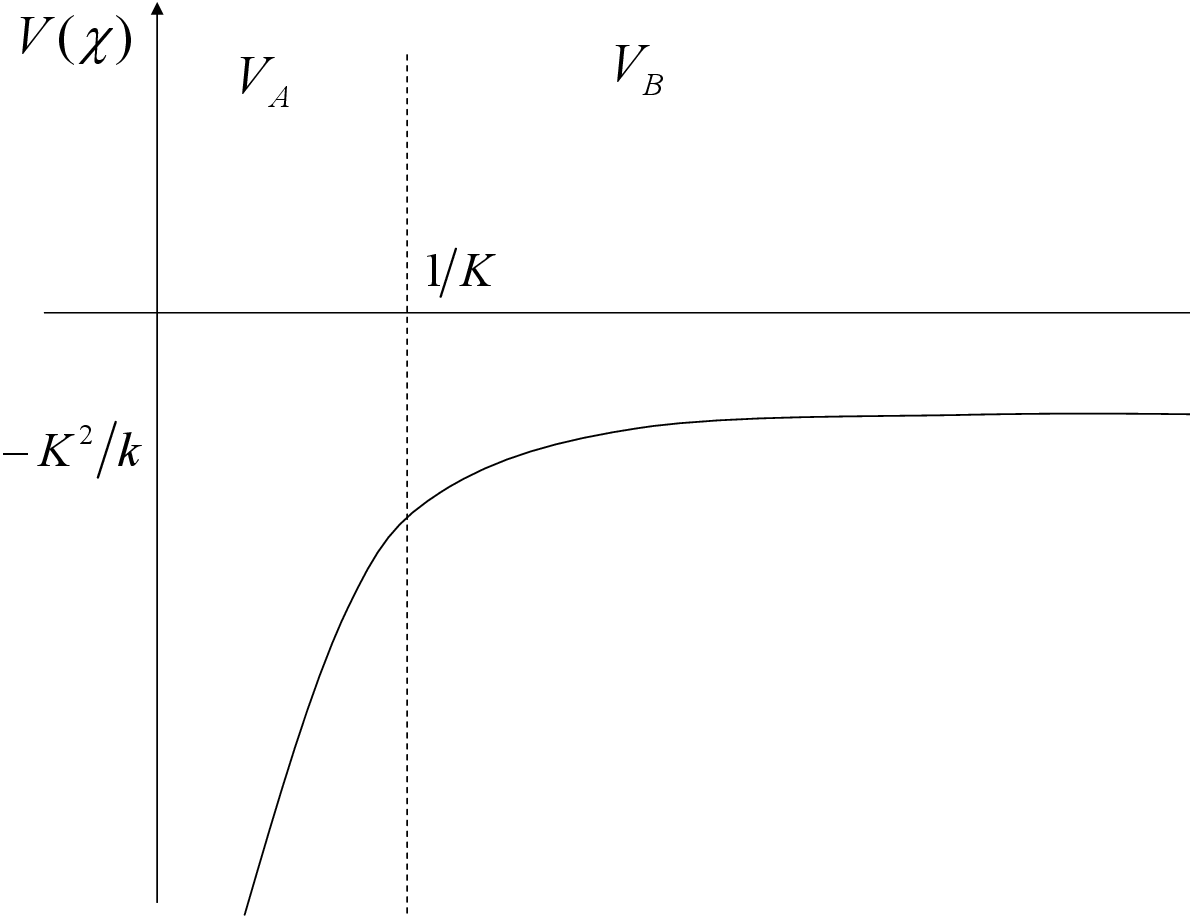}}
\caption{\sl The potential $V(\chi)$ in \eqnum{SchLvac} describing a
Maxwell wave propagating in the $AdS_5$ geometry. Left: the space--like
case ($\omega<k$). Right: the time--like case ($\omega > k$). In these
figures, we denoted $K\equiv |k^2-\omega^2|^{1/2}$ (i.e. $K$ is the same
as the variable $Q$ in the main text). \label{fig:Vvac}} }

\subsection{Space--like current}

For a space--like current ($q^2>0$), one needs to take the upper sign in
front of $Q^2$ in \eqnum{SchLvac}. The general solution is a linear
combination of the modified Bessel functions $\mathrm{K}_0$ and
$\mathrm{I}_0$ :
 \beq
 \psi(\chi)
     \,=\sqrt{\chi}\,\big(c_1\mathrm{K}_0(Q\chi)
     \,+\,c_2\mathrm{I}_0(Q\chi)\big)\,.
     \eeq
For large $x\equiv Q\chi \gg 1$,
 \beq \mathrm{K}_0(x)\,\approx\,\sqrt{\frac{\pi}{2x}}\,\rme^{-x}\,,
 \qquad
  \mathrm{I}_0(x)\,\approx\,\frac{1}{\sqrt{2\pi x}}\,\rme^{x}\,,\eeq
so the requirement that the solution remain regular as $\chi\to\infty$
selects $c_2=0$. The other coefficient $c_1$ is then fixed by the
boundary condition at $\chi=0$, cf. \eqnum{bc}, which becomes
 \beq\label{bcchi}
 \chi\,\frac{\del }{\del\chi}\,
 \frac{\psi}{\sqrt{\chi}}\Bigg |_{\chi\to 0}
 \,=\,2k \Big(kA_0^{(0)}+\omega A_3^{(0)}\Big)\,.\eeq
By also using the expansion $\mathrm{K}_0(x)\approx -\ln(x/2)-\gamma$
when $x\ll 1$, one easily finds $c_1=-2k\big(kA_0^{(0)}+\omega
A_3^{(0)}\big)$. Via \eqnum{change}, the solution $\psi(\chi)$ determines
the longitudinal piece of the classical action density, i.e., the pieces
involving $A_0$ and $A_3$ in Eq.~(\ref{actioncl}). A direct calculation
yields
 \beq\label{Snonren} \mathcal{S}_L =\,-\frac{N^2_c}{64\pi^2}\,
 \big(qA_0^{(0)}+\omega A_3^{(0)}\big)^2
 \,\big[\ln{Q^2}+ 2\ln\chi + {\rm const.}\big]_{\chi=0}\,,\eeq
which however exhibits a logarithmic divergence as $\chi=0$. This might
look disturbing at a first sight, but it has a natural resolution, that
we shall now explain:

Field theories are well known to develop divergences in the limit where
the ultraviolet cutoff (the upper cutoff on the momenta of the virtual
corrections) is sent to infinity. These divergences can generally be
eliminated via {\em ultraviolet renormalization}, i.e., by adding local
`counterterms' to the action, which amounts to (infinite)
renormalizations of what we mean by the fields in the action, their
masses, and their charges. In particular, the perturbative calculation of
the polarization tensor within ${\mathcal N}=4$ SYM meets with
logarithmic divergences of this type, which are then reabsorbed in the
normalization of the ${\mathcal R}$--charge (or of the wavefunction of
the ${\mathcal R}$--photon). But ultraviolet divergences and the need for
renormalization are not restricted to perturbation theory, as shown by
the example of lattice gauge theory. So, they are expected to appear also
in the supergravity calculation, which must somehow encode the effects of
{\em all} the quantum fluctuations of the dual gauge theory, including
those with very high momenta. This discussion makes it plausible to
interpret the logarithmic singularity in \eqnum{Snonren} as $\chi\to 0$
as the `dual counterpart' of the respective ultraviolet divergence in the
gauge theory. This is the content of the {\em holographic
renormalization} \cite{Skend02,Skenderis:2002wp}, which further instructs
us to simply drop out this divergent term, possibly together with
additional finite terms. Here we shall renormalize \eqnum{Snonren} by
replacing
 \beq\label{HOLREN}
 \ln{Q^2}+ 2\ln\chi + {\rm const.}\,\longrightarrow\,
 \ln\frac{Q^2}{\mu^2}\,,\eeq
which features the {\em subtraction scale} $\mu$. Via \eqnum{SPI}, this
finally yields the function $\Pi(q^2)$ displayed in \eqnum{Pivac} below
(for $q^2>0$), and which is real, as expected: a space--like current
cannot decay in the vacuum, by energy--momentum conservation.
Interestingly, the holographic renormalization shows that there is a
connection between the large momentum (more properly, {\em large
virtuality}) limit in the original gauge theory and the limit $\chi\to 0$
(or $r\to\infty$) in the dual supergravity theory. This is a
manifestation of the {\em ultraviolet--infrared correspondence}, that we
shall later discuss in more detail.

\subsection{Time--like current}

The corresponding equation is obtained by takin the lower sign in front
of $Q^2$ in \eqnum{SchLvac}. Then the general solution involves the
oscillating Bessel functions $\mathrm{J}_0$ and $\mathrm{N}_0$ :
 \beq
 \psi(\chi)
     \,=\sqrt{\chi}\,\big(c_1\mathrm{J}_0(Q\chi)
     \,+\,c_2\mathrm{N}_0(Q\chi)\big)\,.\eeq
The condition of regularity as $\chi\to\infty$ is automatically satisfied
by this general solution, so it brings no additional constraint. To fix
the solution, we shall rather require $\psi(t,\chi)=\rme^{-i\omega
t}\psi(\chi)$ to be an {\em outgoing wave} at large $\chi$, as explained
in the previous section. This requires $c_1=-ic_2$ which together with
the boundary condition (\ref{bcchi}) completely fixes the solution as
  \beq \psi(\chi)
     \,=\, -i\pi k\big(kA_0^{(0)}+\omega A_3^{(0)}\big)\sqrt{\chi}
  \,\mathrm{H}_0^{(1)}
   (Q\chi)\,, \label{a0TL} \eeq
where $\mathrm{H}_0^{(1)}=\mathrm{J}_0+i\mathrm{N}_0$ is a Hankel
function encoding the desired outgoing--wave behavior at large $\chi$ :
$\psi(t,\chi)\propto \rme^{-i(\omega t-Q\chi)}$ when $\chi\gg 1/Q$. The
remarkable feature of this solution is that it is {\em complex}, and thus
it encodes dissipation. Specifically, the longitudinal piece of the
action is obtained in the same form as in \eqnum{Snonren} except for an
additional imaginary part. The would--be singular term at the boundary,
which is real, is again removed as in \eqnum{HOLREN}, and the remaining,
finite, part is finally used to compute the function $\Pi(q^2)$.

One can combine together the results for both space--like and time--like
currents in the following expression (recall that $Q^2=|q^2|$):
 \beq\label{Pivac}
 \Pi(q^2)\,=\,\frac{N^2_c Q^2}{32\pi^2}\left(\ln\frac{Q^2}{\mu^2}
 - i\pi\Theta(-q^2)\rm{sgn}(\omega)\right)
 \,,\eeq
where the imaginary part for the time--like case ($q^2<0$) is manifest.
The sign of this imaginary part depends upon the sign of the energy, and
is such as to correspond to {\em retarded boundary conditions}. Hence, as
anticipated, Eqs.~(\ref{R0}) and (\ref{Pivac}) present the exact result
for the retarded, vacuum, polarization tensor of the ${\mathcal
R}$--current in the ${\mathcal N}=4$ SYM theory at infinite 't Hooft
coupling. This has been obtained here via a classical calculation in the
dual supergravity theory, but it also corresponds to an infinite
resummation of (planar) Feynman diagrams of the original gauge theory.
Can we say anything about the {\em physics} encoded in these diagrams ?

The first remarkable observation is that this all--order result in
\eqnum{Pivac} is formally identical to the respective result at {\em zero
order} in the Yang--Mills coupling $g$, i.e., the one--loop polarization
tensor (see, e.g., the left figure in Fig.~\ref{fig:optical}; recall
that, in ${\mathcal N}=4$ SYM, this loop involves both adjoint quarks and
adjoint scalars). This `coincidence' is a consequence of supersymmetry
which protects the conserved ${\mathcal R}$--current
\cite{Anselmi:1997am}; it means that all the higher loop corrections
cancel each other, but it does not tell us much about the physical
interpretation of the final result at strong--coupling. To gain more
physical insight, we shall rely on the {\em ultraviolet--infrared
correspondence}, that we shall first motivate, in the next subsection, on
the basis of our previous results.

\FIGURE[t]{
\centerline{\includegraphics[width=0.5\textwidth]{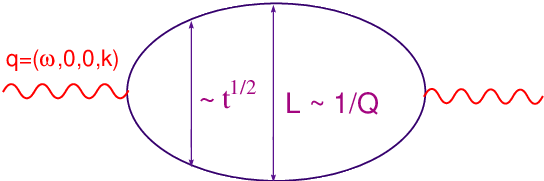}
\includegraphics[width=0.5\textwidth]{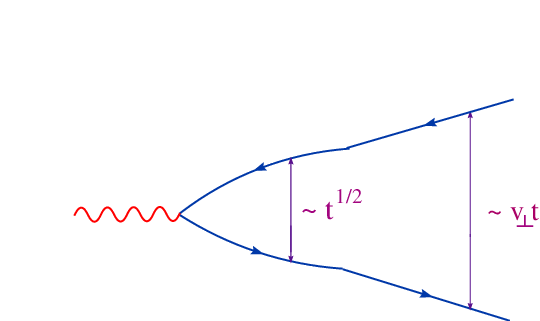}}
\caption{\sl Space--time picture for the ``one--loop'' (one parton pair)
fluctuation of a space--like current (left) and, respectively, time--like
current (right). \label{fig:FLUCT}} }

\subsection{The UV/IR correspondence}

For a {\em space--like} current, we found that the Maxwell wave can
penetrate into $AdS_5$ only up to a distance $\chi \sim 1/Q$ away from
the boundary. This should be put in relation with the fact that, by
energy--momentum conservation, a space--like current cannot decay in the
vacuum, but it generally develops virtual, partonic, fluctuations (see
Fig~\ref{fig:FLUCT} left), with transverse size $L\sim 1/Q$ and lifetime
$\Delta t_{\rm coh}$ which can be estimated from the uncertainty
principle as
 \beq\label{tcoh}
 L\,\sim\,\frac{1}{Q}\,,\qquad
 \Delta t_{\rm coh}\,\sim\,
 \,\frac{1}{Q}\,\times\,\frac{k}{Q}\,\sim
 \,\frac{k}{Q^2}\,.\eeq
As suggested by the above writing, $\Delta t_{\rm coh}$ is obtained as
the product between the lifetime $\sim 1/Q$ of the fluctuation in the
frame in which the current has zero energy (its `rest frame') and the
Lorentz gamma factor $\gamma= k/Q$. We refer to this lifetime as a
`coherence time' since this is the interval during which the current acts
as a color dipole, and hence it can interact via color gauge
interactions. Quantum dynamics also provides us with a {\em space--time
picture} for the fluctuation \cite{Farrar:1988me,Dokshitzer:1991wu}: if
the photon dissociates at $t=0$ into a point--like pair of fermions, or
scalars, then with increasing time the transverse size of this pair
increases diffusively,
 \beq L\sim \sqrt{\frac{t}{k}}\,, \eeq
until it reaches its maximal size $L\sim 1/Q$ at a time $t\sim
{k}/{Q^2}\sim \Delta t_{\rm coh}$.

Remarkably, it turns out that the very same space--time picture applies
for the penetration of the Maxwell wave inside $AdS_5$ \cite{HIM3}. To
see that, let us replace the plane--wave perturbation with a wave--packet
which at $t=0$ is localized near the boundary. Then, as explained at the
end of Sect.~3, the dynamics of the Maxwell wave $\psi(t,\chi)$ is
governed by the time--dependent Schr\"odinger equation (\ref{Schtime}),
which at early times (when the wave remains close to the boundary)
reduces to
 \beq\label{Schdiff}
 i\frac{\del  \psi}{\del t}\,=\,\left(
  -\frac{1}{2k}\,
 \frac{\del^2}{\del \chi^2}\,-\, \frac{1}{8k\chi^2}\right)\psi\,.\eeq
This is valid for $\chi\lesssim 1/Q$, which corresponds to times
$t\lesssim \Delta t_{\rm coh}$, as we shall shortly see. In this region,
the $Q^2$--dependent piece in the potential in \eqnum{SchLvac} is
negligible, so this early--time dynamics is in fact the same for both
space--like and time--like perturbations. \eqnum{Schdiff} admits the
following, exact solution
 \beq\label{psivac}
 \psi(t, \chi)\,=\,-i
  \frac{\sqrt{\chi}}{t}\
  \rme^{i\frac{k\chi^2}{2t}}\,, \eeq
which is such that, as $t\to 0$, the actual field $a(t, \chi) \propto
{\psi(\chi)}/{\sqrt{\chi}}$ (cf. \eqnum{change}) is indeed localized near
$\chi=0$, whereas for $t>0$ it penetrates into the bulk of $AdS_5$
through diffusion (i.e., by undergoing Brownian motion). This implies
that the (typical) position of the center of the wave--packet after a
time $t$ reads
 \beq \chi_{\rm diff}(t)
 \,\sim\, \sqrt{\frac{{t}}{k}}
 \,, \label{dif}
 \eeq
which becomes $\chi\sim 1/Q$ at time $t\sim {k}/{Q^2}$. For the
space--like wave, this is the maximal penetration distance, as clear by
inspection of Fig~\ref{fig:WAVEvac} left (and discussed in Sect.~4.1).

\FIGURE[t]{
\centerline{\includegraphics[width=0.55\textwidth]{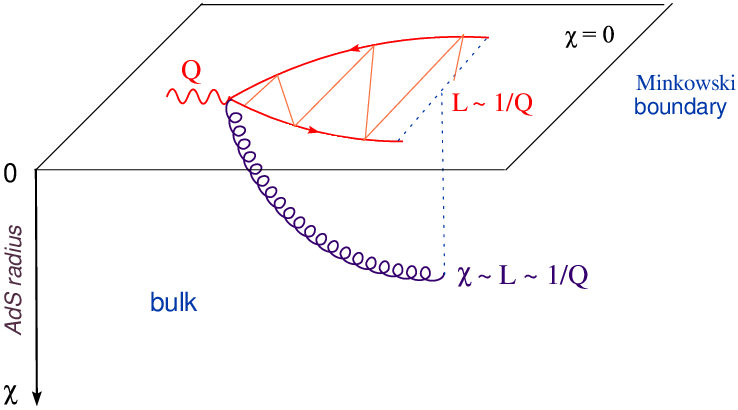}
\includegraphics[width=0.45\textwidth]{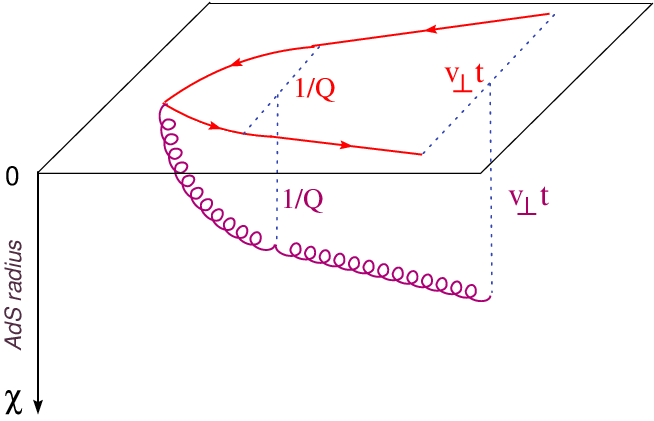}}
\caption{\sl Graphical illustrations of the progression of the Maxwell
wave in the radial dimension $\chi$ (the curly curve represents the
trajectory of the wave--packet) and of the dual partonic fluctuation on
the Minkowski boundary (which can be viewed as the `shadow' of the
Maxwell wave). Left: space--like case. Right: time--like case.
\label{fig:WAVEvac}} }

This precise analogy suggests an identification, or `duality', between
the penetration $\chi=R^2/r$ of the Maxwell wave inside $AdS_5$ and the
transverse size $L$, or inverse virtuality $1/Q$, of the partonic
fluctuation of the current in the gauge theory. This identification holds
in the sense of a proportionality, so like the uncertainty principle:

 \vspace*{.3cm}
 \framebox{Radial penetration $\chi=R^2/r$  in $AdS_5$
 \, $\sim$ \, Transverse size $L\sim 1/Q$ on the boundary}
 \vspace*{.3cm}

This is a specific form of the ultraviolet--infrared correspondence of
AdS/CFT \cite{UVIR,Polchinski} within the context of the high--energy
problem. This is often formulated as a correspondence between the 5th
dimension and the {\em `energy'} in the gauge theory. As such, this is
correct at low energy, but in general the `energy' should be replaced by
the (boost--invariant) {\em virtuality}  \cite{Brodsky:2008pg,HIM3}. As
we shall see, this correspondence is very helpful in reconstructing the
physical interpretation of the AdS/CFT results.

\subsection{Parton branching at strong coupling}

As a first application, consider the case of a {\em time--like} current.
If at $t=0$ we start again with a wave--packet localized near $\chi=0$,
then at early times $ t\lesssim {\omega}/{Q^2}$ the dynamics will be the
same as for a space--like current --- the wave--packet slowly diffuses
into the bulk up to a distance $\chi\sim 1/Q$ --- but then the situation
changes: instead of a potential barrier, the time--like wave meets with a
flat potential, so it can freely propagates towards larger values of
$\chi$ (cf. Figs.~\ref{fig:Vvac} and Fig~\ref{fig:WAVEvac} right). This
is manifest from our previous solution (\ref{a0TL}): by using the
asymptotic form of the Hankel function valid at $\chi Q\gg 1$ and
restoring the exponential dependencies upon $t$ and $z$, one finds that
the late--time solution behaves like
 \beq\label{largetvac}
 \rme^{-i\omega t+ik z}\,
 \frac{\psi}{\sqrt{\chi}}\,\propto\,
  \exp\left\{-i\omega t+ik z +iQ\chi\right\}\,.
 \eeq
This describes a wave--packet\footnote{More precisely a wave--packet
would involve an integration over different values of the energy around
the central value $\omega$; but if the packet is strongly peaked around
$\omega$, the group velocity is indeed given by \eqnum{groupvac}.}
propagating in $AdS_5$ with constant radial velocity $v_\chi=Q/\omega$\,:
 \beq \label{groupvac}
 \frac{\del }{\del \omega}\left(\omega t
 -\sqrt{\omega^2-k^2}\,\chi\right)
 \,=\,0\qquad\Longrightarrow\qquad\chi(t)\,=\,
 \frac{Q}{\omega}\,t\,\equiv\,v_\chi t\,.
 \eeq
At the same time, this wave--packet moves along the $z$ direction with
constant velocity  $v_z={k}/{\omega}$ (as obvious by taking a derivative
in \eqnum{largetvac} w.r.t. $k$ at constant $Q$). $v_z$ is recognized as
the longitudinal velocity of the incoming, time--like current. Notice
that $v_z^2+v_\chi^2=1$, which is the velocity of light in $AdS_5$. We
thus conclude that, for times $t> {\omega}/{Q^2}$, the wave--packet
propagates in $AdS_5$ along a {\em light--like geodesic}.

Via the UV/IR correspondence $\chi\sim L$, these results predict the
following behaviour on the gauge theory side (see Fig~\ref{fig:WAVEvac}
right) : For times $t> {\omega}/{Q^2}$, the partonic system produced via
the dissociation of the time--like current expands in transverse
directions at a constant speed $v_\perp=v_\chi$\,:
 \beq\label{Lfree}
 L(t)\,\sim\,v_\perp t\quad\mbox{with}\quad v_\perp\,=\,\frac{Q}{\omega}\,
 = \sqrt{1-v_z^2}\quad\mbox{where}\quad v_z\,=\,\frac{k}{\omega}\,.
 \eeq
This behaviour admits two different physical interpretations, but as we
shall argue below only the second one is acceptable at strong coupling\,:

\texttt{(i)} {\bf The decay of the current into a pair of partons.}\\ The
time--like current decays into a pair of on--shell, massless partons
(adjoint fermions or scalars) of ${\mathcal N}=4$ SYM theory, which then
move together along the $z$ direction with a longitudinal velocity
$v_z=k/\omega$ inherited from the current, while separating from each
other in transverse directions at velocity $v_\perp=\sqrt{1-v_z^2}$.

This is, of course, the space--time picture of the one--loop
approximation to $\Pi_{\mu\nu}$ and as such it must be {\em consistent}
with the AdS/CFT calculation, since the result of the latter turns out to
be formally the same as the respective one--loop result. But being
`consistent' it not necessarily the same as being {\em correct}. At
strong coupling there is no reason why parton branching should stop at
2--parton level: it takes some time before the original pair of partons
can get on--shell, and during this time they will further radiate, as the
emission time is shorter than the time necessary to evacuate their
virtuality. At weak coupling, such additional emissions are suppressed by
powers of $g$, so they appear as higher--order corrections (cf. the
discussion in Sect. 2). But at strong coupling, there is no such a
suppression, and hence nothing can slow down the branching process, which
is required by the uncertainty principle. Following the same idea, there
is no reason why, at strong coupling, parton branching should favor
special corners of the phase--space, like soft or collinear partons:
phase--space enhancement is not needed when the coupling is strong. Such
considerations suggest a space--time picture for parton evolution at
strong coupling which is quite different from the corresponding one at
weak coupling, and that we now present:

\FIGURE[t]{
\centerline{\includegraphics[width=0.4\textwidth]{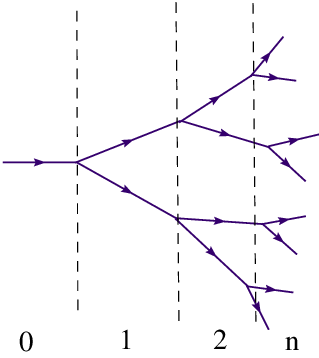}}
\caption{\sl Qualitative picture of the parton cascade generated through
`quasi--democratic' branching at strong coupling. A line (`branch') with
an arrow denotes any of the massless partons (quark, gluon, or scalar) of
the ${\mathcal N}=4$ SYM Lagrangian, except possibly for the first
parton, which initiates the cascade, which can also be a virtual
$\mathcal{R}$--photon. \label{fig:QDBRANCH}} }

\texttt{(ii)} {\bf Quasi--democratic parton branching at strong coupling}
\cite{HIM3}.\\
The virtuality of the current, or of any virtual parton which is
time--like, is evacuated via successive parton branchings which are
`quasi--democratic': at each step in this branching process, the energy
and virtuality are almost equally divided among the daughter partons.
This picture, which is more acceptable at strong coupling, is indeed
consistent with the previous AdS/CFT results, as we now show:

Let $n=0,1,2,\,...$ be the generation index, with $Q_0=Q$ and
$\omega_0=\omega$ (see Fig.~\ref{fig:QDBRANCH}). Then we can write
 \beq\label{BRvacuum}
 \omega_n\,\sim\,\frac{\omega_{n-1}}{2}\,\sim\,\frac{\omega}{2^n}\,,
 \qquad Q_n\,\sim\,\frac{Q_{n-1}}{2}\,\sim\,\frac{Q}{2^n}\,,\qquad
  \Delta t_n\,\sim\,\frac{\omega_{n}}{Q_n^2}\,,\eeq
where the lifetime $\Delta t_n$ of the $n$th parton generation has been
estimated via the uncertainty principle. This implies
 \beq\label{dQdt}
  \frac{Q_n-Q_{n-1}}{\Delta t_n}\, \sim\,-\,\frac{Q}{\omega}\,Q_n^2
 \quad\Longrightarrow\quad \frac{\rmd Q(t)}{\rmd t}\,\simeq\,
 \,-\,\frac{Q^2(t)}{\gamma}\,,\eeq
where $\gamma\equiv \omega/Q = 1/\sqrt{1-v_z^2}$ is the Lorentz factor
for both the incoming, time--like, current and any of the virtual partons
produced via its decay: indeed, the ratio $\omega_n/Q_n \approx \omega/Q$
is approximately constant during the branching process, hence
$\gamma_n\approx \gamma$. This means that each parton generation
progresses along the longitudinal direction at the same speed $v_z$ as
the original current would do. But at the same time the virtuality
decreases from one generation to another, hence the partonic system
expands in transverse directions. Specifically, Eq.~(\ref{dQdt}) together
with the uncertainty principle $L(t)\sim 1/Q(t)$ implies that the
transverse size of the partonic system increases like $L(t)\sim
\sqrt{1-v_z^2}\,t$, in qualitative agreement with the AdS/CFT result in
\eqnum{Lfree}.

By integrating Eq.~(\ref{dQdt}), one can deduce the virtuality $Q(t)$ and
the energy $\omega(t)=\gamma Q(t)$ that a typical parton in the cascade
will have after a time $t\ge \Delta t_0\sim \omega/Q^2$. One thus finds
 \beq\label{QEvac}
 Q(t)\, \simeq\,\frac{\gamma}{t}\,,\qquad \omega(t)\, \simeq\,\frac{\gamma^2}{t}
 \,.\eeq
(For $t= \omega/Q^2$, these equations yield $Q(t)=Q$ and
$\omega(t)=\omega$, as they should.) Of course, the total energy of the
partonic system is conserved and equal to $\omega$ (the energy of the
incoming photon), but with increasing time this energy gets spread among
more and more partons. One can indeed check that the number of partons
within the cascade increases like $N_{\rm part}(t)\simeq t/\Delta t_0$.

For how long will this branching process last ? Within the conformal
${\mathcal N}=4$ SYM theory, the partons will keep branching for ever,
thus producing more and more partons, with lower and lower energies. But
if one introduced a infrared cutoff $\Lambda$ in the theory, as a crude
model to mimic confinement and ensure the existence of hadron--like
states, then the branching will continue until the parton virtualities
degrade down to values of order $\Lambda$; then hadrons will form and the
particle distribution will get frozen. The total duration of the
branching process is essentially the same as the lifetime $\Delta t_N$ of
the last generation, the one with $Q_N\sim\Lambda$. (Indeed the parton
lifetime increases down the cascade: $\Delta t_n\simeq 2 \Delta
t_{n-1}\simeq 2^n \Delta t_0$.) This yields $\Delta t_N\sim 2^N(\omega
/Q^2)\sim \gamma/\Lambda$, where we have used $2^N=Q/\Lambda$ and
$\omega/Q=\gamma$. The final partons produced in this process are
relatively numerous ($N_{\rm part}\sim 2^N=Q/\Lambda\gg 1$) and have
small transverse momenta $k_\perp\sim Q_N\sim \Lambda$, so they will be
{\em isotropically distributed in transverse space}, within a disk with
area $\sim 1/\Lambda^2$ around the longitudinal axis.

\FIGURE[t]{
\centerline{
\includegraphics[width=.9\textwidth]{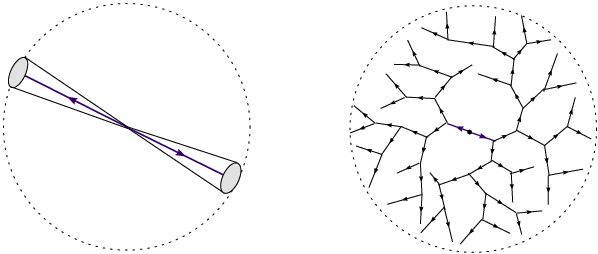}}
\caption{\sl Final state produced in $e^+e^-$ annihilation. Left: weak
coupling. Right: strong coupling. }\label{Fig:ISOTROPY} }

This picture of `quasi--democratic branching' --- that one should think
off as a kind of mean field approximation to the actual dynamics in the
gauge theory at strong coupling --- has intriguing consequences for
processes like $e^+e^-$ annihilation at high energy ($\sqrt{s}\gg
\Lambda$). Consider the respective final state as seen in the center of
mass frame. Unlike what happens in QCD at weak coupling, where this state
involves only a few, well collimated, jets (cf. Sect. 2 and
Fig.~\ref{Fig:ISOTROPY} left), at strong coupling there will be {\em no
jets at all} ! Rather, the final hadrons will be relatively soft --- they
all carry energies and momenta of order $\Lambda$ ---, numerous and
isotropically distributed in space, as illustrated in the r.h.s. of
Fig.~\ref{Fig:ISOTROPY}. (See also Refs. \cite{Hofman,Shuryak,Mat,Csaki}
for different arguments leading to similar conclusions.) Such a structure
for the final state is clearly inconsistent with what is actually seen in
the high--energy experiments, and this should not come as a surprise: as
argued in Sect. 2 (see, e.g., \eqnum{sigmaepem}), the decay of a
highly--energetic time--like current in QCD is rather controlled by {\em
weak coupling}, because of asymptotic freedom. One may nevertheless hope
that strong--coupling techniques like AdS/CFT could be more useful when
the current propagates through a finite--temperature plasma, where the
relevant coupling is believed to be stronger. This is the topics that we
shall discuss in the next section.

\section{${\mathcal R}$--current in the ${\mathcal N}=4$ SYM plasma
at strong coupling}

We are now prepared to address the problem which is our main physical
interest, namely the propagation of the ${\mathcal R}$--current through
the strongly coupled ${\mathcal N}=4$ SYM plasma. As we shall see, the
corresponding AdS/CFT results are again suggestive of a
`quasi--democratic branching' picture, which is now generalized to
accommodate the effects of the plasma.

We focus on a current with large virtuality, $Q\gg T$ (`hard probe'),
which therefore explores the structure of the plasma on distances short
as compared to the thermal wavelength $1/T$. We shall perform our
calculations in the plasma rest frame, but then interpret the results in
the plasma infinite momentum frame, in order to unveil the partonic
structure of the plasma. It is moreover interesting to choose this
current to have a relatively high longitudinal momentum in the plasma
rest frame, such that $k\gg Q\gg T$ (which in turn implies a high energy:
$\omega\sim k$; recall that $Q^2 \equiv |k^2-\omega^2|$). Indeed, below
\eqnum{tcoh} we have argued that the interactions of the current with an
external target extend over a time $\Delta t_{\rm coh}\sim k/Q^2$, i.e.,
the lifetime of its partonic fluctuation. For the current to explore
medium properties in the plasma, we would like this time to be much
larger than $1/T$ --- so that the current explores a relatively large
longitudinal slice $\Delta z\sim \Delta t_{\rm coh}\gg 1/T$. This implies
$k\gg Q^2/T$ (and hence $k\simeq\omega \gg Q$), which is tantamount to
the condition that the associated Bjorken--$x$ variable be very small:
$x\ll 1$. (This variable will be introduced in \eqnum{xT} below.) In
fact, as we shall later discover, for a space--like current to
significantly interact with the plasma we need an even higher energy
$\omega\gtrsim Q^3/T^2$ \cite{HIM2}.

For an ordinary plasma at weak coupling, this physical set--up would
probe the parton evolution of the individual thermal quasiparticles, so
the plasma structure functions would be simply the sum of the structure
functions for those quasiparticles weighted by the respective densities
in thermal equilibrium. For instance, the gluon distribution per unit
volume in the weakly--coupled quark--gluon plasma is given by
 \beq
 x g(x,Q^2)\,\approx\,n_q(T)x g_q(x,Q^2)\,+\,n_g(T)x g_g(x,Q^2)\,,\eeq
where $n_q(T)\sim N_cT^3$ and $n_g(T)\sim N_c^2T^3$ are the thermal
densities for (anti)quarks and gluons, and $x g_q(x,Q^2)$ and $x
g_g(x,Q^2)$ are gluon distribution functions generated by the evolution
of a single quark, or gluon, respectively. However, at strong coupling,
the quasiparticle structure of the plasma is not known (if any !) and,
moreover, we expect the evolution of the plasma as whole to be different
from that of its individual constituents taken separately (once again,
assuming that such individual constituents exist in the first place,
which may not be true !). It then becomes interesting and meaningful to
compute directly the plasma structure functions. This calculation refers
to a space--like current, but a time--like current is interesting too,
since this decays into jets, which then interact with the plasma. In what
follows we shall consider both space--like and time--like currents, but
we shall skip most technical details and focus on the results and their
physical interpretation.

\subsection{Space--like current: DIS off the strongly coupled plasma}

Let us start with some kinematics. The polarization tensor in the plasma,
as defined in \eqnum{Rdef}, involves two independent scalar functions,
$\Pi_1$ and $\Pi_2$, and admits the following decomposition in a generic
frame:
 \beq\label{PiTten}  \Pi_{\mu\nu}(q,T)=
  \left(\eta_{\mu\nu}-\frac{q_\mu q_\nu}{Q^2} \right)\Pi_1(x,Q^2)+
  \left(n_\mu-q_\mu \frac{n\cdot q}{Q^2}\right)
  \left(n_\nu-q_\nu \frac{n\cdot q}{Q^2}\right)\Pi_2(x,Q^2)\,.\eeq
Here $n^\mu$ is the four--velocity of the plasma, with $n^\mu=(1,0,0,0)$
corresponding to the plasma at rest. Also, the Bjorken--$x$ variable for
the current--plasma scattering is defined as
 \beq\label{xT}
       {x}\,\equiv\,\frac{Q^2}{-2(q\cdot n)
       T}\,=\,\frac{Q^2}{2\omega T}\,,\eeq
with the second expression valid in the plasma rest frame, where
$q^\mu=(\omega,0,0,k)$. The plasma structure functions are obtained as
 \beq  \label{F12T} F_1(x,Q^2)\,=\,\frac{1}{2\pi}\, {\rm Im} \,\Pi_1,\qquad
  F_2(x,Q^2)\,=\,\frac{-(n\cdot q)}{2\pi T}\, {\rm Im}\, \Pi_2\,. \eeq
This tensorial structure is similar to that introduced in Sect. 2 for DIS
off a hadron, and the above formulae correspond indeed to
Eqs.~(\ref{Pitensor}) and (\ref{F12}) up to the replacement $P^\mu\to
n^\mu$. But unlike the hadronic polarization tensor, or structure
functions, which are dimensionless, their plasma counterparts in
Eqs.~(\ref{PiTten})-(\ref{F12T}) have dimensions of momentum squared.
This difference is related to their physical interpretation that we shall
later discuss.

In order to compute $\Pi_1$ and $\Pi_2$ from classical supergravity, we
need to solve the equations for both the transverse and longitudinal
Maxwell waves, that is, Eqs.~(\ref{ai}) and (\ref{key}), respectively.
There is an important simplification which simplifies this analysis: the
most important dynamics takes plays relatively far away from the BH
horizon, at $\chi\ll \chi_0$, where $f(\chi)\approx 1$ (cf. \eqnum{def}).
Of course, the absorbtion of the wave by the BH takes place around the
horizon, but the effects of the interactions with the BH makes themselves
felt already well above $\chi_0$, because of the long range nature of the
gravitational interactions (see below); in turn, these long--range
interactions uniquely determine the classical solution near the Minkowsky
boundary ($\chi\to 0$), which is all that we need in order to compute the
polarisation tensor (cf. Eq.~(\ref{actioncl})). Because of that, we can
replace $f\to 1$ (i.e., ignore the effects of the BH) everywhere except
in the terms where the difference $1-f=(\chi/\chi_0)^4$ is amplified by
the large longitudinal momentum of the current. To be more specific, let
us consider the longitudinal sector and use the form (\ref{SchL}) of the
respective EOM. The third term in this equation involves
 \beq
 \omega^2-k^2f(\chi)\,=\,\omega^2-k^2+k^2\,\frac{\chi^4}{\chi_0^4}\,=\,
 \mp Q^2 + \big(\pi^2 k T^2 \chi^2)^2\,,\eeq
(as usual, the upper/lower sign in front of $Q^2$ corresponds to a
space--like/time--like current, respectively), where the last term
$\propto k^2T^4$ becomes comparable with $Q^2$ for any $\chi$ greater
than a `critical' value $\chi_{\rm cr}= \chi_0\sqrt{Q/k}$. Note that, in
the high--energy of interest here ($k\gg Q$), this value $\chi_{\rm cr}$
is much smaller than $\chi_0$ and in fact it can be arbitrarily small.
Hence this piece of the gravitational interactions --- which describes
the Newton potential created by the BH (or one graviton exchange) --- can
be important even far away from the horizon, including in the vicinity of
the boundary. This is the piece of the interaction that we must keep. But
all the other factors of $f$ appearing in \eqnum{SchL} can be safely
replaced by 1 so long as we restrict ourselves to $\chi\ll\chi_0$, which
we shall do indeed in what follows. Then the respective equation of
motion (including time--dependence) takes indeed the form of a
Schr\"odinger equation, as anticipated at the end of Sect. 3:
  \beq
   i\frac{\del  \psi}{\del t}\,=\,
  -\frac{1}{2k}\,
 \frac{\del^2 \psi}{\del \chi^2}\,+\,V(\chi)\psi\,,\qquad
  V(\chi)\,=\,
 -\,\frac{1}{8k\chi^2}\,\pm\,
 \frac{Q^2}{2k}\,-\,\frac{k}{2}\,\frac{\chi^4}{\chi_0^4}\,. \label{SchT}
 \eeq
The respective roles of the three pieces in the potential should be clear
by now: {\em (a)} the first piece ($V_A$), which is independent of both
the virtuality and the temperature, describes the diffusive penetration
of the wave at early times (or, in the dual gauge theory, the diffusive
growth of the partonic fluctuation of the current in transverse
directions); {\em (b)} the second one ($V_B$), which is flat and
proportional to $Q^2$, is the potential barrier which prevents a
space--like current to decay into the vacuum, and {\em (c)} the third
piece ($V_C$) is the one--graviton exchange interaction between the
current and the BH. We shall latter attempt to provide a physical
interpretation for this last piece on the gauge theory side. The balance
between these three pieces depends upon the energy $\omega\simeq k$ and
the virtuality $Q^2$ of the current, and upon the temperature. There are
two important physical regimes, a {\em low energy} one and a {\em high
energy} one, which for the space--like current are illustrated in
Fig.~\ref{fig:Vsl}.  The transition between these two regimes occurs at
an energy $\omega\sim Q^3/T^2$, as we now explain:

\FIGURE[t]{
\centerline{\includegraphics[width=0.45\textwidth]{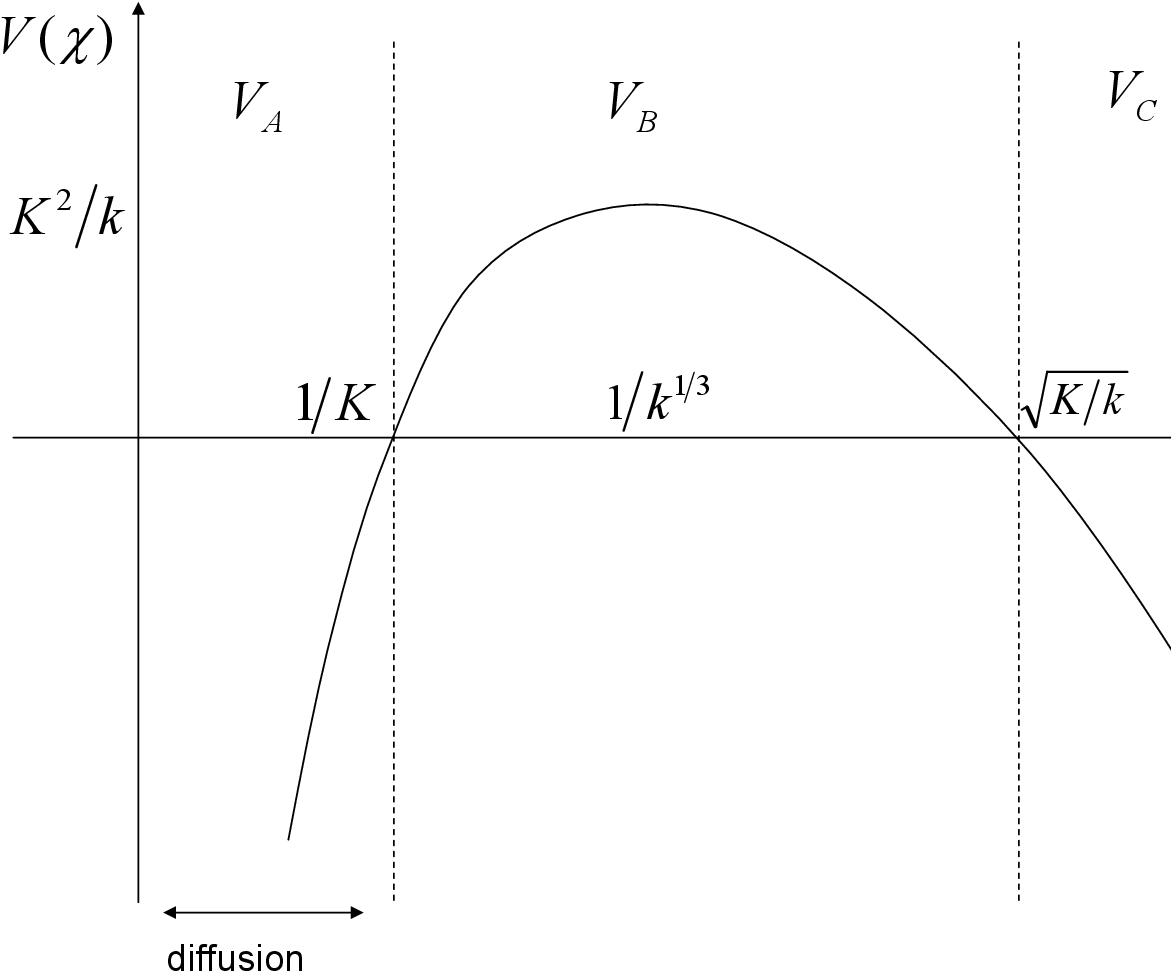}
\qquad
\includegraphics[width=0.45\textwidth]{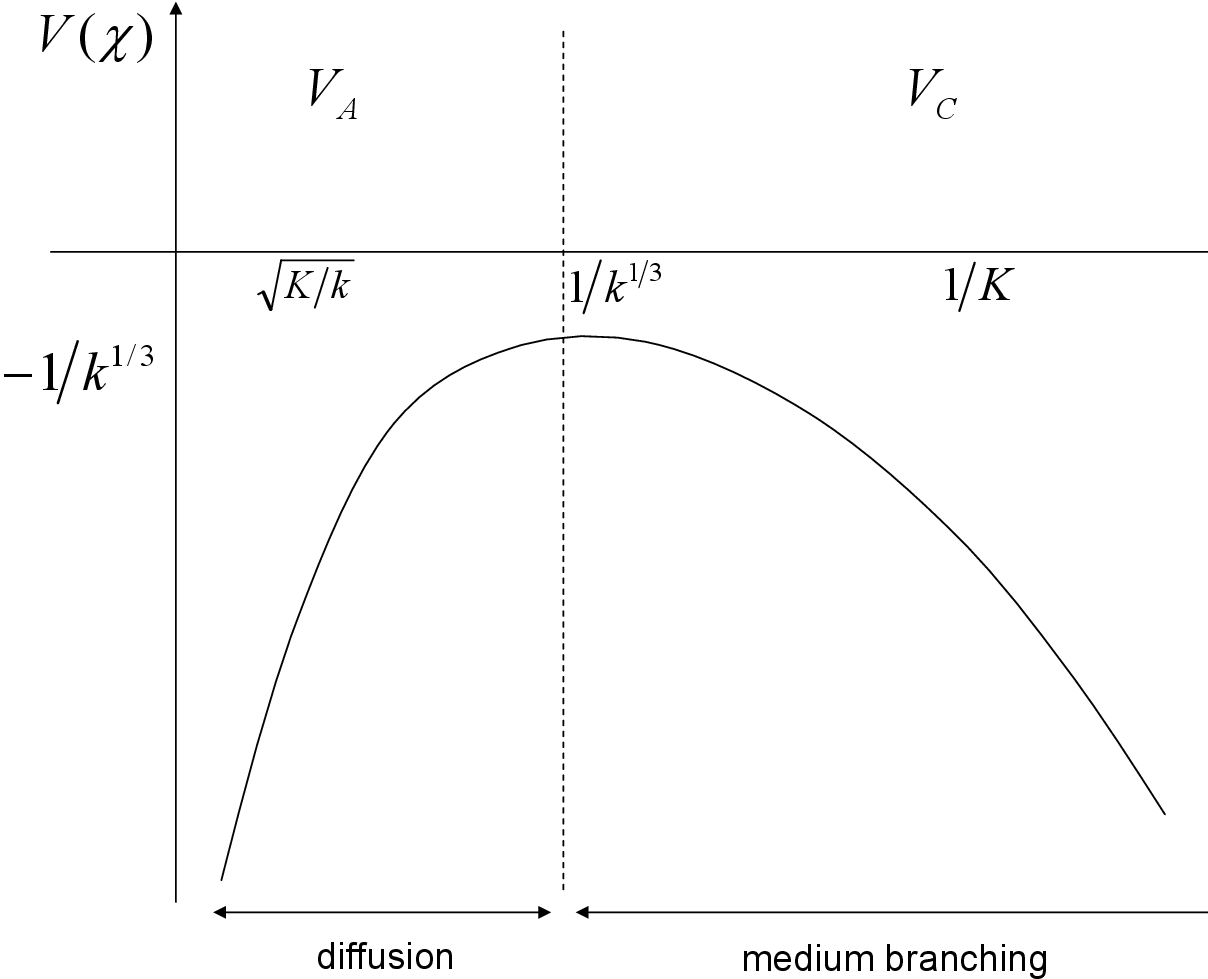}}
\caption{\sl The potential $V(\chi)$ in \eqnum{SchT} in the space--like
case (upper sign in front of $Q^2$). Left: low--energy case ($\omega\ll
Q^3/T^2$). Right: high--energy case ($\omega\gtrsim Q^3/T^2$). Note that,
in the high--energy case the potential looks qualitatively similar for a
time--like current as well. (In these figures, $Q$ is denoted as $K$, and
all variables have been made dimensionless by multiplying with
appropriate powers of $T$.) \label{fig:Vsl}} }

\texttt{(i)} {\bf Low energy:} $\omega\ll \omega_s\equiv Q^3/T^2$
(see Fig.~\ref{fig:Vsl} left)\\
So long as the energy is relatively low (with $\omega\gg Q^2/T\gg Q$
though), the potential shows a barrier corresponding to energy--momentum
conservation, so like in the vacuum (compare to Fig.~\ref{fig:Vvac}
left). However, and unlike in the vacuum, this barrier has now only a
finite width --- it extends over the interval $1/Q\lesssim \chi \lesssim
\chi_{\rm cr}$, with $\chi_{\rm cr}= \chi_0\sqrt{Q/k}$ ---, and for
larger $\chi\gg \chi_{\rm cr}$ we have $V\simeq V_C$ which describes
attraction by the BH. Hence, there is a small, but non--vanishing,
probability for the wave to penetrate through the barrier via tunnel
effect, and once that this happens the wave will fall into the BH. This
tunnel effect will provide an exponentially small contribution to the
imaginary part of the polarization tensor\footnote{The corresponding real
part remains the same as in the vacuum up to exponentially small terms.},
and hence to the structure functions, which can be estimated in the WKB
approximation as (parametrically) \cite{HIM2}
  \beq
    F_{2}\,\sim\,xF_1\,\sim\,
    x{N^2_cQ^2}D\,,\qquad  D\,\sim \,
     \exp\Big\{-c (\omega_s/\omega)^{1/2}\Big\}\,=\,
    \exp\left\{-c\,\frac{Q}{\sqrt{\gamma}\, T}\right\}\,,
   \label{Ftunnel} \eeq
where $c$ is some undetermined numerical coefficient and we have used
$\gamma=\omega/Q$. Interestingly, the exponential attenuation factor $D$
which is generated through tunneling looks formally like a Boltzmann
thermal factor $\exp(-Q/T_{\rm eff})$ with an effective temperature
$T_{\rm eff}=\sqrt{\gamma}\, T$. One can understand $T_{\rm eff}$ as the
temperature of the plasma in a boosted frame in which the current has
zero longitudinal momentum (and hence the plasma has a large global
velocity $\omega/k\simeq 1$): indeed, the energy density of the plasma,
which in the plasma rest frame scales like\footnote{Its precise value in
this strong coupling limit can be deduced from \eqnum{pstrong} as
$\mathcal{E}=3p=(3\pi^2/8) N_c^2 T^4$.} $\mathcal{E}\equiv T_{00}\sim
N_c^2 T^4$, becomes $\mathcal{E}'=\gamma^2\mathcal{E}$ in the boosted
frame; this is the same energy density as for a plasma at rest but with
an effective temperature $\sqrt{\gamma}\, T$.

We conclude that the low--energy space--like current can decay inside the
plasma, albeit very slowly. We shall later interpret this decay as pair
production induced by a uniform background force --- that is, a kind of
Schwinger mechanism.

\texttt{(ii)} {\bf High energy:} $\omega\gg \omega_s\equiv Q^3/T^2$ (see
Fig.~\ref{fig:Vsl} right)\\
With increasing energy at fixed $T$ and $Q^2$, $\chi_{\rm cr}$ becomes
smaller and smaller, so the barrier becomes narrower and it eventually
disappears: this happens when $\chi_{\rm cr}\sim 1/Q$, or $\omega\sim
\omega_s$. For even higher energies we are in the situation illustrated
in Fig.~\ref{fig:Vsl} right, where the wave can move all the way up to
the horizon, where it is ultimately absorbed with probability one. From
the point of view of DIS, this situation corresponds to the unitarity, or
`black disk', limit (the strongest possible scattering).

In this high energy regime, the virtuality--dependent term $V_B$ in the
potential is comparatively small at any $\chi$ and thus can be neglected.
We conclude that, for such a high energy, the dynamics is in fact the
same for both space--like and time--like currents; and, of course, it
would be the same also for a light--like current ($Q^2=0$) with high
energy $\omega \gg T$. Then $V(\chi)\simeq V_A+V_C$ has a maximum at
$\chi=\chi_s$ with
 \beq \chi_s\,\sim\,\frac{1}{T}\,\left(\frac{T}{\omega}\right)^{1/3}
 =\,\frac{1}{Q_s}\,,\eeq
which is far away from the horizon: $\chi_s\ll 1/T\sim \chi_0$. Above, we
have introduced the {\em plasma saturation momentum}
 \beq\label{QsatT}
 Q_s(\omega, T)\,\sim\, ({\omega} T^2)^{1/3}\,,\qquad
 \mbox{or}\qquad Q_s(x, T)\,\sim\,\frac{T}{x}\,,\eeq
which is the virtuality which separates between the (almost)
no--scattering regime at $Q \gg Q_s$ and the strong scattering regime at
$Q\lesssim Q_s$. In other terms, the strong--scattering condition
$\omega\sim Q^3/T^2$ can be solved either for $\omega$, thus yielding
$\omega\sim \omega_s$, or for $Q$, which gives $Q\sim Q_s$. We shall
later argue that, also in this context at strong coupling, the scale
$Q_s$ is associated with the phenomenon of parton saturation, so like in
QCD at weak coupling (cf. Sect. 2).

\FIGURE[t]{
\centerline{\includegraphics[width=0.5\textwidth]{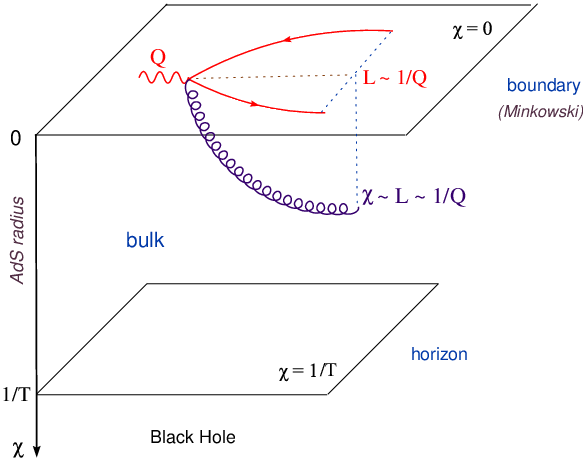} 
\includegraphics[width=0.5\textwidth]{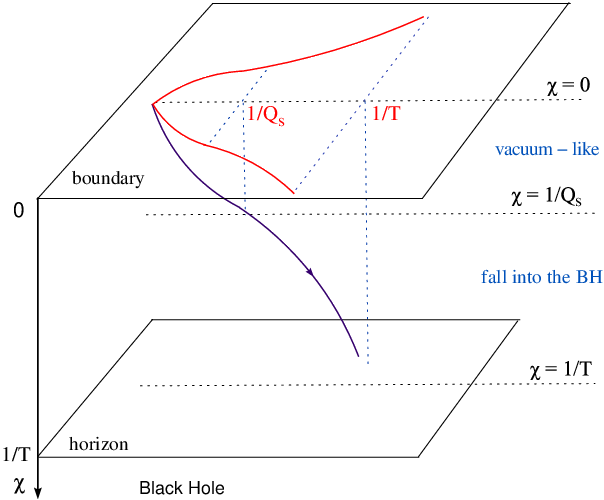}}
\caption{\sl  Space--like current in the plasma: the trajectory of the
wave packet in $AdS_5$ and its `shadow' on the boundary. Left: the
(relatively) low energy case ---  the Maxwell wave gets stuck near the
boundary up to tunnel effect. Right: the high energy case --- the wave
has an accelerated fall into the BH. \label{fig:Maxsl}} }

The high--energy dynamics thus proceeds as follows (for either
space--like or time--like current; see also Fig.~\ref{fig:Maxsl} right):
Starting at $t=0$ with a wave--packet localized near the boundary
($\chi=0$), this will slowly diffuse inside the bulk, so like in the
vacuum (cf. \eqnum{dif}), up to a distance $\chi\sim \chi_s$ where it
starts feeling the BH. This takes a time $t_s$ determined as
 \beq\label{tsat}
 \chi(t)\,\sim\, \sqrt{\frac{{t}}{\omega}}\quad{\rm \&}\quad
 \chi(t_s)\,\sim\,\chi_s\quad\Longrightarrow\quad
 t_s\,\sim\,\frac{\omega}{Q_s^2}.\eeq
Then, the wave falls towards the BH following an accelerated trajectory
which, interestingly, brings the wave--packet in the vicinity of the
horizon ($\chi(t)\sim \chi_0$) at a time $t_f$ which is parametrically of
the same order as $t_s$. This can be understood as follows: in the
semi--classical, WKB, approximation,  the center of the wave--packet
moves in the same way as a classical particle with mass $k$ in the
potential $V\simeq V_c$ and with zero total (kinetic plus potential)
energy. The last condition reads
 \beq\label{part}
 T\,+\,V_c\,=\,\frac{k\dot\chi^2}{2}
 \,-\,\frac{k}{2}\,\frac{\chi^4}{\chi_0^4}\,=\,0
 \quad\Longrightarrow\quad \frac{\dif \chi}{\dif t}
 \,=\,\frac{\chi^2}{\chi_0^2}\,,
 \eeq
which is easily integrated starting at time $t=t_s$ to yield
 \beq\label{chifall}
  \chi(t)\,=\,\frac{\chi_s}{1-
 \frac{\chi_s}{\chi_0^2}(t-t_s)}\,.\eeq
This $\chi(t)$ approaches $\chi_0\gg\chi_s$ when the denominator is
almost vanishing, which implies
 \beq\label{tf}
 t_f-t_s\,\simeq\,\frac{\chi_0^2}{\chi_s}\,\sim\,\frac{Q_s}{T^2}
 \,\sim\,\frac{\omega}{Q_s^2}\,\sim\,t_s\,.
 \eeq
Thus, as anticipated, the total fall time (defined as the time after
which the wave packet arrives in the vicinity of the horizon) reads,
parametrically,
 \beq\label{tfall}
 t_s\,\sim\,
 \frac{\omega}{Q_s^2}\,\sim\,\frac{1}{T}\,
 \left(\frac{\omega}{T}\right)^{1/3}
 \,.\eeq
From the perspective of the dual gauge theory, this time $t_s$ is the
{\em lifetime} of the high energy current before being absorbed by the
plasma. Since, moreover, the current propagates essentially at the speed
of light (at least, before it starts to feel the plasma), $t_s$ also
gives also the {\em penetration length} for the high energy current,
i.e., longitudinal distance $\Delta z$ traveled by the current before
disappearing in the plasma. As shown by the above estimate, $\Delta z$
scales like ${\omega}^{1/3}$, which is also the law found for a falling
open string (the dual of a `massless gluon') in Refs.
\cite{GubserGluon,Chesler3}. This similarity points towards the
universality of the mechanism for energy loss in the strongly coupled
plasma, that we shall describe in Sect. 5.3.

To compute the plasma structure functions in this high--energy regime, it
is enough to consider the time--independent version of the `Schr\"odinger
equation' (\ref{SchT}) with the simplified potential $V=V_A+V_C$
(together with a similar equation for the transverse waves \cite{HIM2}).
The details of the geometry near the BH horizon are again irrelevant,
since the outgoing--wave boundary condition can be enforced already at
relatively small distances $\chi\ll\chi_0$, namely at any $\chi \gg
\chi_s$. (Recall that $\chi_s= 1/Q_s\ll \chi_0$.) Then the classical
solutions are fixed at all smaller values of $\chi$ and, in particular,
near the Minkowski boundary. One thus obtains the following parametric
estimates\footnote{A similar result was found in Ref.
\cite{CaronHuot:2006te} in a study of real photon production in the
strongly coupled plasma, where the equation corresponding to the
zero--virtuality case $Q^2$ has been solved exactly.}  for $F_{1,2}$
\cite{HIM2}
 \beq\label{Flow}
    F_{2}\,\sim\,xF_1\,\sim\,
    x{N^2_cQ^2}\left(\frac{T}{xQ}\right)^{2/3}\quad
    \mbox{for}\quad Q\,\lesssim\,Q_s(x,T)\,=\,\frac{T}{x}\,.\eeq
A physical interpretation for this result will be presented in Sect. 5.4.

Note finally that the lifetime (\ref{tfall}) of the high--energy current
is formally the same as the coherence time for a current with virtuality
equal to $Q_s$ (and not to $Q$ !). Since $Q_s\gg Q$ in this regime, it is
clear that $t_s\ll \Delta t_{\rm coh}(Q)$: that is, the current
disappears in the plasma before having the time to develop a normal
partonic fluctuation with size $L\sim 1/Q$, as it would do in the vacuum.
This has interesting consequences for the survival of a `meson' state in
the plasma:

A high--energy space--like current is the simplest device to create a
`meson', i.e., a partonic excitation which is overall color neutral but
has a non--zero color dipole moment. This is, of course, a virtual
excitation and not a truly bound state, but its lifetime $\Delta t_{\rm
coh}\sim \omega/Q^2$ can be made arbitrarily large by increasing the
energy $\omega$ of the current (for a given transverse size $L\sim 1/Q$).
At least, this is the situation in the vacuum. But what about the
strongly--coupled plasma ? There, a similar situation holds too, but only
so long as the energy of the current is not {\em too} high: namely, when
$\omega\ll \omega_s \sim Q^3/T^2$, the `mesonic' fluctuation lives nearly
as long as in the vacuum, since its interactions with the plasma are
exponentially suppressed. But for higher energies $\omega\gtrsim
Q^3/T^2$, the current is absorbed already before having the time to
create a meson. This puts an upper limit on the `rapidity' $\gamma\equiv
\omega/Q$ of the meson\footnote{More precisely, the rapidity is the
quantity $\eta$ defined by $\cosh \eta\equiv\gamma$.} (with a given size
$L$) that can be created by a high energy process occurring within the
plasma (`limiting velocity') :
 \beq
 \gamma_{\rm max}\,\sim\,\frac{\omega_s}{Q}\,\sim\,\frac{Q^2}{T^2}
 \,\sim\,\frac{1}{(LT)^2}\,,
 \eeq
or, alternatively, an upper limit on its transverse size for a given
value of $\gamma$ (`screening length') :
 \beq
 L_{\rm max}\,\sim\,\frac{1}{Q_s}\,\sim\,\frac{Q^2_s}{\omega T^2}
 \,\sim\,\frac{1}{\gamma L_{\rm max} T^2}\quad
 \Longrightarrow\quad L_{\rm max}
 \,\sim\,\frac{1}{\sqrt{\gamma}\,T}\,.
 \eeq
(Notice the emergence of the effective temperature $T_{\rm
eff}=\sqrt{\gamma}\, T$.) Similar limits have been found in a different
approach
\cite{Peeters:2006iu,Liu:2006nn,Chernicoff:2006hi,Caceres:2006ta,Ejaz:2007hg},
in which the `meson' is viewed as a quark--antiquark pair (with heavy
quarks), whose string dual is an open string with endpoints attached to a
D7--brane embedded in the $AdS_5$--BH geometry. This similarity between
seemingly different physical problems and approaches ---
$\mathcal{R}$--current vs. open string, heavy quarks vs. massless quanta
of ${\mathcal N}=4$ SYM --- points, once again, towards an universal
mechanism for energy loss at strong coupling. We shall present our
conjecture \cite{HIM3} for this mechanism in Sect. 5.3, after the
discussion of the time--like current in the plasma. Before concluding,
let us also mention a difference between our results and those based on
the open string picture for the meson: in the low--energy/small--size
regime where the meson can form in the plasma, our approach predicts that
the meson can decay, albeit very slowly, via tunneling (the corresponding
width is exponentially small, cf. \eqnum{Ftunnel}), whereas in the
approach of Refs. \cite{Peeters:2006iu,Liu:2006nn,Chernicoff:2006hi} one
finds that the width is strictly zero (the lifetime of the meson is
infinite). Very recently, finite--width effects have been added to the
string picture in Ref. \cite{Faulkner:2008qk}, as string worldsheet
instantons; it would be interesting to clarify the relation between these
new results and those in \eqnum{Ftunnel}.

\subsection{Time--like current: $e^+e^-$ annihilation in a strongly
coupled plasma}

The evolution of a time--like current in the plasma should in principle
teach us about the behaviour of nearly on--shell partonic jets which are
produced by a high--energy process, so like $e^+e^-$ annihilation, taking
place within the plasma. From the previous discussion, we know already
that, at strong coupling, the situation is in fact more subtle. First,
even in the vacuum, the partons created by the decay of the time--like
current are far from being on--shell, at least in the early stages of the
branching process. Second, if the energy $\omega$ is high enough, such
that $\omega/Q\gtrsim (Q/T)^2$, then the current disappears so fast into
the plasma that it cannot even create the kind of partonic fluctuation
that it would develop into the vacuum. In other terms, the virtual
partons that the current fluctuates into have even larger virtualities,
of order $Q_s \gg Q$.

This last case, that of a highly--energetic current, has been already
covered in the previous subsection: indeed, the respective dynamics is
insensitive to the virtuality $Q^2$, and hence it is the same for
time--like, space--like, or even light--like, currents. Before we propose
a physical interpretation for this dynamics, let us first consider the
only remaining case, that of a {\bf time--like current at relatively low
energy}:  $\omega\ll \omega_s\equiv Q^3/T^2$ (with $\omega\gg Q^2/T\gg Q$
though).

\FIGURE[t]{ \centerline{
\includegraphics[width=0.46\textwidth]{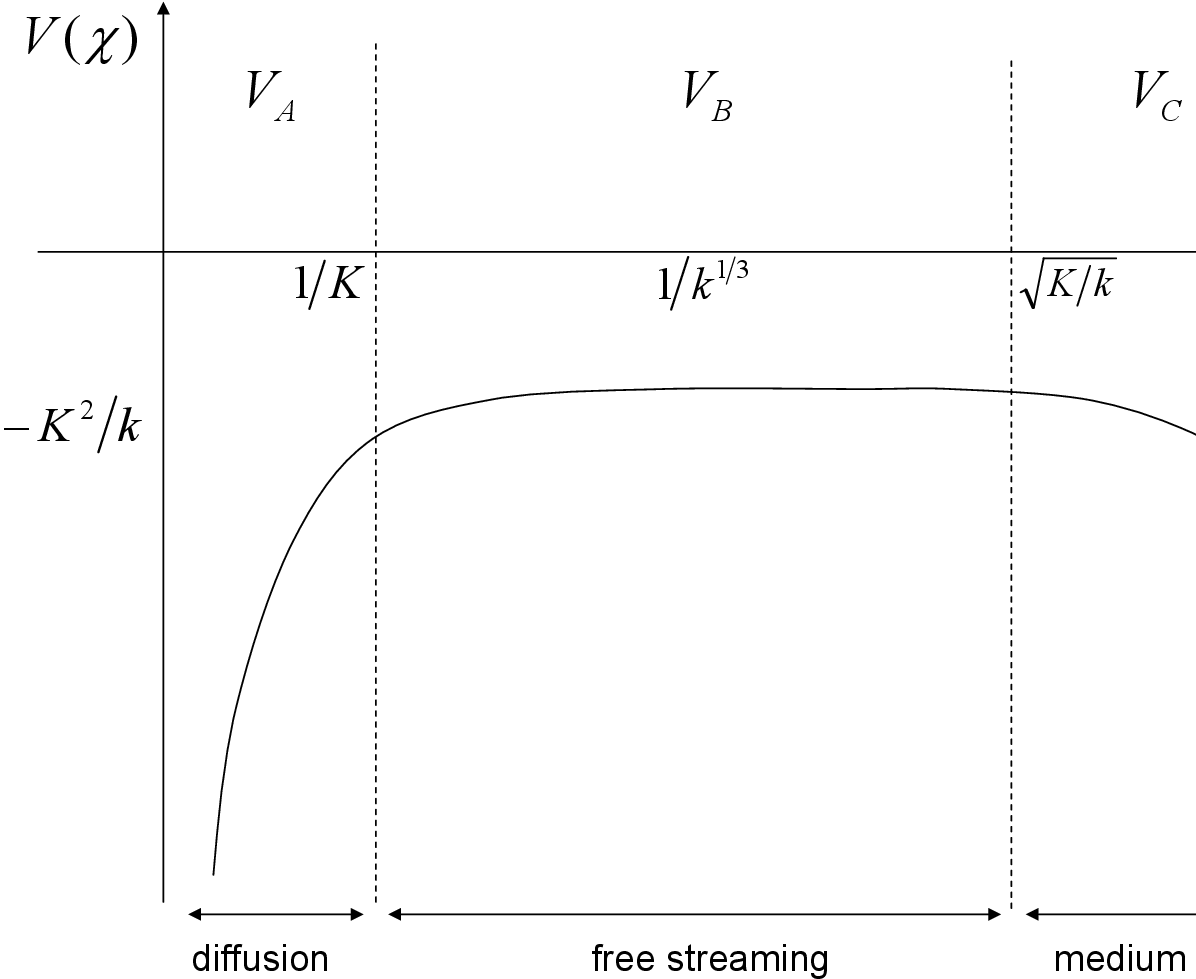}\qquad\qquad
\qquad
\includegraphics[width=0.5\textwidth]{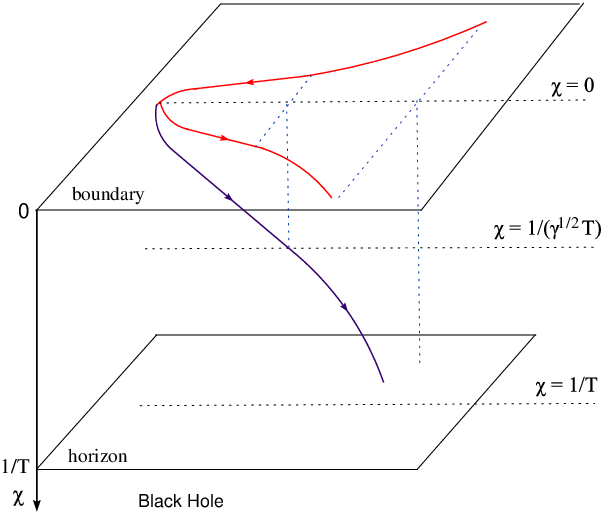}}
\caption{\sl Left: the potential $V(\chi)$ in \eqnum{SchT} in the
time--like case (lower sign in front of $Q^2$). Right: the trajectory of
the Maxwell wave--packet in $AdS_5$ and its `shadow' on the Minkowski
boundary. In the left figure one uses dimensionless variables together
with the notation $K\equiv Q$; e.g. $\chi\sim \sqrt{K/k}$ in the left
figure is the same as $\chi\sim 1/(\gamma^{1/2}T)$ in the right figure.}
\label{fig:Vtl}}

The respective potential, as obtained by taking the lower sign in front
of $Q^2$ in \eqnum{SchT}, is displayed in Fig.~\ref{fig:Vtl} left, which
should be compared to the respective potential in the vacuum, cf.
Fig.~\ref{fig:Vvac} right, and also to the space--like potential in
Fig.~\ref{fig:Vsl} left. Similarly to the space--like case, there is a
critical radial distance $\chi_{\rm cr}= \chi_0\sqrt{Q/\omega}$ below
which the Maxwell wave does not feel the plasma. However, unlike in tat
case, now there is no potential barrier anymore, so even for a relatively
low energy the time--like wave can propagate up to this critical distance
and then start its fall into the BH. Note that one can also write
$\chi_{\rm cr}= \chi_0/\sqrt{\gamma}$ since $\gamma= \omega/Q$.

For $\chi \ll \chi_{\rm cr}$, meaning at early times, the dynamics is the
same as for a time--like current in the vacuum (cf. Sect. 4) :
\texttt{(i)} The Maxwell wave--packet first diffuses inside the bulk up
to a distance $\chi_1\sim 1/Q$; this takes a time $t_1\sim \omega/{Q^2}$
(the coherence time for the virtual photon). \texttt{(ii)} Then, the
potential becomes flat, so the wave--packet propagates at constant radial
speed $v_\chi=Q/\omega$ up to a distance $\chi_{\rm cr}\gg 1/Q$; this
takes an additional time
  \beq\label{tg}
 t_c-t_1\,\simeq\,\frac{\chi_{\rm cr}}{v_\chi}
  \,\sim\,\frac{1}{T}\,
 \sqrt{\frac{\omega}{Q}}\ \gg\ t_1\,\sim\, \frac{\omega}{Q^2}\,,
 \eeq
which is much larger than $t_1$; therefore, $t_c\simeq t_c-t_1\sim
 \sqrt{\gamma}/T$. \texttt{(iii)} For $\chi \gtrsim \chi_{\rm cr}$,
$V\simeq V_c$ and the wave--packet falls towards the horizon according to
the same law as for the high--energy regime discussed in the previous
subsection, cf. Eqs.~(\ref{part})--(\ref{chifall}). The only difference
is that, now, this fall begins at a different time ($t_c$ instead of
$t_s$) and at a different radial location ($\chi_{\rm cr}$ instead of
$\chi_s$). Hence the trajectory of the center of the wave--packet now
reads (compare to \eqnum{chifall})
 \beq\label{chifalltl}
  \chi(t)\,=\,\frac{\chi_{\rm cr}}{1-
 \frac{\chi_{\rm cr}}{\chi_0^2}(t-t_c)}\,,\eeq
so that the traveling time $t_f-t_c$ down to the vicinity of the horizon
is now of order $t_c$ :
 \beq
 t_f-t_c\,\simeq\,\frac{\chi_0^2}{\chi_{\rm cr}}\,\sim\,\frac{1}{T}\,
 \sqrt{\frac{\omega}{Q}}\,\sim\,t_c\,.
 \eeq

\eqnum{part} shows that the radial velocity $v_\chi\equiv {\dif
\chi}/{\dif t}$ increases with time, due to the attraction exerted by the
BH, and becomes of $\order{1}$ when $\chi\sim\chi_0$. One can similarly
show that, at times $t> t_c$, the motion of the wave--packet along the
$z$ axis is decelerated according to\footnote{Recall that for $t < t_c$,
$v_z(t)=k/\omega\simeq 1$ since we always assume $\omega\simeq k\gg Q$.}
 \beq\label{vzfall}
 v_z(t)\,\equiv \,\frac{\dif z}{\dif t}\,\simeq\,1\,-\,
 \frac{\chi^4}{\chi_0^4}\,,
 \eeq
and thus it approaches to zero when the wave--packet approaches the
horizon. Note that, within the present approximations (which are valid so
long as $\chi\ll\chi_0$), the total velocity remains luminal,
$v_z^2+v_\chi^2=1$, even during the phase of fall towards the BH. This is
suggestive of a partonic interpretation in terms of massless partons
which are eventually stopped into the plasma. We shall elaborate on this
interpretation in Sect.~5.3.

These results also show that the wave--packet falls into the BH along a
light--like geodesic. More precisely, the trajectory of the light--like
geodesic in the $AdS_5$--BH geometry reads (see, e.g.,
\cite{GubserGluon,Chesler3}) :
 \beq\label{geodesic} \frac{\dif z}{\dif t}\,=\,v f(\chi(t))\,,\qquad
 \frac{\dif \chi}{\dif t}\,=\,f\sqrt{1-v^2f}\,, \eeq
with $f(\chi)$ as defined in \eqnum{def} and $v\le 1$ the longitudinal
velocity near the boundary. We have indeed: $(\dif z)^2 +
(1/f)(\dif\chi)^2 = f(\dif t)^2$, as it should for a light--like geodesic
in this particular geometry. When $v\simeq 1$ and $\chi\ll \chi_0$, the
equations (\ref{geodesic}) are fully consistent with our previous results
for the propagation of the wave--packet at times $t>t_1$ (i.e., after the
early diffusion). For instance the second equation (\ref{geodesic})
implies (with $\gamma\equiv 1/\sqrt{1-v^2}\gg 1$)
 \beq
 \frac{\dif \chi}{\dif t}\,\simeq\,\sqrt{1-v^2+ v^2\,
 \frac{\chi^4}{\chi_0^4}}\,\simeq\,\frac{\chi^2}{\chi_0^2}\qquad
 \mbox{for}\qquad \frac{1}{\sqrt{\gamma}}\,\ll\,
 \frac{\chi}{\chi_0}\,\ll\,1\,,
 \eeq
which is the same as \eqnum{part} and holds within the same range of
values for $\chi$ as the latter (in this time--like case). On the other
hand, for ${\chi}/{\chi_0}\ll 1/{\sqrt{\gamma}}$, one finds ${\dif
\chi}/{\dif t}\simeq \sqrt{1-v^2}$ and ${\dif z}/{\dif t}\simeq v$, in
agreement with \eqnum{groupvac} (and the discussion after it). This
agreement can be related to the fact that, for $t>t_1$, the solution to
the effective `Schr\"odinger equation' (\ref{SchT}) is well reproduced by
the WKB approximation.

To summarize, a time--like current with relatively low energy disappears
into the plasma after a time of order $t_c$, which scales with the energy
like $\omega^{1/2}$ (rather than $\omega^{1/3}$ for the high--energy
current; compare to \eqnum{tfall}). This lifetime yields also the
penetration length in the longitudinal direction (since the longitudinal
velocity is $v_z\simeq 1$ at least during the free--streaming part of the
dynamics) : $\Delta z= t_c\sim \sqrt{\gamma}/T$.

The polarization tensor for the time--like current in this `low--energy'
regime is essentially the same as in the vacuum, cf. \eqnum{Pivac} with
$q^2<0$, since this is determined by the classical solution in the region
of small $\chi \ll \chi_{\rm cr}$. In particular, the rate for the
dissipation of the current, as given by Im $\Pi_{\mu\nu}(q)$, is the same
as in the vacuum: this simply tells that the current disappears via
branching into the partons of ${\mathcal N}=4$ SYM, and this branching
proceeds in its early stages in the same way as it does in the vacuum (as
it should be obvious from the previous discussion). Of course, the
late--time evolution of the partons will be different at finite
temperature as compared to the zero temperature case, but the inclusive
cross--section for the decay of the current is insensitive to this
late--time evolution, and also to the details of the final state. The
situation is more interesting in that respect in the high--energy regime
(cf. Sect. 5.1), since there the branching process is affected by the
temperature already in its early stages, thus yielding
temperature--dependent decay rates. (These can be obtained from the
structure functions (\ref{Flow}), via the relations (\ref{F12T}); one
finds, e.g., Im $\Pi_1\sim N_c^2 Q_s^2(\omega,T)\,$.) A physical picture
for the plasma effects in the branching process will be presented in the
next subsection.

\subsection{Physical interpretation: Medium--induced parton branching}

Now, that we have presented the AdS/CFT results for an
$\mathcal{R}$--current in the plasma in all the interesting kinematical
regimes, it is important to try and understand the physical meaning of
these results in the original gauge theory. To that aim, we shall heavily
rely on the IR/UV correspondence (cf. Sect. 4.3) together with the
previously developed physical picture for the evolution of the current in
the vacuum (cf. Sect. 4.4).

The new dynamics that we have to understand is the fall of the Maxwell
wave--packet towards the BH horizon. Let us first `translate' the
corresponding laws via the IR/UV correspondence: after identifying
$\chi(t)$ with the inverse $1/Q(t)$ of the virtuality of the evolving
partonic system, the equation of motion (\ref{part}) for the center of
the wave packet  can be rewritten as
 \beq\label{force}
 \frac{\dif Q(t)}{\dif t}\,\sim \,-T^2\,.
 \eeq
The l.h.s. of this equation is the rate for the change in the parton
transverse momentum, hence the r.h.s. should be interpreted as a {\em
transverse force}. This force $F_T\sim (-T^2)$
--- the simplest one that one can built with the unique scale $T$ offered
by the plasma in this strong coupling regime where the coupling
disappears from all formulae ! --- is uniform and independent of the
parton momentum\footnote{This is strictly true only so long as the plasma
has an infinite extent or, in any case, a longitudinal extent which is
much larger than the coherence length $\sim\omega/Q^2$ of the photon. For
a finite--size medium, the force becomes proportional to $1/x$, as it
will be argued in the Appendix \cite{Alpers}.}, and it acts towards
decreasing the parton virtuality. That is, it favors the parton evolution
towards lower virtualities, meaning that it speeds up the branching
process.

Another way to recognize this force within AdS/CFT is via the condition
that a space--like current has strong interactions with the plasma. In
the context of the supergravity calculation of Sect. 5.1, this was simply
the condition that the gravitational potential due to the BH, $V_C \sim
\omega(T\chi)^4$, when evaluated at the position $\chi\sim 1/Q$ of the
wave packet, be strong enough to balance the potential barrier $V_B\sim
Q^2/\omega$ expressing energy--momentum conservation:
 \beq
 \omega\big(T\chi \big)^4\bigg|_{\chi=1/Q}\ \sim \ \frac{Q^2}{\omega}
 \quad\Longrightarrow\quad Q \ \sim \ \frac{\omega}{Q^2}\,T^2\,.\eeq
In the last condition, the r.h.s. $({\omega}/{Q^2})\times T^2$ can be
recognized as the product between the coherence time $\Delta t_{\rm coh}
\sim {\omega}/{Q^2}$ of the current and the plasma force $F_T\sim T^2$.
This suggests the following interpretation: the interaction between the
current and the plasma becomes strong when the lifetime of the partonic
fluctuations of the current becomes large enough for the mechanical work
done by the plasma force on these partons to compensate their virtuality.
Once this happens, the partons can move away from each other and
eventually disappear into the plasma, so that the current decays.

This interpretation can be promoted into a qualitative and
semiquantitative physical picture for parton branching in the presence of
the strongly--coupled plasma. As we shall see, this picture is consistent
with all the results of the supergravity calculations in Sects. 6.1 and
6.2. This picture involves again a parton cascade like the one shown in
Fig.~\ref{fig:QDBRANCH} (where the `parton' which initiates the cascade
is chosen as the $\mathcal{R}$--photon), but the branching law is now
modified by plasma effects. We focus on the more interesting case at high
energy, $\omega\gg \omega_s\equiv Q^3/T^2$, where the plasma effects are
important even in the early stages. Recall that, in this regime, the
initial virtuality $Q^2$ of the current plays no dynamical role, so the
subsequent discussion applies equally well to space--like, time--like, or
even light--like, current. (In the latter case, we simply require
$\omega\gg T$.) Starting with a point--like current at $t=0$, this will
develop a partonic fluctuation which grows diffusively like
$L(t)\sim\sqrt{t/\omega}$, but at the same time feels the effects of the
plasma force, which reduces the partons virtuality at the rate shown in
\eqnum{force}. During this phase, the effective virtuality of the
partonic system is set by the uncertainty principle as $Q(t)\sim 1/L(t)$.
After some time $t_s$, the mechanical work $t_sT^2$ done by the plasma
force becomes of the order of the system virtuality at that time,
$1/Q(t_s)$, and then the system can further decay. The corresponding
values $t_s$ and $Q_s\equiv Q(t_s)$ are easily found as
  \beq\label{ts}
 t_s\,\sim\,\frac{1}{T}\,
 \left(\frac{\omega}{T}\right)^{1/3}\,\sim\,
 \frac{\omega}{Q_s^2}\,,\qquad  Q_s(\omega, T)\,=\, ({\omega} T^2)^{1/3}
 \,,\eeq
in agreement with \eqnum{tfall}. This first branching produces (in
general) two new partons, each of them roughly carrying half of the
energy of the original current: $\omega_1\simeq \omega/2$. Thus, the new
partons are themselves very energetic, so their intrinsic virtuality is
irrelevant for their subsequent evolution, so like for the original
photon. Therefore, they undergo an evolution similar to that in the
previous step, but at the lower energy $\omega_1$. This argument
generalizes to the $n$th step in the evolution, where $\omega_n\simeq
\omega/2^n$ : a parton from this generation, whose intrinsic virtuality
is still negligible (which is indeed the case so long as $\omega_n\gg
T$), grows up a partonic fluctuation whose effective virtuality $Q_n\sim
1/L_n$ is of the order of the mechanical work $\Delta t_n T^2$ done by
the plasma during the lifetime $\Delta t_n\sim \omega_n/Q_n^2$ of the
fluctuation. This condition implies
  \beq\label{Qn}
   Q_n\,\sim\,Q_s(\omega_n,T)\,=\, ({\omega_n} T^2)^{1/3}
 \,.\eeq
Note that the virtuality and lifetime of a given parton generation are
now dynamically established, via the action of the plasma force, and they
are independent of the intrinsic virtuality of the partons in the
previous generation (unlike what happens in the vacuum, where we have
seen that $Q_n\sim Q_{n-1}/2$, cf. Sect. 4.4). The process stops when
$Q_n$ and $\omega_n$ become both of order $T$, since by then the partonic
system has extended over a transverse distance $L_n\sim 1/T$ and hence
the partons originating from the current cannot be distinguished anymore
from the degrees of freedom of the plasma: they become a part of the
thermal bath.

To evaluate the overall lifetime of the cascade, we now study the
evolution of the virtuality $Q(t)$ and of the energy $\omega(t)$ of a
typical parton in the cascade. By integrating \eqnum{force} starting at
time $t=t_s$ (when $Q(t_s)=Q_s$) and using $\omega(t)\simeq Q^3(t)/T^2$,
one easily finds
  \beq\label{QET}
  Q(t)\,\simeq\,Q_s -T^2 (t-t_s)\,,\qquad \omega(t)\,\simeq\,
  \frac{1}{T^2}\,\big[Q_s -T^2 (t-t_s)\big]^3\,.\eeq
These quantities become simultaneously of order $T$ after a time $t_f$
such that (recall that $Q_s\gg T$)
 \beq
 t_f-t_s\,\simeq\,\frac{Q_s-T}{T^2}\,\simeq\,\frac{Q_s}{T^2}\,\sim\,t_s\,,\eeq
in agreement with the respective AdS/CFT result, \eqnum{tf}. \eqnum{QET}
should be compared to the corresponding equations in the vacuum, cf.
\eqnum{QEvac}: like in the vacuum, the energy of a typical parton
decreases with time because the total energy gets spread among an
increasing number of partons. So long as $Q(t)\gg T$, these partons can
be still distinguished from the thermal bath, and thus the energy
$\omega$ brought in by the virtual photon remains within the parton
cascade. This energy is transmitted to the plasma only in the last stages
of the branching process, i.e., in a relatively short lapse of time $\sim
1/T$. This may explain the final, explosive, burst of energy seen in
numerical simulations for the energy loss of a `light quark' (a null
string falling in the $AdS_5$ BH geometry) in Ref. \cite{Chesler3}.

It is finally interesting to study the stopping of the partons in the
plasma and, related to this, the shape of the parton cascade. As we shall
see, this study will provide an interesting connection to the `trailing
string' constructed in Refs. \cite{Herzog:2006gh,Gubser:2006bz}. Consider
the `rapidity' $\gamma_n=\omega_n/Q_n$ of the partons in the $n$th
generation; in continuous notations, this becomes
$\gamma(t)=\omega(t)/Q(t)$ and it decreases with time (unlike for a
branching process taking place in the vacuum, for which we have seen, in
Sect. 4.4, that $\gamma_n$ was constant along the cascade). This means
that the partons in each new generation move slower along the $z$
direction than their predecessors in the previous generations; this
deceleration continues until $t\sim t_f$, when $\gamma(t)$ decreases to a
value of $\order{1}$. If $z(t)$ denotes the longitudinal position of the
partons existing at time $t$, then the previous argument implies that
$z(t)< vt\approx t$ and, moreover, the separation $\zeta(t)\equiv
vt-z(t)$ is increasing with time. ($v\equiv k/\omega\approx 1$ is the
velocity of the incoming photon.) At this stage, it is convenient to
recall that $L(t)\sim 1/Q(t)$ represents the transverse size of the
partonic system at time $t$. If we eliminate the variable $t$ between the
functions $L(t)$ and $\zeta(t)$, then the resulting function $L(\zeta)$
describes the {\em enveloping curve of the partonic cascade}, i.e., the
curve which characterizes the shape of the parton distribution within the
cascade. To construct this function, we start with the longitudinal
velocity of the partons at time $t$ (recall that $\omega(t)\simeq
Q^3(t)/T^2$) :
 \beq\label{vzbranch}
 v_z(t)\,\equiv\,\frac{\rmd z}{\rmd t}\qquad\Longrightarrow\qquad
 1-v_z^2(t)\, =\frac{1}{\gamma^2(t)}
 \,=\,\frac{Q^2(t)}{\omega^2(t)}\,\sim\,(TL(t))^4
 \,.\eeq
Via the UV/IR correspondence $L\to\chi$, this result is consistent with
\eqnum{vzfall}, thus providing a consistency check for the proposed
physical interpretation. The difference $1-v_z^2(t)$ is small so long as
$L(t)\ll 1/T$, and it is parametrically of the same order as $v_\perp^2$,
where $v_\perp$ is the transverse velocity: $v_\perp \equiv {\rmd
L}/{\rmd t}\sim (TL(t))^2$. This is consistent with the fact that the
highly--energetic, massless, partons are nearly on--shell.
\eqnum{vzbranch} implies
 \beq
 z(t)- t \,=\,- \zeta(L(t))\qquad\mbox{with}\qquad
 \frac{\rmd \zeta}{\rmd t}\,
 \equiv  \,\frac{\rmd \zeta}{\rmd L} \, \frac{\rmd L}{\rmd t}
 \,\sim\,(TL(t))^4\,.\eeq
After also using ${\rmd L}/{\rmd t}\simeq (TL)^2$, cf. \eqnum{force}, we
finally deduce
 \beq\label{trail}
 \frac{\rmd \zeta}{\rmd L}\,\sim\,(TL(t))^2 \qquad\Longrightarrow\qquad
 \zeta(L)\,\sim\,T^2L^3\,.\eeq
This function $\zeta(L)$ represents the enveloping curve of the partonic
cascade in the regime where $L\ll 1/T$ (and hence $\zeta\ll 1/T$ as
well), and is illustrated in Fig.~\ref{Figtrail}. What is remarkable
about this curve is that it is `dual' --- via the standard replacement
$L\to\chi$ with $\chi=R^2/r$ (the radial coordinate on $AdS_5$) --- to
the `trailing string' solution constructed in Refs.
\cite{Herzog:2006gh,Gubser:2006bz}. The trailing string is the
supergravity dual of a heavy quark propagating at constant speed $v_z$
through the strongly--coupled $\mathcal{N}=4$ SYM plasma; roughly
speaking, this is the trajectory of the energy flow from the heavy quark
to the BH horizon. This string moves solidary with the heavy quark and it
is parameterized as $z(t,\chi)=v_zt-\zeta(\chi)$, where the function
$\zeta(\chi)$ describes the shape of the string in the comoving frame.
For $\chi\ll 1/T$, this function has the parametric form\footnote{The
restriction to $\chi\ll 1/T$ is necessary since our physical discussion
of parton branching is too qualitative to capture the dynamics of the
parton cascade at late times, where $L(t)\sim 1/T$. One should however
emphasize that the exact function $\zeta(\chi)$, as valid for any
$\chi\le\chi_0$, appears also in the context of the AdS/CFT calculation
for the  $\mathcal{R}$--current, as a line of stationary phase for the
Maxwell wave at $\chi\gg \chi_{\rm cr}$ \cite{HIM2}.} shown in
\eqnum{trail} where we identify $L\equiv \chi$. This strongly suggests
that the heavy quark looses energy to the plasma via the same mechanism
as the $\mathcal{R}$--current, that is, through parton branching: the
heavy quark radiates quanta (massless partons of $\mathcal{N}=4$ SYM),
which in turn radiate other such quanta, so that the energy originally
encoded in the heavy quark is progressively spread among many partons.
Then the piece of the trailing string located at radial distance $\chi$
is `dual' (via the UV/IR correspondence) to that part of the parton
distribution which has a transverse extent $L\sim\chi$; hence, the shape
$\zeta(\chi)$ of the string should be described by the same function as
the enveloping curve $\zeta(L)$, which is what we found indeed.

\FIGURE[t]{
\includegraphics[height=9.cm]{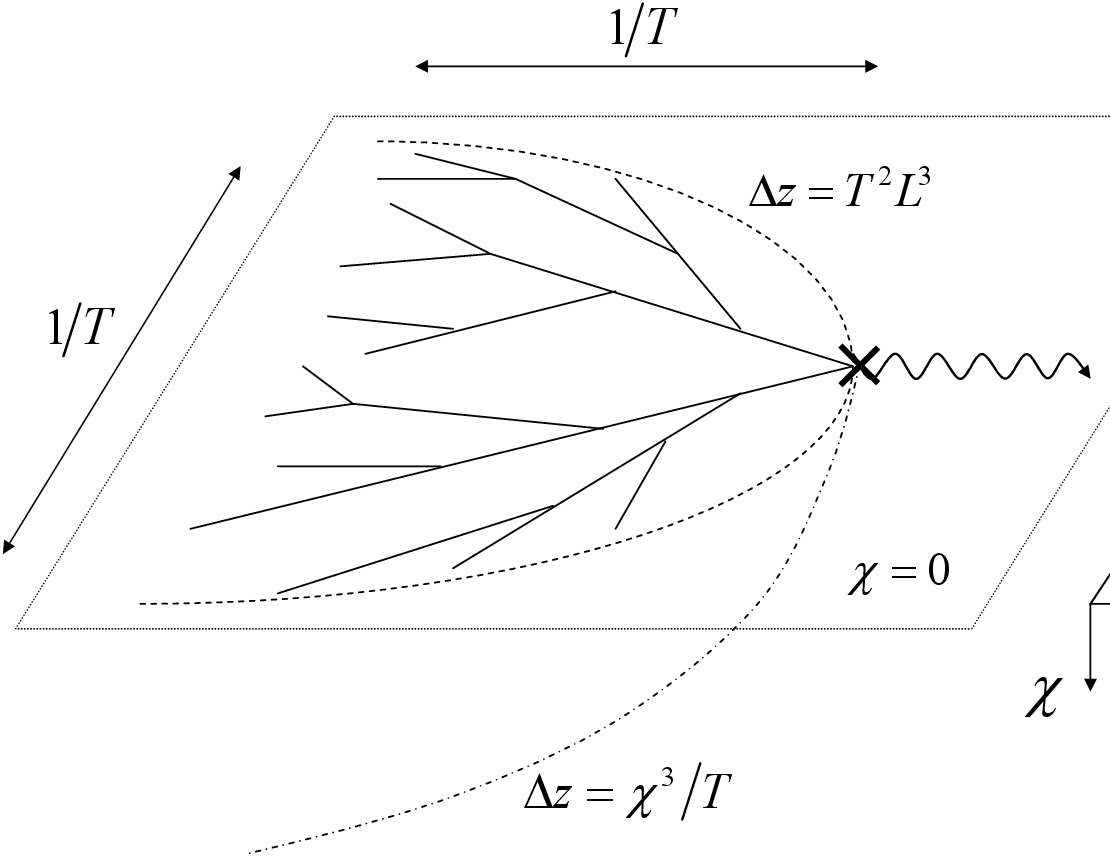}
\caption{\sl The parton cascade generated by the current via
medium--induced branching in the physical Minkowski space (represented
here as the boundary of $AdS_5$ at $\chi=0$) and the trailing string
attached to the leading photon (represented for $\chi\ll 1$). The latter
is `dual' to the enveloping curve of the former. \label{Figtrail} }}

As another check of this physical interpretation, let us compare the
`drag force' computed in Refs. \cite{Herzog:2006gh,Gubser:2006bz} --- the
force which is required to pull the heavy quark at constant speed through
the plasma --- to the rate of energy degradation for partons in our
parton cascade. (In the case of the heavy quark, this is also the rate at
which the heavy quark looses energy, since this quark can be traced
during the branching process, due to its conserved baryonic charge.)
Namely, in Refs. \cite{Herzog:2006gh,Gubser:2006bz} one found
 \beq\label{drag}
 -\,\frac{\dif E}{\dif t}\,=\,\frac{\pi}{2}
 \sqrt{\lambda}\,v_z^2 \,\gamma\, T^2
 \,,\eeq
with $v_z$ the (constant) velocity of the heavy quark and
$\gamma=1/\sqrt{1-v^2_z}$. On the other hand, for the branching process
described in this section we can write (this follows from \eqnum{force}
by using $\omega(t)\sim Q^3(t)/T^2$, or directly from \eqnum{QET})
 \beq\label{Fl} -\,
 \frac{\rmd \omega(t)}{\rmd
 t}\,\sim\,
 Q_s^2(\omega(t), T) \,\sim\,
 (\omega(t) T^2)^{2/3}
 \,\sim\,\gamma(t) T^2\,.
 \eeq
where $\gamma(t)=\omega(t)/Q(t)$ is now time-dependent, because we are
not in a stationary situation (there is no drag force). But except for
this time--dependence, Eqs.~(\ref{drag}) and (\ref{Fl}) are indeed
consistent with each other\footnote{Recall that our calculation applies
to a ultrarelativistic particle with $v_z\simeq 1$. Also the factor of
$\sqrt{\lambda}$ in \eqnum{drag} is the coupling between the heavy quark
and the quanta of $\mathcal{N}=4$ SYM; for the $\mathcal{R}$--current,
this coupling is rather the electric charge, that we have implicitly
chosen to be one.} : in both cases, the rate for energy loss is
proportional to $\gamma T^2$.

What happens with this energy after being transferred to the plasma,
i.e., at times $t > t_f$ (corresponding to large distances $L,\,\zeta >
1/T$) ? This question cannot be answered on the basis of the branching
picture alone, nor within our previous approximations for the
supergravity solution, which are valid only sufficiently far away from
the horizon. The proper answer to this question within AdS/CFT would
require a more precise solution, valid near the horizon, and also a study
of the backreaction of the Maxwell wave on the $AdS_5$ BH geometry
\cite{Skend02,Skenderis:2002wp}. Such a study has not been done so far
for the problem of the $\mathcal{R}$--current, but similar analyses for
the `heavy quark' problem \cite{Gubser:2007xz,Chesler:2007sv} show that
the transfer of energy from the hard probe to the plasma induces
collective motion in the latter, in the form of sound waves or Mach
cones.

\subsection{Physical interpretation: Parton saturation at strong
coupling}

We now return to the AdS/CFT results for a space--like current, cf. Sect.
5.1, and show that these can be naturally interpreted in terms of {\em
parton distributions in the strongly--coupled plasma.} From Sect. 2.2 we
recall that the structure function $F_2(x,Q^2)$ is a measure of the
hadron (here, plasma) constituents with longitudinal momentum fraction
$x$ and transverse momenta $k_\perp\lesssim Q$ (i.e., which occupy a
transverse area $\sim 1/Q^2$). Hence, by inspection of the corresponding
results in Sect. 5.1, one can immediately deduce that there are {\em no
partons} at sufficiently large values of $Q^2\gg Q^2_s(x)=(T/x)^2$ for a
given value of $x$, or, equivalently, at sufficiently large values of $x$
for a given $Q^2$. Indeed, the structure functions are exponentially
suppressed in this high--$Q^2$ (or `low--energy', or `large--$x$')
regime, as shown in \eqnum{Ftunnel} that one can rewrite as
  \beq\label{FhighQ} F_2/(x{N^2_cQ^2})\,\sim\,
\exp\big\{-c(Q/Q_s)^{1/2}\big\}\,= \,\exp\big\{-c(x/x_s)^{1/2}\big\}\quad
{\rm for}\quad x
> x_s\equiv\frac{T}{Q}\,.
 \eeq
The saturation line $Q_s(x)=T/x$ is shown as the straight line $\ln
Q_s^2(Y)=2Y$ in the kinematical plane for DIS, in Fig.~\ref{fig:SatADS}
left. As also indicated in that figure, there are no partons on the right
side of the saturation line: the respective structure functions are so
small that the scattering can be characterized as quasi--elastic. The
absence of partons from the wavefunction of a hadron at strong coupling
(and for relatively large $Q^2$) was anticipated by Polchinski and
Strassler \cite{Polchinski}, via the following argument based on the
operator product expansion (OPE) :

\FIGURE[t] {\centerline{
\includegraphics[width=0.52\textwidth]{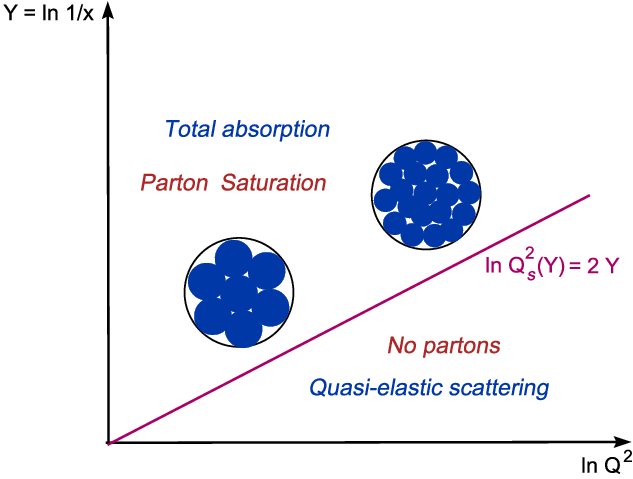}\quad
\qquad
\includegraphics[width=0.45\textwidth]{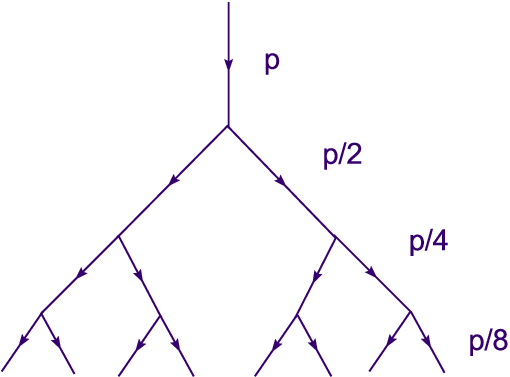}}
\caption{\sl Left: The `phase diagram' for DIS off a ${\mathcal N}=4$ SYM
plasma at high energy and strong coupling. Right: Parton cascades through
which the partons fall at small values of $x\lesssim x_s$.}
\label{fig:SatADS} }

At weak coupling, the parton picture for DIS is meaningful because the
OPE for the structure functions at high $Q^2$ is dominated by the
operators with {\em leading twist} --- i.e., those having the minimal
value for the difference $\tau_{j,n}\equiv \Delta_{j,n}-j$ between their
mass dimension $\Delta_{j,n}$ and their spin $j$ ($n$ is an operator
index) ---, which have a manifest interpretation in terms of quark and
gluon number densities (see, e.g., \cite{Peskin}). In the classical, or
zero--order, approximation, this value is $\tau_{\rm cl}=2$ for all the
`leading--twist' operators. But in general this classical value receives
quantum corrections in the form of the `anomalous dimensions'
$\gamma_{j,n}\equiv \Delta_{j,n}-d_{j,n}$ (with $d_{j,n}$ the respective
classical dimension). At weak coupling, these anomalous dimensions start
at $\order{g^2}$ and can be computed in perturbation theory; in
particular, for a theory with conformal symmetry, so like the ${\mathcal
N}=4$ SYM theory, and for large $N_c$, these are pure numbers
$\gamma_{j,n}\sim g^2N_c\equiv \lambda$ which turn out to be positive, or
zero in some exceptional cases. The `exceptional cases' refer to the
operators which are `protected' against quantum corrections by some
symmetry, so like the energy--momentum tensor $T^{\mu\nu}$ (for which
$j=2$ and $\gamma=0$). The fact that $\gamma_{j,n}$ is positive means
that the respective contribution to the OPE of $F_2(x,Q^2)$ decreases
with increasing $Q^2$, according to the power law
$(1/Q^2)^{\gamma_{j,n}/2}$. (In pQCD, due to asymptotic freedom, this
decrease is slower, as an inverse power of $\ln Q^2$ \cite{Peskin}.) But
at weak coupling, the exponent $\gamma_{j,n}/2\sim g^2N_c\ll 1$ is small,
so in spite of their positive anomalous dimensions the `leading--twist'
operators still dominate over those with a higher (classical) twist
$\tau\ge 3$.

The situation however changes when moving to strong coupling, as
Polchinski and Strassler have emphasized: there, the anomalous dimensions
for the leading--twist operators are very large, of
$\order{\lambda^{1/4}}$ \cite{gubser02}, and they are still positive
(whenever non--vanishing), so the respective contributions die away very
fast with increasing $Q^2$. As a consequence, the DIS structure functions
at high $Q^2$ and strong coupling are rather dominated by special {\em
higher--twist} operators which are protected by symmetries, and which in
general can be of two types: operators which describe the scattering off
the hadron {\em as a whole} \cite{Polchinski} (as opposed to the
scattering off its partonic constituents), and multiple insertions of the
protected leading--twist operator $T^{\mu\nu}$, which can be interpreted
as diffractive scattering \cite{HIM1}. (In the dual string theory, such
diffractive processes appear as multiple graviton exchanges which can be
resummed in the eikonal approximation \cite{cornalba,tan}.) These
higher--twist operators provide contributions to $F_2(x,Q^2)$ which at
high $Q^2$ fall off as large, but finite (i.e., independent of
$\lambda$), powers of $1/Q^2$.

Yet, for the strongly--coupled {\em plasma}, the structure functions in
\eqnum{FhighQ} show an even faster decrease at high $Q^2$ --- exponential
rather than power--like. This is so since these results correspond to the
strict limit $N_c\to\infty$, in which the multiple graviton exchanges
{\em with a single constituent of the plasma} are naturally suppressed:
indeed, each such an exchange is of order $1/N_c^2$; this suppression can
be compensated by the large number $\propto N^2_c$ of degrees of freedom
in the plasma (this explains why the potential for
one--graviton--exchange in \eqnum{SchT} is independent of $N_c$), but
this is not possible when the scattering involves only a single
constituent of the plasma. This explains why the plasma structure
functions have no power tail at high $Q^2$. As for the fact that this
tail is exponential, this can be understood as follows: In the previous
subsection, we have argued that the gravitational interactions in the
supergravity problem correspond, in the dual gauge theory, to a constant
force $F_T\sim T^2$ acting on a colored particle within the plasma. Then
the plasma--induced decay of a space--like current with high $Q^2$ can be
understood as a version of the Schwinger mechanism for pair production by
a uniform electric field.

We now turn to the more interesting case at small--$x$, $x\lesssim
x_s=T/Q$, or relatively small virtuality $Q^2\lesssim Q_s^2$, where the
plasma structure functions are significantly large, cf. \eqnum{Flow},
which is suggestive of a parton picture. To develop such a picture, we
need two additional ingredients:

\texttt{(i)} {\em The Breit frame}\\
Recall that the concept of `parton' makes sense only in a frame in which
the plasma has a large longitudinal momentum (an `infinite momentum
frame'). It is convenient to choose the {\em Breit frame}, which is the
frame in which the $\mathcal{R}$--current is a standing wave, with
4--momentum $q^{\prime\mu}=(0,0,0,Q)$. This frame is obtained from the
plasma rest frame by performing a boost in the negative $z$ direction
with a boost factor equal to that of the original current, i.e.,
$\gamma=k/Q$. A typical quanta in the plasma (whatever is its nature) has
energy and momentum of order $T$ in the plasma rest frame, hence it will
have a longitudinal momentum $\sim \gamma T$ in the Breit frame. However
the current is not absorbed directly by such a typical, thermal, quanta,
rather by a partonic constituent of it, which carries only a small
fraction $x=Q^2/(2\omega T)\ll 1$ of its longitudinal momentum; hence,
this parton has a longitudinal momentum $p'_z\sim x(\gamma T)\simeq Q$.
We see that, in this particular frame, the current acts as a probe of the
plasma with both longitudinal and transverse resolutions equal to $Q$.

\texttt{(ii)} {\em The energy sum rule}\\
We have previously mentioned the fact that the spin--2 operator
$T^{\mu\nu}$ receives no anomalous dimension, because of energy--momentum
conservation. This can be used to demonstrate the sum--rule
(\ref{energysum}) for the longitudinal momentum fraction inside a hadron
wavefunction \cite{Peskin}: this sum--rule is proportional to the
expectation value $\langle T^{\mu\nu}\rangle$, and hence it is
independent of $Q^2$. There are similar sum--rules constraining the
plasma structure functions \cite{HIM2}; for instance,
 \beq \mathcal{E}\,= \,18T^2\int_0^1\rmd
 x\,F_2(x,Q^2)\,, \label{SRT} \eeq
where $\mathcal{E}$ is the energy density in the ${\mathcal N}=4$ SYM
plasma at infinite coupling:
\beq \mathcal{E}\,\equiv \,\langle
T^{00}\rangle\,=\,\frac{3\pi^2}{8}\,N^2_cT^4\,.
\eeq
Let us first check that \eqnum{SRT} is indeed verified, at least
parametrically, by our present approximation for $F_2$. Clearly, the
region at (relatively) large $x\gg x_s$ yields only a negligible
contribution to the integral, since $F_2$ is exponentially suppressed
there. Using \eqnum{Flow} for $x\lesssim x_s=T/Q$, one can check that the
integral is in fact dominated by the upper limit $x\simeq x_s$ of this
`small--$x$' region, i.e., by points in the {\em vicinity of the
saturation line}; moreover, these points yield a contribution with the
right order of magnitude: $T^2\,xF_2(x,Q^2)\sim N_c^2 T^4$ for $x\sim
T/Q$ (for any $Q^2$ !). Note that, with increasing $Q^2$, the support of
the structure function $F_2(x,Q^2)$ shrinks to smaller and smaller values
of $x\lesssim T/Q$ (cf.  Fig.~\ref{fig:SatADS} left).

One can furthermore rely on \eqnum{SRT} to deduce a physical
interpretation for $F_2(x,Q^2)$ valid in the Breit frame. In this frame,
the energy density reads $\mathcal{E}'=\gamma^2 \mathcal{E}$ and the
current explores a region of the plasma with longitudinal width $\Delta
z'\sim 1/Q$. Note that $\Delta z'$ is the same as the coherence time of
the current, cf. \eqnum{tcoh}, when measured in the Breit frame. Hence,
the quantity
 \beq \frac{\rmd E'}{\rmd^2b}\,\equiv\,\mathcal{E}'\times
 \Delta z'\,\sim\,\gamma\times\frac{\gamma}{Q}\times \mathcal{E}\eeq
represents the energy per unit transverse area in the region of the
plasma explored by the current. Using ${\gamma}/{Q}\sim 1/(xT)$ together
with \eqnum{SRT}, we deduce
  \beq  \label{89} \frac{\rmd E'}{\rmd^2b}
 \sim xT\gamma \left(\frac{1}{x}\,F_2(x,Q^2)\right)_{x=T/Q}\,.
 \eeq
As before mentioned, the quantity $xT\gamma\sim Q$ in the r.h.s. is the
longitudinal momentum of the constituent (parton) which interacts with
the ${\mathcal R}$--current. It is therefore natural to interpret
 \beq \frac{1}{x}\,F_2(x,Q^2)\Big|_{x=T/Q} \sim
 \frac{\rmd N}{\rmd Y\rmd^2b_\perp}\Big|_{x=T/Q}\,,  \label{90} \eeq
as the {\em number of partons in the plasma per unit area per unit
rapidity} as `seen' by a virtual photon with resolution $Q$. The
occurrence of the rapidity $Y\equiv \ln(1/x)$ can be understood in the
same as way as for the hadron structure functions in Sect. 2.2 (cf.
\eqnum{rapid}): namely, the partons explored by the virtual photon have
longitudinal momentum $p'_z\sim Q$ and occupy a longitudinal distance
$\Delta z'\sim 1/Q$, hence they are distributed within one unit of
rapidity: $\Delta Y= \Delta z'\Delta p'_z\sim 1$.

On the other hand, the AdS/CFT calculation in Sect. 5.1 implies, cf.
\eqnum{Flow},
  \beq
 \frac{1}{x}\,F_2(x,Q^2)\Big|_{x=T/Q}\,\sim \,N_c^2Q^2\qquad\mbox{for}\qquad
 x\sim T/Q\,. \label{91} \eeq
By comparing Eqs.~(\ref{90}) and (\ref{91}), we finally deduce
 \beq\frac{\rmd N}{\rmd Y\rmd^2b_\perp}\,\equiv\,\int^{Q}\!\! \rmd^2k_\perp\
\frac{\rmd N}{\rmd Y \rmd^2b_\perp \rmd^2k_\perp}
 \,\sim \,N_c^2Q^2\qquad\mbox{for}\qquad
 x\sim T/Q\,. \label{nsat}
 \eeq
This result is naturally interpreted as saying that the partons with
$x\sim T/Q$ are distributed in phase--space in such a way that, at all
transverse momenta $k_\perp\lesssim Q$, there is roughly one parton of
each color per unit phase--space. Alternatively, one can say that, for a
given value of $x\ll 1$, the partons occupy the phase--space on the left
of the saturation line, i.e., at $k_\perp\lesssim Q_s(x)=T/x$, with
occupation numbers of $\order{1}$ :
 \beq
 \frac{1}{N^2_c}\, \frac{\rmd N}{\rmd Y \rmd^2b_\perp \rmd^2k_\perp}
 \,\simeq\,1\qquad\mbox{for}\qquad k_\perp\lesssim Q_s(x)\,=\,
 \frac{T}{x}\,. \label{phisat} \eeq
This phase--space distribution is reminiscent of that produced by gluon
saturation in weakly--coupled QCD (cf. Sect. 2.1 and Fig.~\ref{fig:QSAT})
--- the occupation numbers are maximal and uniform (i.e., independent of
$Y$ and $k_\perp$) on the left of the saturation line, and they decrease
rapidly when moving to its right --- but there are also interesting
differences: \texttt{(i)} the occupation numbers at saturation are of
$\order{1}$ at strong coupling, while they were much larger, $\sim
{1/\lambda}\gg 1$, in the perturbative regime at $\lambda\ll 1$;
\texttt{(ii)} in the dilute region at $k_\perp\gg Q_s(x)$ there are
essentially no partons in the strongly--coupled plasma, while in pQCD the
respective occupation numbers decrease rather slowly, roughly like
$(Q_s/k_\perp)^2$ (cf. \eqnum{ngdef}); \texttt{(iii)} for a given,
`hard', resolution $Q^2$, the energy of a hadron in pQCD is carried
mostly by its large--$x$ partons, while at strong coupling this is rather
concentrated in the vicinity of the saturation line, i.e., at small $x$;
\texttt{(iv)} the rise of the saturation momentum with $1/x$ is much
faster at strong coupling than at weak coupling: the saturation exponent
$\omega_s$ (introduced in \eqnum{Qsat}) is estimated as $\omega_s\sim
0.2\div 0.3$ in pQCD, and as $\omega_s=2$ for the strongly--coupled
plasma (cf. \eqnum{QsatT}).

The fact that, at strong coupling, all partons lie at small values of $x$
is in fact quite natural \cite{Polchinski,HIM1}, and can be heuristically
explained via the `quasi--democratic branching' scenario previously
introduced for the ${\mathcal R}$--current (cf. Sect.~4.4). Already at
weak coupling, we noticed in Sect.~2.2 the natural tendency of the parton
evolution to increase the number of partons with small values of $x$. In
that case, however, the evolution was biased towards the emission of
small--$x$ gluons, which carry only a small fraction $x\ll 1$ of the
longitudinal momentum of their parent parton; hence, after emission, the
latter could emerge with a relatively large momentum, which explains why
most of the total energy was still carried by the large--$x$ partons. By
contrast, at strong coupling there is no reason why the energy and
momentum should not be `democratically' divided among the daughter
partons. Then the energy is rapidly degraded along the parton cascade (as
illustrated in Fig.~\ref{fig:SatADS} right), and no partons can survive
at large $x$. The fact that this branching process stops when $x$ becomes
as small as $x_s\sim T/Q$ can be `understood' as a consequence of energy
conservation, \eqnum{SRT}, together with the condition that the
occupation numbers at strong coupling cannot be much larger than one.
However, we have no intuitive explanation for this last condition, except
for the fact that it looks natural.

Note that there is nothing specific to the finite--temperature plasma in
the above argument, and indeed it turns out that a similar partonic
picture holds also for a single {\em hadron at strong coupling}. This was
studied in Refs. \cite{Polchinski,HIM1}, with the `hadron' being a bound
state (a kind of glueball) of the ${\mathcal N}=4$ SYM theory `deformed'
by the introduction of an infrared cutoff $\Lambda$, to mimic
confinement. Via AdS/CFT, this `glueball' is dual to a dilaton state in
supergravity. The respective DIS process is then computed as the
graviton--mediated scattering between the dilaton and the Maxwell wave
induced in $AdS_5$ by the ${\mathcal R}$--current. For sufficiently high
$Q^2$, the inelastic scattering is mainly `diffractive'
--- it proceeds via multiple graviton exchanges ---, and its study requires
going beyond the classical supergravity approximation --- in the sense
that $N_c$ must be kept finite, although large, to allow for multiple
scattering. (The large--$N_c$ limit and the high--energy limit are then
correlated with each other \cite{HIM1}.) The main conclusion in Ref.
\cite{HIM1} is that the hadron wavefunction at strong coupling can be
given a partonic interpretation which is quite similar to that for the
plasma: all partons are concentrated, with occupation numbers of
$\order{1}$, at transverse momenta below the respective saturation
momentum, which reads
 \beq
  Q_s^2(x)\,=\, \frac{\Lambda^2}{xN_c^2} \qquad
  \mbox{(hadron at strong coupling)}\,. \label{QsatD}
  \eeq
This is suppressed in the large $N_c$ limit since so is the scattering
amplitude. (In the case of the plasma, this suppression is compensated by
the number of thermal degrees of freedom, which is proportional to
$N_c^2$.) The fact that $Q_s^2(x)$ rises as $1/x$ is the expected `Regge
behaviour' $\propto 1/x^{j-1}$ for an amplitude mediated by the exchange
of a `particle' with spin $j$ --- here, a graviton with $j=2$. The
corresponding rise appears to be even faster in the case of the plasma,
where we have seen that $Q_s^2(x)= T^2/x^2$ (cf. \eqnum{QsatT}). This
difference can be easily understood: $Q_s^2(x)$ is proportional to the
density of partons per unit transverse area, as obtained after
integrating over the longitudinal extent of the interaction region
(recall, e.g., \eqnum{ngdef}). For a hadron, this longitudinal extent is
simply the hadron width, and is independent of $x$. But for the plasma
this is set by the coherence time of the virtual photon, that is, $\Delta
t_{\rm coh}\sim 1/xT$ (cf. \eqnum{tcoh}); this explains the additional
factor of $1/x$ in \eqnum{QsatT}.

\comment{ \FIGURE[t]{
\centerline{
\includegraphics[width=0.7\textwidth]{ev2_side.eps}}
\caption{\sl A final event as actually seen in a Au-Au collision at RHIC
(here, by the STAR collaboration). }\label{Fig:EVENTRHIC} }}

The above argument also suggests an heuristic way to generalize our
previous results to a {\em plasma with finite longitudinal extent} (a
situation which may be relevant to phenomenology): Namely, so long as
this extent is much larger than the photon coherence time, then
everything proceeds like for an infinite plasma, and the saturation
momentum is given by \eqnum{QsatT}. On the other hand, if the plasma has
a longitudinal width $L_z\ll \Delta t_{\rm coh}$, the corresponding value
for $Q_s^2$ can be obtained by rescaling the result in \eqnum{QsatT} by a
factor $L_z/\Delta t_{\rm coh}\sim xTL_z$. This yields
 \beq\label{QsatTL}
 Q_s^2(x, T,L_z)\,\sim\, \frac{T^3 L_z}{x}\qquad
  \mbox{(plasma with longitudinal extent $L_z\ll 1/xT$)}\,.\eeq
A more detailed argument supporting this conclusion will be presented in
the Appendix.

\FIGURE[t]{
\centerline{
\includegraphics[width=0.9\textwidth]{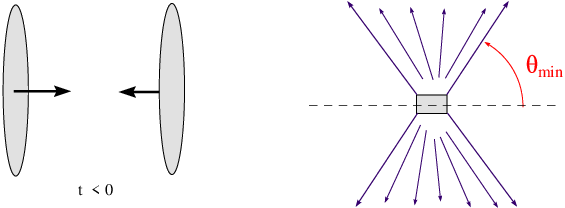}}
\caption{\sl A picture of a hypothetical hadron--hadron collision at
strong coupling: there is no particle production within an angle
$\theta_{\rm min}$ around the collision axis, which is determined by
$x_s$. }\label{Fig:HHSTRONG} }

The peculiar partonic picture has striking consequences for a
(hypothetical) nucleus--nucleus collision at strong coupling. Such a
collision allows us to visualise the partons in the incoming nuclear
wavefunctions: they are first liberated by the collision and then
hadronise on their way towards the detector. Those hadrons originating in
large--$x$ partons have large longitudinal momenta and thus appear in the
detector at either forward, or backward, `rapidities', i.e., at small
angles relative to the collision axis. (Here, by `rapidity' we mean the
space--time rapidity $\eta$ related to the collision angle by $\eta=
-\ln\tan({\theta}/{2})$; for a massless particle, $\eta$ coincides with
the momentum rapidity.) By contrast, the small--$x$ partons give rise to
hadrons which appear at `central rapidities' $\eta\approx 0$, i.e., at
large scattering angles $\theta\simeq \pi/2$. In the actual $AA$
collisions at RHIC, one clearly sees the hadrons being produced at both
forward, and central, rapidities, and the latter are more numerous than
the former\footnote{See, e.g., the image of the final state for a Au+Au
collision at RHIC as recorded by the STAR experiment on
\href{http://www.star.bnl.gov/public/imagelib/collisions2001/ev2_side.jpg}{\tt
http://www.star.bnl.gov/public/imagelib/collisions2001/}.}. This
observation is in agreement with the parton balance in the nuclear
wavefunction as predicted by pQCD (cf. Sect.~2.2). But the situation
would be very different at strong coupling: the absence of large--$x$
partons in the incoming wavefunction would then imply that there is no
particle production at small angles, so the final event would exhibit
`rapidity gaps' $\eta_{\rm gap}(Q)\simeq \ln(1/x_s(Q))$ (for jets with
transverse momentum $Q$) in both forward and backward directions (see
Fig.~\ref{Fig:HHSTRONG}).

\section{Concluding remarks}

The main lesson of these lectures may be summarized as follows: The
physical picture of a plasma as revealed by hard probes and, more
generally, the overall picture of scattering at high energy appear to be
quite different at strong coupling as compared to the respective
predictions of perturbative QCD, and also to the actual experimental
observations. At strong coupling, there are no jets in $e^+e^-$
annihilation, no forward/backward particle production in hadron--hadron
collisions, no partons in the hadron wavefunctions except at very small
$x$. Also, phenomena like the energy loss or the transverse momentum
broadening of a partonic jets travelling into the plasma are controlled
by different mechanisms at strong coupling  --- where, as we have seen,
the dominant mechanism at work is medium--induced parton branching ---,
as compared to weak coupling --- where the momentum broadening is mainly
due to thermal rescattering, and the energy loss to the emission of a
hard gluon (as made possible by thermal rescattering, once again)
\cite{BDMPS,Kovner:2003zj,CasalderreySolana:2007zz}.

\comment{\FIGURE[t] {\centerline{
\includegraphics[width=0.5\textwidth]{Broad_QCD.eps}
\qquad
\includegraphics[width=0.5\textwidth]{HeavyQ_Broad.eps}}
\caption{\sl The dominant mechanism for transverse momentum broadening
for a heavy quark propagating through the plasma is thermal rescattering
(i.e., scattering off the plasma constituents) at weak coupling (left)
but medium--induced parton branching at strong coupling (right).}
\label{fig:Broad} } }

Such differences should not come as a surprise: they reflect the fact
that the corresponding processes involve large momentum transfers, so in
QCD they are naturally controlled by small values of the coupling,
because of asymptotic freedom. Accordingly, much caution should be taken
when trying to extrapolate results from AdS/CFT to QCD in this particular
context of high--energy scattering and hard probes. But this does not
exclude the possibility that long--range processes in the quark--gluon
plasma (so like transport and screening phenomena, or the approach
towards thermalization) be still strongly coupled, precisely because they
involve smaller energies and momenta. This may explain the RHIC data for
elliptic flow which, as explained in the Introduction, are consistent
with a small value for the viscosity--to--entropy ratio, as expected for
a strongly--interacting fluid. For a theory with asymptotic freedom and
confinement, so like QCD, it is natural and necessary to use different
effective theories (or descriptions) at different energy--momentum
scales, as well known from the experience with nuclear theory, chiral
perturbation theory, heavy--quark effective theory, hard thermal loops,
color glass condensate, etc. From this perspective, the gauge/gravity
duality is so far the unique effective theory which allows us to address
long--range and time--dependent phenomena in a QCD--like plasma in the
regime of strong coupling. This method has already produced some very
interesting results, so like the lower bound on the $\eta/s$ ratio
mentioned in the Introduction, and has the potential to explain some
outstanding open questions, so like the rapid thermalization of the
quark--gluon matter observed at RHIC, which seem to transcend
perturbation theory. This is explained in the lecture notes by M. Heller,
R. Janik and R. Peschanski, included in this volume \cite{Pesch}.

But even in the context of hard probes, which has been our main concern
throughout these lectures, the gauge/gravity duality may turn out to be
useful. Some of the observables measured by hard probes (so like jet
quenching) receive contributions from both short--range and long--range
phenomena, and thus combine perturbative and non--perturbative aspects. A
possible strategy to deal with such phenomena, as suggested in Refs.
\cite{Liu:2006ug,Mueller08}, is to distinguish between the respective
`hard' and `soft' momentum contributions, and then use string--inspired
techniques in the soft sector alone, while the hard sector is still
treated in perturbation theory.

\subsection*{Acknowledgments}
I would like to thank the organizers of the 48th Cracow School of
Theoretical Physics {\em Aspects of Duality} for their warm hospitality
at Zakopane during the School and for their patience with my slow writing
of these lectures notes. I am grateful to Al Mueller for suggestions on
the manuscript (in particular, for the argument developed in Appendix)
and to Gr\'egory Giecold for a careful reading. This work is supported in
part by Agence Nationale de la Recherche via the programme
ANR-06-BLAN-0285-01.

\appendix
\section{Saturation momentum for a finite--size plasma}

Towards the end of Sect.~5 we have conjectured a formula, \eqnum{QsatTL},
for the saturation momentum of a plasma whose longitudinal extent $L_z$
is smaller than the coherence length ${\omega}/{Q^2}$ of the incoming
virtual photon. In this Appendix, we shall present an argument\footnote{I
would like to thank Al Mueller for bringing this argument to my
attention.} based on the previous AdS/CFT calculations which supports
this formula. This requires a more precise identification of the physical
force acting on the virtual photon in the strongly coupled plasma, and
thus is a little ambiguous --- it involves subtle differences of physical
interpretation which cannot be fully justified without an explicit
calculation.

As discussed in Sect.~5.3, the physical transverse force exerted by the
plasma can be identified, via the UV/IR correspondence, with the
attraction exerted by the black hole on the Maxwell wave. For
definiteness, we focus on the high-energy case $\omega\gtrsim Q^3/T^2$,
where the virtuality drops out from the potential felt by the Maxwell
wave. Then, as explained after \eqnum{tsat}, the radial fall of the
wave--packet at late times can be described as the motion of a classical
particle in the potential $V\simeq V_c$. The corresponding Newton law is
shown in \eqnum{part}, which can be rewritten as
 \beq
k\, \frac{\dif^2 \chi}{\dif t^2}
 \,=\,2k\,\frac{\chi^3}{\chi_0^4}\,.
 \eeq
Via the UV/IR correspondence $\chi\sim L$, this is naturally interpreted
as a transverse force acting on the virtual photon (or, more precisely,
on its partonic fluctuation) in the plasma:
 \beq\label{FT} F_T\,\equiv\,k \ddot{L}\,\sim\,kL^3T^4 
  \,\sim\,T^3\,\frac{L}{x}\,,
  \eeq
where the last estimate follows after recalling that $L\sim 1/Q$ (by the
uncertainty principle) and $x\sim Q^2/kT$. The emergence of the
longitudinal momentum $k$ as an inertial mass for the transverse dynamics
is natural for a classical particle moving with very high momentum.

\eqnum{FT} is the most general prediction of the present AdS/CFT
calculation for the plasma force. One might interpret the various factors
in this equation as follows: $T^3$ is the density of thermal excitations
in the plasma\footnote{There should be also a factor of $N_c^2$ counting
the number of thermal degrees of freedom, but this is compensated by the
$1/N_c^2$ dependence of the amplitude for graviton exchange, cf.
\eqnum{QsatD}.}, $L$ (the transverse size of the partonic fluctuation)
appears because this is a dipolar force, and $1/x$ reflects the
scattering via graviton exchange. For the case of the {\em infinite}
plasma, \eqnum{FT} is equivalent with the force $F_T\sim T^2$ argued in
Sect.~5.3. For instance, it gives rises to the same estimate for the
saturation momentum, \eqnum{QsatT}, as we show now: In the infinite
plasma, a space--like current interacts with the medium over a time
$\Delta t_{\rm coh} \sim {\omega}/{Q^2}\sim 1/xT$; for the current to
decay, it must receive a mechanical work $F_T\times\Delta t_{\rm coh}$
from the plasma that compensate for its virtuality. This condition
implies
 \beq
 T^3\,\frac{1}{xQ} \times \frac{1}{xT}\,\sim\,Q\ \Longrightarrow\ Q^2\,
 \sim\,Q_s^2(x,T)\,\equiv\,\frac{T^2}{x^2}\,.\eeq
Furthermore, in the high--energy regime, the effective virtuality of the
current is set by the saturation momentum (since the maximal size of its
partonic fluctuation is $1/Q_s\ll 1/Q$). With $L\sim 1/Q_s=x/T$,
\eqnum{FT} yields $F_T\sim T^2$, as anticipated.

On the other hand for a {\em finize--size} medium, with longitudinal
extent $L_z\lesssim \Delta t_{\rm coh}$, the mechanical work is
$F_T\times L_z$, and the saturation condition becomes
 \beq
 T^3\,\frac{1}{xQ} \times L_z\,\sim\,Q\ \Longrightarrow\ Q^2\,
 \sim\,Q_s^2(x,T,L_z)\,\equiv\,\frac{T^3 L_z}{x}\,,\eeq
in agreement with \eqnum{QsatTL}.

What is particularly appealing about \eqnum{FT} is that it exhibits the
$1/x$ rise at high--energy associated with one graviton exchange, whose
appearance is natural in the context of the supergravity calculation, but
which remains a bit mysterious back in the original gauge theory.

\end{document}